\documentclass{aa}

\usepackage[T1]{fontenc}
\usepackage{ae,aecompl}
\usepackage{graphicx}
\usepackage{txfonts}
\usepackage{amsmath}
\usepackage{amssymb}
\usepackage{threeparttable}
\usepackage{sidecap}
\usepackage{xcolor}

\usepackage{appendix}
\usepackage{rotating}
\bibpunct{(}{)}{;}{a}{}{,} % to follow the A&A style
\def \msun{\rm \, M_\odot}
\newcommand\kms{{\,km\,s$^{-1}$}}
\usepackage{multirow}
\usepackage[switch]{lineno}

\usepackage{hyperref}
\usepackage{upgreek}

\hypersetup{
  colorlinks   = true, %Colours links instead of boxes
  urlcolor     = blue, %Colour for external hyperlinks
  linkcolor    = red, %Colour of internal links
  citecolor    = blue %Colour of citations  
}
\DeclareUnicodeCharacter{2212}{-}

\titlerunning{The galaxy group NGC~507: Newly detected AGN remnant plasma}
\authorrunning{M. Brienza, L. Lovisari, K. Rajpurohit et al.}

\begin{document}

%%%%%%%%%%%%%%%%%%%%%%%%%%%%%%%%%%%%%%%%%%%%%%%%

\title{The galaxy group NGC~507: Newly detected AGN remnant plasma transported by sloshing\thanks{Fits files of the radio maps are available at the CDS via anonymous ftp to cdsarc.u-strasbg.fr (130.79.128.5) or via http://cdsweb.u-strasbg.fr/cgi-bin/qcat?J/A+A/}}

\author{M. Brienza\inst{1,2}\fnmsep\thanks{m.brienza@ira.inaf.it},
L. Lovisari\inst{3,4},
K. Rajpurohit\inst{1},
A. Bonafede\inst{1,2},
F. Gastaldello\inst{5},
M. Murgia\inst{6},
F. Vazza\inst{1,2,7},
E.~Bonnassieux\inst{1},
A. Botteon\inst{8},
G. Brunetti\inst{2},
A. Drabent\inst{9}
M. J. Hardcastle\inst{10},
T. Pasini\inst{7},
C.~J. Riseley\inst{1},
H.~J.~A.~R\"ottgering\inst{8},
T. Shimwell\inst{11,8}
A. Simionescu\inst{12,8,13},
R. J. van Weeren\inst{8}
}

\institute{Dipartimento di Fisica e Astronomia, Università di Bologna, via P. Gobetti 93/2, I-40129, Bologna, Italy
\and 
INAF - Istituto di Radioastronomia, Bologna Via Gobetti 101, I-40129 Bologna, Italy
\and
INAF - Osservatorio di Astrofisica e Scienza dello Spazio di Bologna, via Piero Gobetti 93/3, I-40129 Bologna, Italy
\and
Center for Astrophysics | Harvard \& Smithsonian, 60 Garden Street, Cambridge, MA 02138, USA
\and
INAF- Istituto di Astrofisica Spaziale e Fisica Cosmica (IASF) - Milano, Via A. Corti 12, I-20133 Milano, Italy
\and
INAF - Osservatorio Astronomico di Cagliari, Via della Scienza 5 - I-09047 Selargius (CA), Italy
\and
Hamburger Sternwarte, Universit\"at Hamburg, Gojenbergsweg 112, 21029, Hamburg, Germany
\and
Leiden Observatory, Leiden University, PO Box 9513, 2300 RA Leiden, The Netherlands
\and
Th\"uringer Landessternwarte, Sternwarte 5, D-07778 Tautenburg, Germany
\and
Centre for Astrophysics Research, University of Hertfordshire, College Lane, Hatfield AL10 9AB, UK
\and
ASTRON, Netherlands Institute for Radio Astronomy, Oude Hoogeveensedijk 4, 7991 PD, Dwingeloo, The Netherlands
\and
SRON Netherlands Institute for Space Research, Niels Bohrweg 4, 2333 CA Leiden, The Netherlands 
\and
Kavli Institute for the Physics and Mathematics of the Universe (WPI), The University of Tokyo, Kashiwa, Chiba 277-8583, Japan
}

\date{Accepted ---; received ---; in original form \today}

\abstract{
Jets from active galactic nuclei (AGN) are known to recurrently enrich their surrounding medium with mildly relativistic particles and magnetic fields. Here, we present a detailed multi-frequency analysis of the nearby (z=0.01646) galaxy group NGC~507. In particular, we present new high-sensitivity and high-spatial-resolution radio images in the frequency range 144-675\,MHz obtained using Low Frequency Array (LOFAR) and upgraded Giant Metrewave Radio Telescope (uGMRT) observations. These reveal the presence of previously undetected diffuse radio emission with complex, filamentary morphology likely related to a previous outburst of the central galaxy. Based on spectral ageing considerations, we find that the plasma was first injected by the AGN 240-380\,Myr ago and is now cooling. Our analysis of deep archival X-ray Multi-Mirror Mission (\textit{XMM-Newton}) data confirms that the system is dynamically disturbed, as previously suggested. We detect two discontinuities in the X-ray surface-brightness distribution (towards the east and south) tracing a spiral pattern, which we interpret as cold fronts produced by sloshing motions. The remarkable spatial coincidence observed between the newly detected arc-like radio filament and the southern concave X-ray discontinuity strongly suggests that the remnant plasma has been displaced by the sloshing motions on large scales. Overall, NGC~507 represents one of the clearest examples known to date in which a direct interaction between old AGN remnant plasma and the external medium is observed in a galaxy group. Our results are consistent with simulations that suggest that filamentary emission can be created by the cluster or group weather, disrupting AGN lobes and spreading their relativistic content into the surrounding medium. }

\keywords{galaxies : active - radio continuum : galaxy - individual: NGC~507 - galaxies:
clusters: general - intracluster medium - acceleration of particles - X-rays: galaxies: clusters}
\maketitle

\section{Introduction}
\label{intro}

Jets from active galactic nuclei (AGN) are a recurrent phenomenon in the lifetime of a galaxy. This is clearly demonstrated by observations of jetted AGN exhibiting multiple pairs of radio lobes (the so-called ‘restarted radio galaxies’, \citealp{saikia2009, schoenmakers2000, morganti2017, jurlin2021}), as well as by the presence of multiple generations of X-ray cavities in the intracluster and intragroup medium (ICM and IGrM) surrounding the central AGN of galaxy clusters and groups (e.g. \citealp{randall2011, vantyghem2014, biava2021a}). 

During their active phases, jets inject relativistic particles into the surrounding environment, inflating lobes of plasma, which emit synchrotron radiation (characterised by a power law with S $\rm \propto \nu^{-\alpha}$, where $S$ is the flux density and $\alpha$ the spectral index). In this process, jets can significantly influence the thermal history of the galaxy cluster or group they reside in; indeed they can drive shocks and sound waves, induce turbulence, and uplift the coolest, lowest entropy, metal-enriched material out of the central galaxy (e.g.  \citealp{mcnamara2007,fabian2012,mcnamara2012, eckert2021}).

As the jets switch off, the so-called {AGN remnant lobes} expand into the surrounding medium until pressure equilibrium is reached, and radiate their energy away via synchrotron and inverse Compton scattering. In this phase, their spectrum becomes steep ($\rm \alpha >1.2$) and curved \citep{jaffe1973}, and, in the absence of any re-energisation phenomena, after a few tens of millions of years they become invisible even at the lowest observable frequencies (e.g. \citealp{brienza2017, hardcastle2018, english2019}). 

\begin{figure*}[!htp]
\centering
\includegraphics[width=0.65\textwidth]{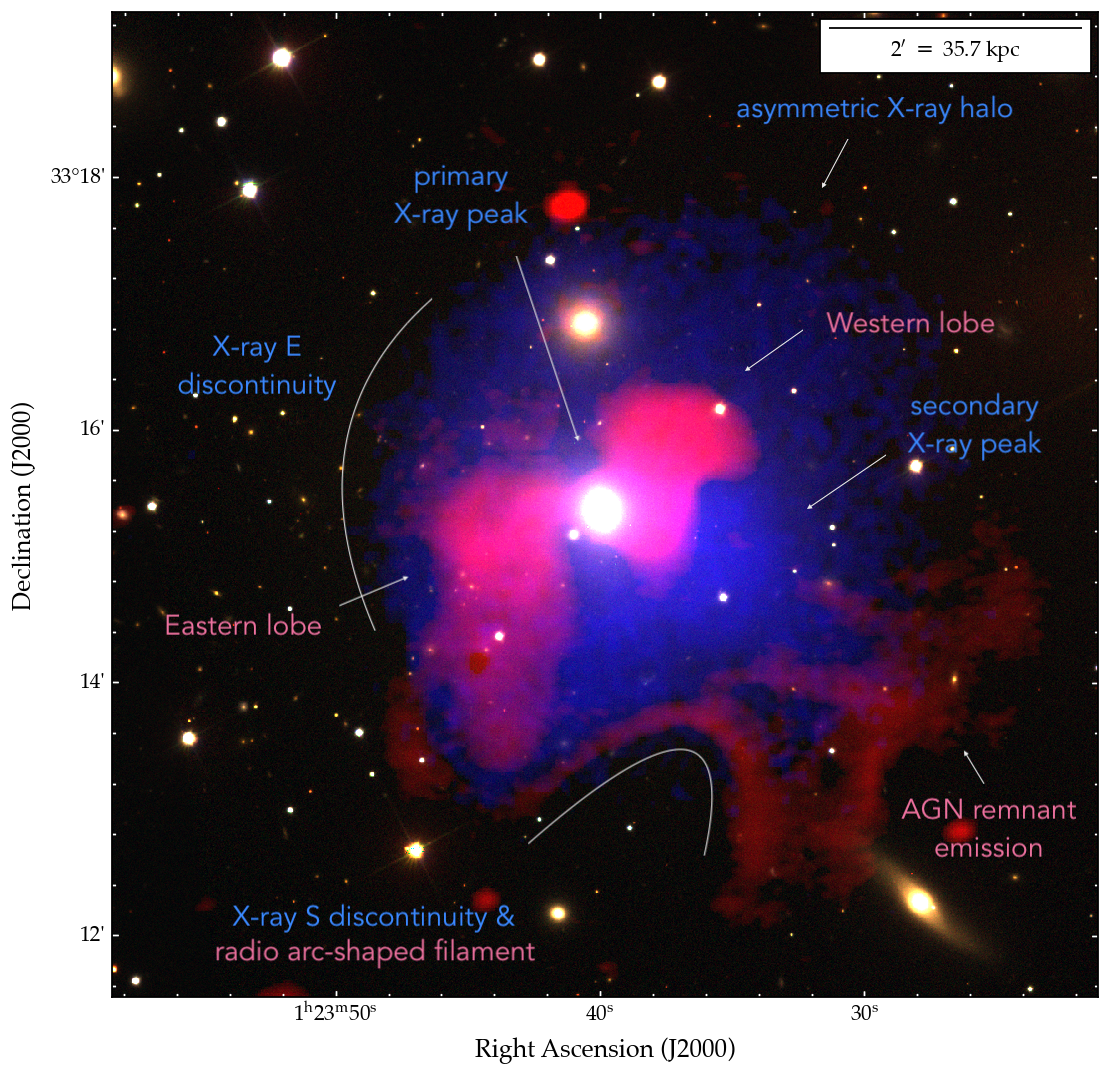}
\caption{Composite image of the galaxy group NGC\,507. Optical emission is shown in background (r-band, g-band, and i-band images from the Sloan Digital Sky Survey, SDSS), radio emission is shown in red (LOFAR image presented in this work at 144\,MHz with 5.57 arcsec $\times$ 8.09 arcsec resolution), and X-ray emission is shown in blue (\textit{XMM-Newton} image at 0.7-2 keV). The most significant features of the system are labelled and a reference scale is shown in the top-right corner.}
\label{fig:rgb}
\end{figure*}

If located in a galaxy group or cluster, the evolution of this remnant AGN plasma can be further affected by the dynamics of the system. For example, any encounters with smaller subclusters might initiate gas sloshing, which is mild, long-lasting (several gigayears) gas oscillations within the gravitational potential of the  system \citep{ascasibar2006, markevitch2007}. These can distort and disrupt AGN radio lobes, spreading their relativistic content around the surrounding environment to up to hundreds of kiloparsecs from the original position \citep{zuhone2013, fabian2021, zuhone2021, vazza2021}. Should the remnant plasma be subjected to a shock, such as those induced by cluster or group mergers, it can also be compressed. This compression leads to a boost in its synchrotron luminosity and a shift of the spectrum towards higher frequencies, allowing for the possibility to detect it again \citep{ensslin2001, ensslin2002, markevitch2007}. These sources are referred to as `phoenices' in the literature, and despite the first discovery two decades ago \citep{slee2001}, they are only now starting to be unveiled in larger numbers thanks to radio observations below a few hundred MHz (e.g. \citealp{nuza2017, kale2018, duchesne2020, manadal2020}).

AGN remnant plasma has also been invoked by many authors to explain the formation of diffuse non-thermal radio sources in galaxy clusters, such as radio mini-halos, halos, and relics (\citealp{vanweeren2019} for a review). These large-scale (up to several megaparsecs) non-thermal sources are not directly associated with individual galaxies but are created when shock waves or turbulence are injected into the ICM by cluster mergers, which are able to (re)-accelerate particles to relativistic energies \citep[e.g.][]{brunetti2014}. In a number of cases, shock acceleration of particles from the cluster thermal pool has been shown to require unrealistically high efficiencies to match the observed radio relic luminosities (e.g. \citealp{kang2016, botteon2020C}). The presence of a pre-existing mildly energised particle population provided by AGN remnant plasma could therefore solve the apparent inconsistency. The same AGN seed particle population is also thought to be essential for the formation of mini-halos in particular \citep{richardlaferrire2020}.

Especially interesting cases are those where a direct interaction between the AGN remnant plasma and the surrounding medium is observed in galaxy groups (e.g. \citealp{gastaldello2013, osullivan2014}). These systems guide us towards a deeper understanding of how the non-thermal plasma gets transported and distributed in the IGrM on long timescales and large spatial scales. While groups are subject to less energetic mergers and interactions than the better-studied massive clusters ---implying that they are less likely to host in situ re-acceleration---, they represent the most abundant systems in the Universe and the building blocks of massive clusters, and for this reason deserve special attention.

In this paper, we present deep, low-frequency radio observations performed with the LOw Frequency ARray (LOFAR, \citealp{vanhaarlem2013}) and the upgraded Giant Metrewave Radio Telescope (uGMRT, \citealp{gupta2017}) of the radio galaxy B2~ 0120+33, associated with the galaxy NGC\,507 located at the centre of the homonymous galaxy group (see Fig. \ref{fig:rgb}). Archival X-ray Multi-Mirror Mission (\textit{XMM-Newton}) X-ray observations are also re-analysed to investigate the interaction between the radio-emitting plasma and the surrounding thermal medium.

The outline of the paper is as follows. In Sect. \ref{sec:ngc507}, we provide an overview of the properties of the galaxy NGC\,507 and its host group. Radio and X-ray observations and data reduction procedures are described in Sect. \ref{sec:data}. In Sect. \ref{sec:results}, we present our radio and X-ray results and discuss their implications for our understanding of the system. A discussion and summary of our findings is finally reported in Sect. \ref{sec:concl}. The cosmology adopted throughout the paper assumes a flat Universe with the following parameters: $\rm H_{0} = 70$ $\rm km$ $\rm s^{-1}$ $\rm Mpc^{-1}$, $\rm \Omega_{\Lambda} =0.7$, $\rm \Omega_{M} =0.3$.

\section{NGC~507: system overview}
\label{sec:ngc507}

\begin{figure}
\centering
\includegraphics[width=0.48\textwidth]{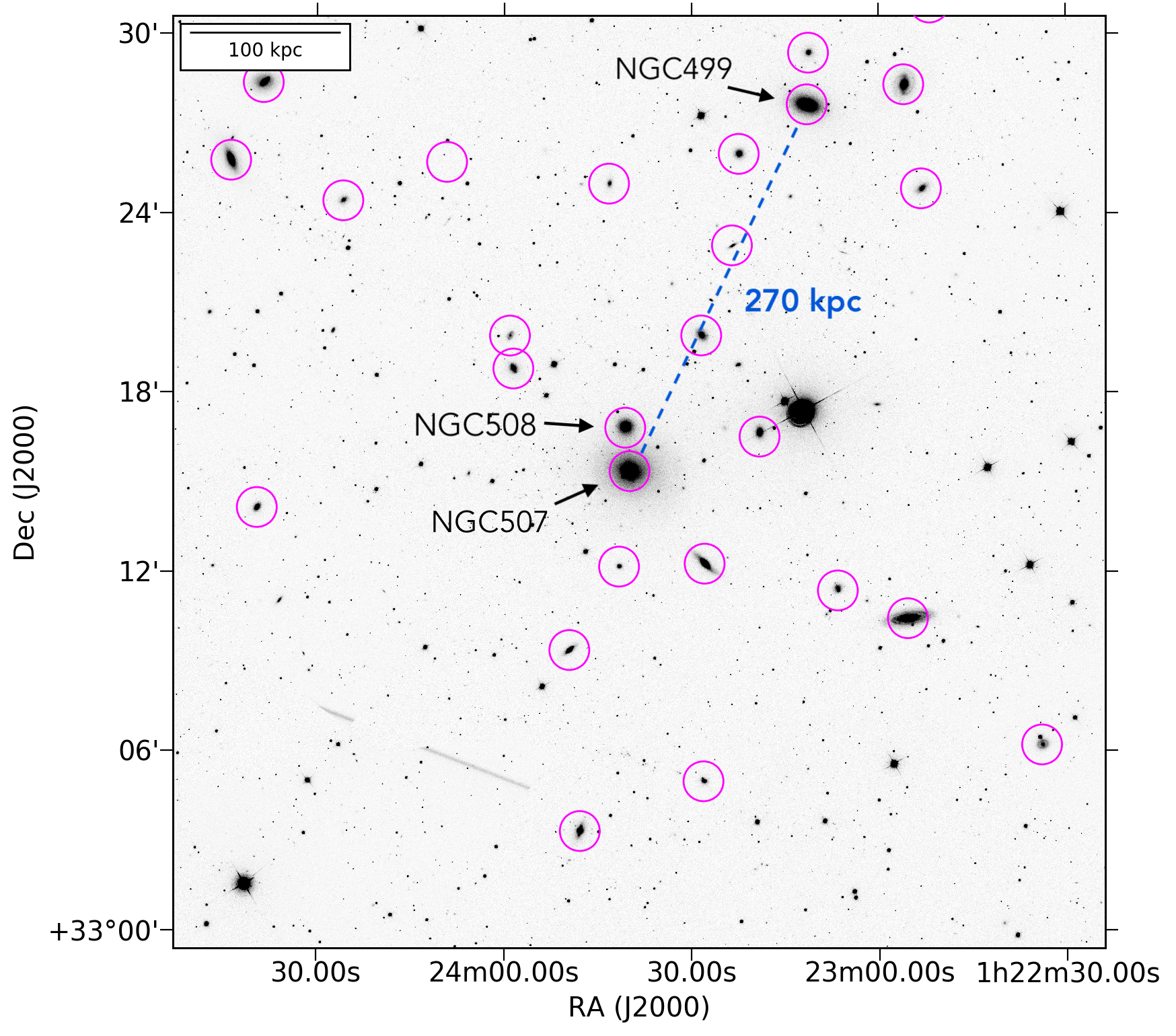}
\caption{\small SDSS r-band image of the galaxy group NGC~507. Magenta circles mark the most central members of the galaxy group. A blue dashed line marks the distance between NGC~507 and the neighbouring group NGC~499.}
\label{fig:507sdss}
\end{figure}

NGC\,507 is a massive elliptical galaxy located at the centre of the nearby ($z$=0.01646, 0.33 kpc/arcsec), optically rich galaxy group (see Fig.\,\ref{fig:507sdss}) also called NGC~507 (or ZwCl\,0107.5+3212), which is part itself of the Pisces supercluster \citep{mulchaey2003, osullivan2003, jeltema2008}. According to \cite{barton1998}, the group consists of 76 galaxies with magnitude $\rm m_{Zw}{<} 16.4$ distributed within a radius of $\sim$1~Mpc from the group centre and has a total velocity dispersion of $658\pm31$~km~$\rm s^{-1}$.

The mass of the system, taken from \cite{piffaretti2011}, is $M_{500}=6.11\times10^{13} \ \msun$ (corresponding to a radius of R$_{500}=596$ kpc). However, this estimate likely includes a second small group called NGC~499, which is located at 13.7 arcmin (270\,kpc) to the northwest (see Fig. \ref{fig:507sdss}) and is probably in the process of merging with NGC~507 \citep{kim2019}.

The galaxy NGC~507 hosts a low-power ($\rm P_{150\,MHz}=1\times 10^{24}\,\rm W Hz^{-1}$)  Fanaroff-Riley I \citep{fanaroff1974} radio galaxy named B2~0120+33 \citep{parma1986, giacintucci2011, murgia2011}. The radio source shows two asymmetric lobes with a total extension of $\sim$70 kpc (see Fig. \ref{fig:rgb}) and a faint, unresolved central component that has been detected at frequencies $\geq610$~MHz with power equal to $\rm P_{core,1400\,MHz}=1.3\times 10^{21}\,\rm W Hz^{-1}$ \citep{giacintucci2011, murgia2011}. The absence of large-scale jets and the pronounced curvature of the integrated radio spectrum of the entire source in the range 150-4850\,MHz (spectral curvature, SPC=1.5; \citealp{murgia2011}) suggest that the lobes represent remnants of a past phase of jet activity. In particular, using spectral ageing models \citep{komissarov1994}, \cite{murgia2011} estimated the total age of the plasma to be in the range $70{-}140$\,Myr and that the jets switched off $\sim$50 Myr ago. In this context, the comparably faint radio nuclear emission could either be related to restarting jets or to a leftover nuclear activity at very low levels. Assuming the lobes to be bubbles of plasma rising in the surrounding medium at the local sound speed, \cite{allen2006} also derived a first-order  age for the source of $\sim$25 Myr.

In the X-rays, the NGC\,507 group is one of the most luminous low-mass systems of the local Universe ($ L_{X}=1.43\rm\times10^{43}\rm \,erg\,s^{-1}$ in the $0.1{−}2.05$~keV band, \citealp{piffaretti2011}) and has been extensively studied in the literature. 
The morphological parameters  (i.e. high-surface-brightness concentration and high centroid shift) determined by \cite{lovisari2019} indicate that, although the group NGC~507 hosts a bright core, the system is not fully relaxed, presenting significant large-scale inhomogeneities. The extremely complex morphology of the diffuse X-ray associated with the galaxy group NGC\,507 has been discussed in several studies (e.g. \citealt{paolillo2003, kim2004, kraft2004, sato2009, kim2019, islam2021}) and consists of two main components: an inner core associated with the stellar distribution of the galaxy, and a bright outer halo, tracing the larger potential associated with the galaxy group. The inner core mostly extends in the southwest direction and shows multiple emission peaks. The most prominent of those peaks is co-spatial with the nucleus of the optical galaxy, while the secondary peaks are likely due to gas compression produced by the expansion of the radio galaxy. In apparent support of this interpretation, the western lobe of the radio galaxy appears to be compressed in that direction and then bent towards the north. Overall, the presence of a central cusp in the X-ray surface brightness profile suggests the presence of a cool core. A tentative detection of X-ray cavities on scales $<10$ kpc ---likely related to the central AGN activity--- has also been reported \citep{paolillo2003,dong2010,osullivan2011,kim2019}.

The outer halo is also disturbed and is more extended towards the north than to the south. This might be the effect of sloshing motions induced by the interaction with the companion NGC\,499 \citep{kim2019}. \cite{paolillo2003} and \cite{kraft2004} also reported two surface-brightness discontinuities in the northeast and southwest directions, respectively. Based on the analysis of \textit{Chandra} data, \cite{kraft2004} exclude the possibility that the northeast discontinuity represents a cold front or that it is supported by an unseen remnant radio lobe from an earlier epoch of galaxy activity located beyond the discontinuity. These authors instead propose that there is low surface brightness radio plasma interior to the discontinuity, expanding south-southeast with a Mach number of $\mathcal{M}<1.2$, which has gently pushed material with higher metallicity from the central regions of NGC\,507 into the group halo, creating an `abundance front'. No detailed analysis of the southwest discontinuity is present in the literature to date.

\section{Observations and data processing}
\label{sec:data}

\subsection{LOFAR 144 MHz}
\label{lofar}

The target has been observed as part of the LOFAR Two-meter Sky Survey (LoTSS; \citealp{shimwell2017, shimwell2019}) using the High Frequency Band Antennas (HBA) at a central frequency of 144\,MHz. 

\begin{table*}[!htp]

 \caption{Summary of the observations used in this work.
 }
\centering
                \begin{tabular}{c c c c}
                \hline
                \hline
                Telescope & Frequency & TOS$^1$ & Date \\
                \hline
                \hline
                LOFAR HBA$^*$ & 120-168 MHz & 24h & 14-23-28/10/2016\\
                uGMRT & 300-500 MHz & 7h & 12/09/2020\\
                uGMRT & 550-950 MHz& 7.2h & 11/09/2020\\
                %Chandra & \mb{?} & \mb{?} & \mb{?}\\
                \textit{XMM-Newton} & 0.3-14 keV & 113ks & 14/07/2013\\

                \hline
                \hline  
                \end{tabular}
                
\begin{tablenotes}

\item \centering $^1$Time On Source. $^*$Observations not centred on the target (see Sect. \ref{lofar} for details). 
\end{tablenotes}
     \label{tab:data}
\end{table*}

\begin{table*}[!htp]

\caption{Summary of the radio images of NGC~507.
 }
\centering

\begin{tabular}{c c c c c}
\hline
\hline
Frequency & Beam & Weighting & UV-taper & RMS \\

$\rm [MHz]$ & [arcsec$\times$arcsec]& & [arcsec] & [mJy/beam]\\
\hline

144  & 8.1$\times$5.6 & Briggs $-0.5$ & - & 0.17 \\
144  & 21.2$\times$19.5 & Briggs $-0.5$  & 20 & 0.35 \\
400  & 6.9$\times$6.9 & Briggs 0.0  & - & 0.036 \\
400  & 22.5$\times$21.3 & Briggs 0.0  & 20 & 0.13\\
675  & 10.8$\times$10.4 & Briggs 0.0   & 10 & 0.04 \\
675  & 21.3$\times$18.8 & Briggs 0.5  & 15 & 0.11 \\
                      
\hline
\hline  

\end{tabular}
\label{tab:imageparam}
\end{table*}

In particular, we used the three 8h datasets related to the LoTSS pointings P019+34, P022+34, and P021+31, whose centres lie at 0.99, 1.71, and 2.18 degrees away from the target position, respectively (the full width at half maximum, FWHM, of the field of view at 144\,MHz being 3.96 degrees using the LOFAR Dutch array in HBA\_DUAL\_INNER configuration).

Following the LoTSS observational setup, the datasets have a frequency coverage of 48\,MHz, in the range 120-168\,MHz and a frequency resolution of 12.2\,kHz. The sampling time was set to 1s and all four correlations were recorded (XX, XY, YX and YY). For each dataset, a primary calibrator was observed for 10 min at the beginning and end of each observing run. Both Dutch and international stations were used in the observations.  However, for this work, only data collected by the Dutch stations (baselines $\lesssim$120\,km) were analysed.

Before being stored in the LOFAR long-term archive, the data were flagged for radio frequency interference (RFI) and then averaged by a factor four in frequency by the observatory. The data were then processed using the standard LoTSS procedures. The {\tt PreFactor} pipeline\footnote{\url{https://github.com/lofar-astron/prefactor}} \citep{vanweeren2016, williams2016} was used to correct the data for direction-independent effects such as ionospheric Faraday rotation, offsets between XX and YY phases, and clock offsets (see \citealt{degasperin2019}). The {\tt DDF}-pipeline\footnote{\url{https://github.com/mhardcastle/ddf-pipeline}} v2.2 was then used to perform a direction-dependent self-calibration to correct for ionospheric distortions and errors in the beam model. This pipeline is described in \cite{tasse2021} and \cite{shimwell2019}, and it uses {\tt kMS} (\citealt{tasse2014} and \citealt{smirnov2015}) to derive direction-dependent calibration solutions and {\tt DDFacet} for imaging with the solutions applied (\citealt{tasse2018}). To further improve the calibration quality and for easier reimaging, all sources outside a square region of $\sim$30-arcmin per side and centred on the target were subtracted from the uv-data, and several loops of self-calibration were performed (see \citealp{vanweeren2020} for a detailed description of the procedure). Baselines below 350$\rm\lambda$ were not considered in the calibration to avoid including any very large-scale structures, which are not related to the target and are difficult to model and calibrate.

\begin{figure*}[!htp]

\centering
\minipage{0.48\textwidth}
\centering
\includegraphics[width=1\textwidth]{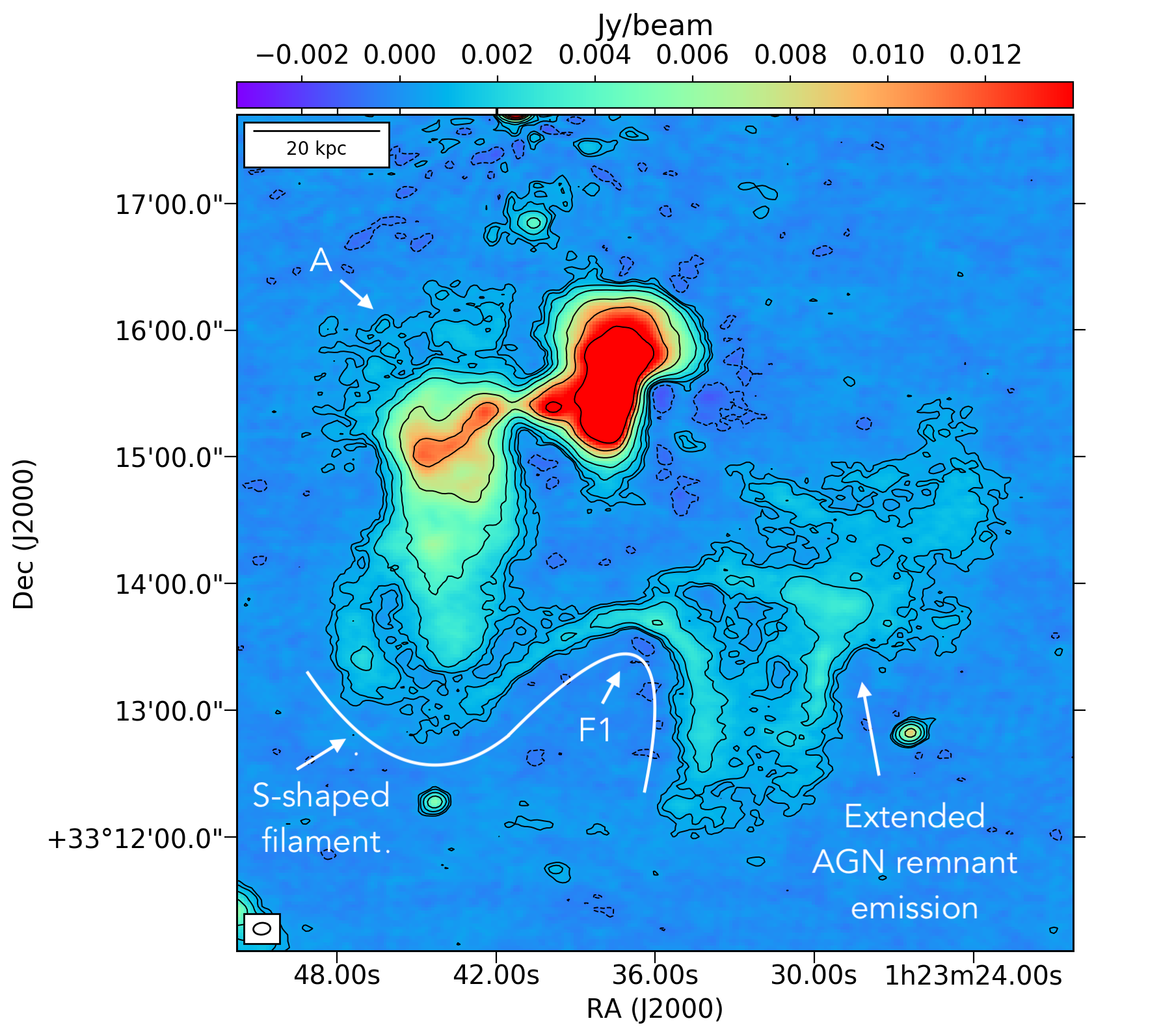}
\endminipage\hfill
\centering
\minipage{0.48\textwidth}
\centering
\includegraphics[width=1\textwidth]{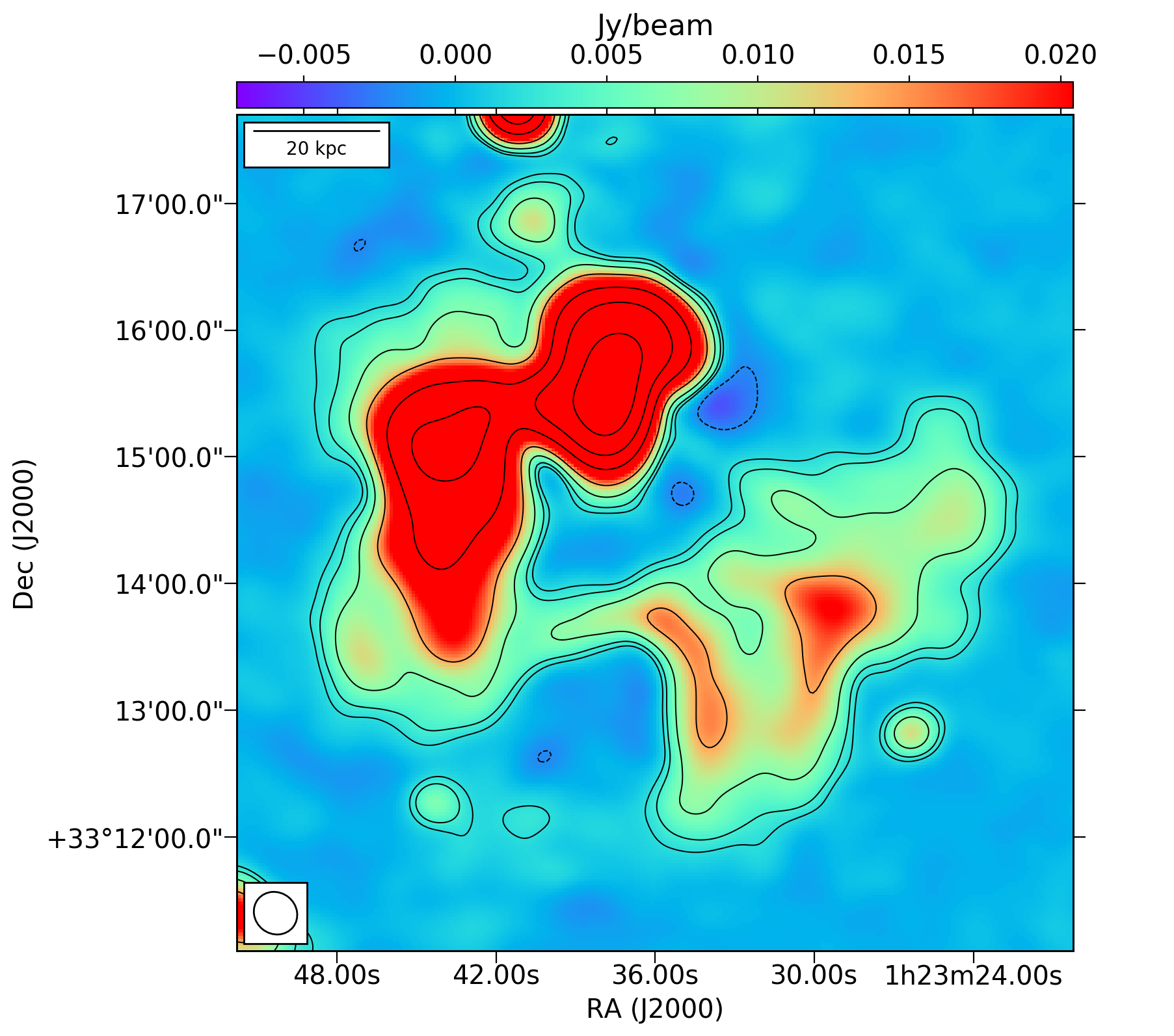}
\endminipage\hfill
\centering
\minipage{0.48\textwidth}
\centering
\includegraphics[width=1\textwidth]{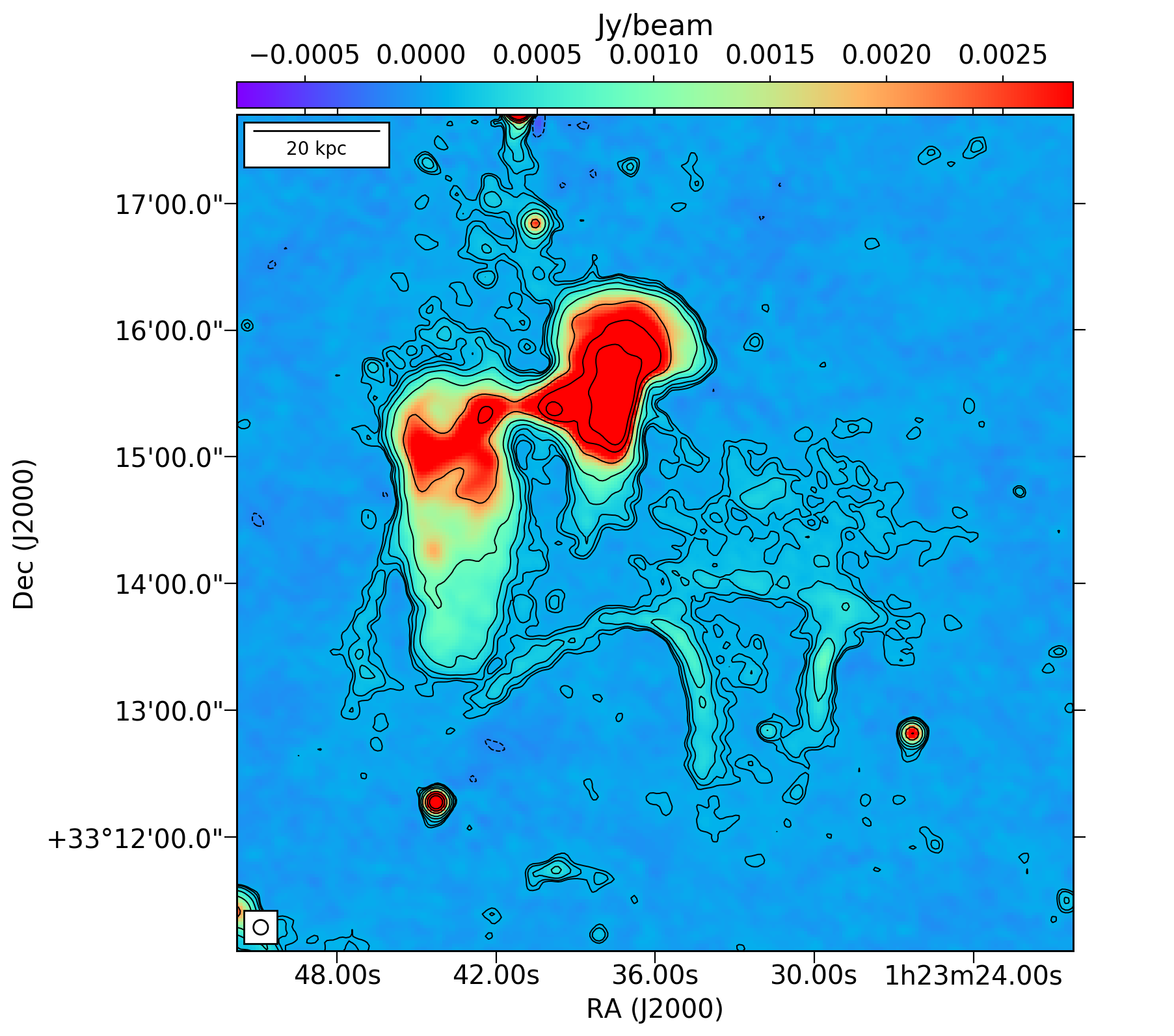}
\endminipage\hfill
\centering
\minipage{0.48\textwidth}
\centering
\includegraphics[width=1\textwidth]{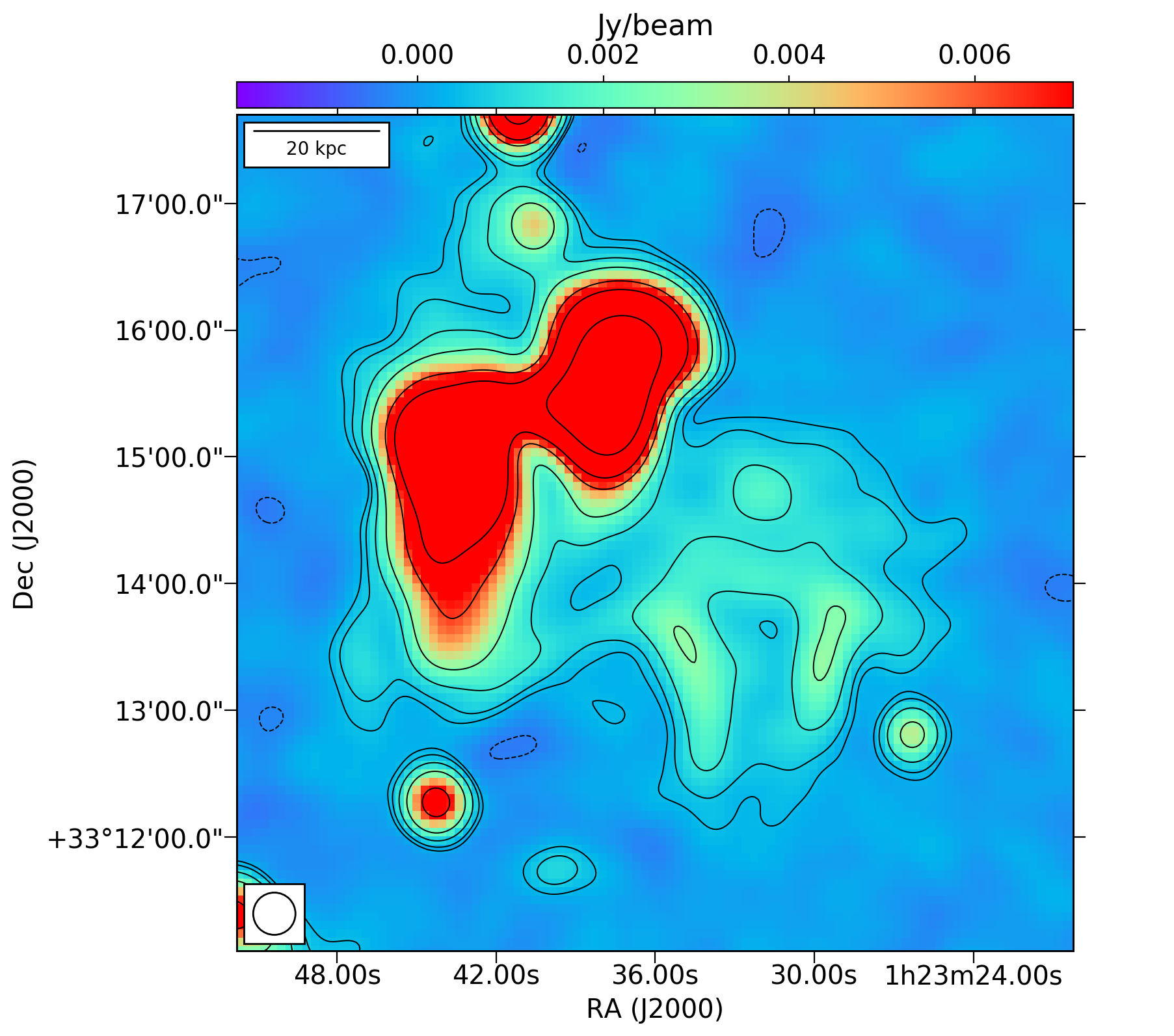}
\endminipage\hfill
\centering
\minipage{0.48\textwidth}
\centering
\includegraphics[width=1\textwidth]{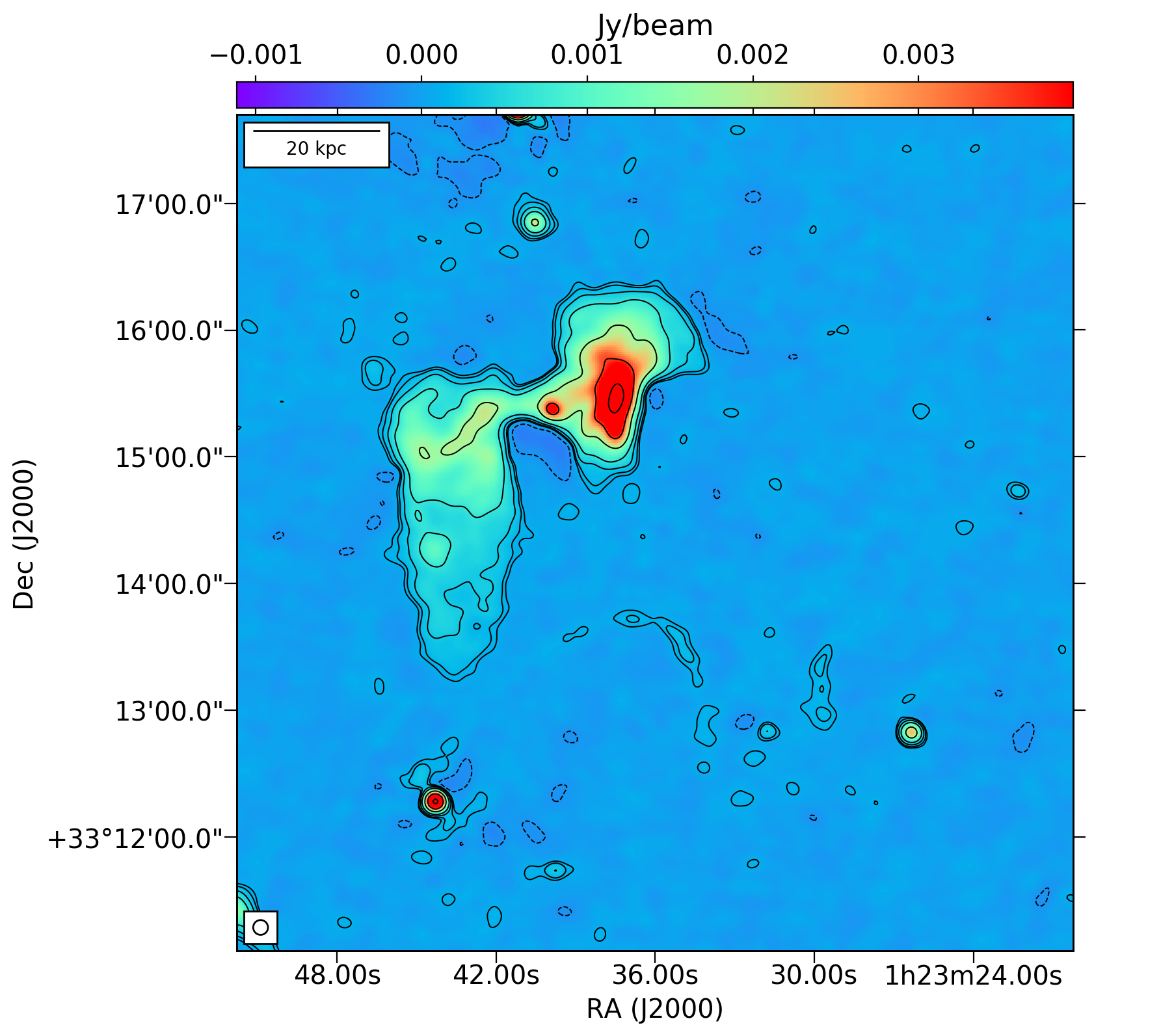}
\endminipage\hfill
\centering
\minipage{0.48\textwidth}
\centering
\includegraphics[width=1\textwidth]{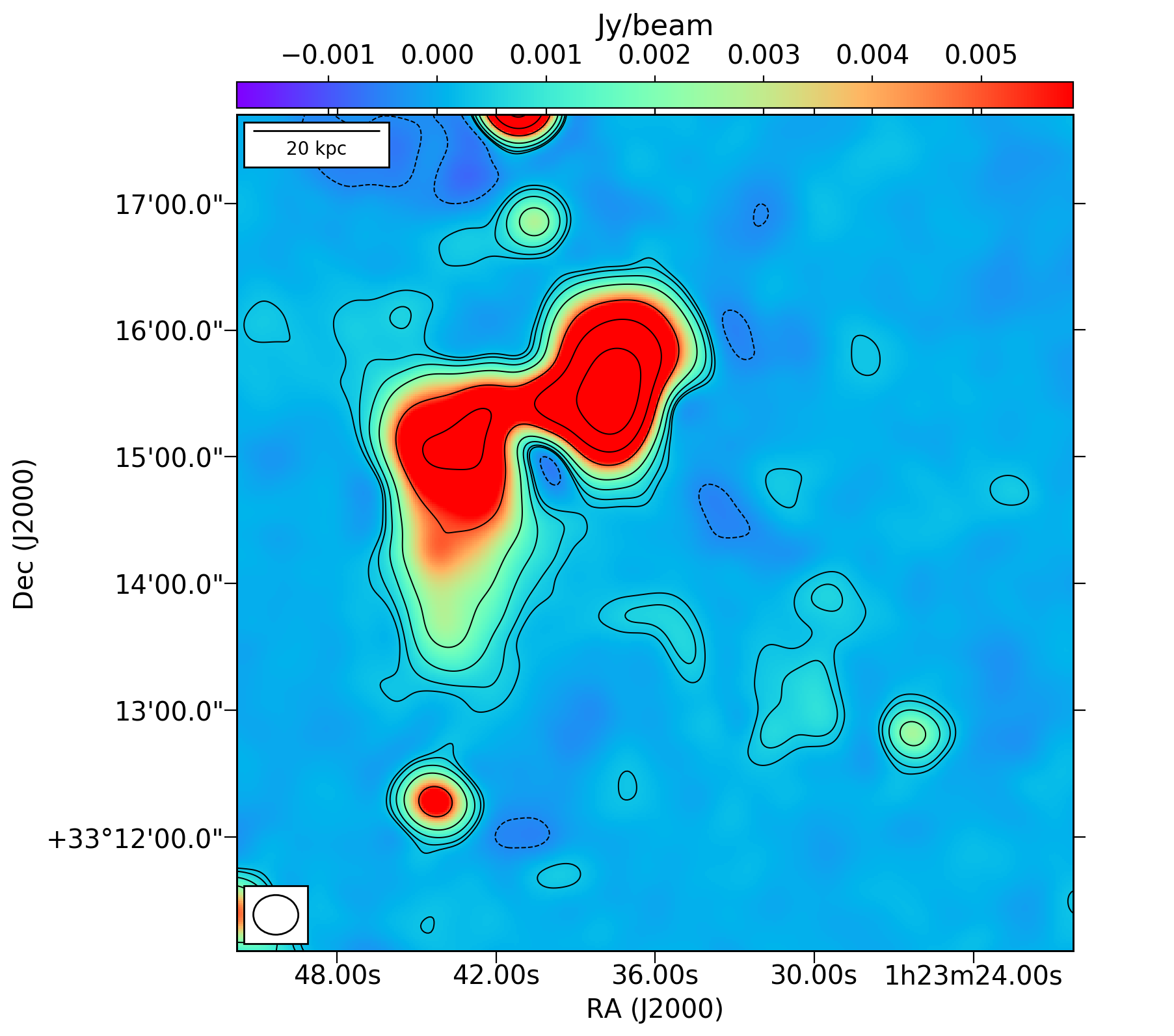}
\endminipage\hfill
\caption{Radio images of the source NGC\,507 at 144\,MHz (top row), 400\,MHz (central row), and 675\,MHz (bottom row). The left column shows images at high resolution and the right column at low resolution, as reported in Table \ref{tab:imageparam}. Contours are drawn at -3, 3, 5, 10, 25, and 40$\sigma$. The beam is shown in the bottom-left corner and a reference physical scale is shown in the top-left corner.}
\label{fig:507maps}
 \end{figure*}

The final images were produced using multi-scale, multi-frequency cleaning in {\tt WSClean} (version 2.8, \citealp{offringa2014}) and are presented in the first row of Fig. \ref{fig:507maps}. The two different sets of imaging parameters were chosen as a compromise between enhancing the large-scale diffuse emission and recovering the small morphological features. In all images, visibilities from baselines shorter than 80$\lambda$ were excluded, again in order to discard very large-scale emission,
which is unrelated to the target and usually challenging to calibrate. The image properties and associated parameters are given in Table\,\ref{tab:imageparam}. 

The flux density scale of all final images was checked by comparing the flux density of the target and the brightest point-like sources in the field, with values published in previous papers and surveys (such as the Very Large Array Low-Frequency Sky Survey, VLSS, \citealp{cohen2007}; the NRAO VLA Sky Survey, NVSS, \citealp{condon1998}) and the NVSS-TIFR GMRT Sky Survey catalogue, \citealp{degasperin2018}). The flux densities measured in our images were found to be consistent with the expectations within the uncertainties.

\subsection{uGMRT 400 MHz and 675 MHz}

We followed up NGC\,507 with the uGMRT during September 2020 in both Band\,3 (300-500\,MHz) and Band\,4 (550-950\,MHz). The observational details are summarised in Table \ref{tab:data}. The total on-source observing time was $\sim$7 hours in both bands. 3C48 was used as flux density calibrator and was observed for 8 minutes at the beginning or end of each observing run. At both frequencies, the total bandwidth was divided into 4096 channels and the integration step was set to 5.3 seconds.

We calibrated the data using the Source Peeling and Atmospheric Modeling pipeline ({\tt SPAM}, \citealp{intema2009}) upgraded for handling new wideband uGMRT\footnote{\url{http://www.intema.nl/doku.php?id=huibintemaspampipeline}} data and we set the absolute flux density scale according to \cite{scaife2012}. Due to severe RFI, all visibilities above 800\,MHz in Band\,4 were discarded. Using the output calibrated data, we created the final images with {\tt WSClean} using multi-scale, multi-frequency cleaning. As for the LOFAR data, we imaged both datasets with two sets of imaging parameters, which are reported in Table\,\ref{tab:imageparam} together with the final image properties. The final images are shown in the second and third rows of Fig. \ref{fig:507maps}. The flux density scale of all final images was checked using the approach
used for the LOFAR images and no flux density corrections were applied.

\begin{figure*}[!htp]
\centering
\includegraphics[width=0.98\textwidth]{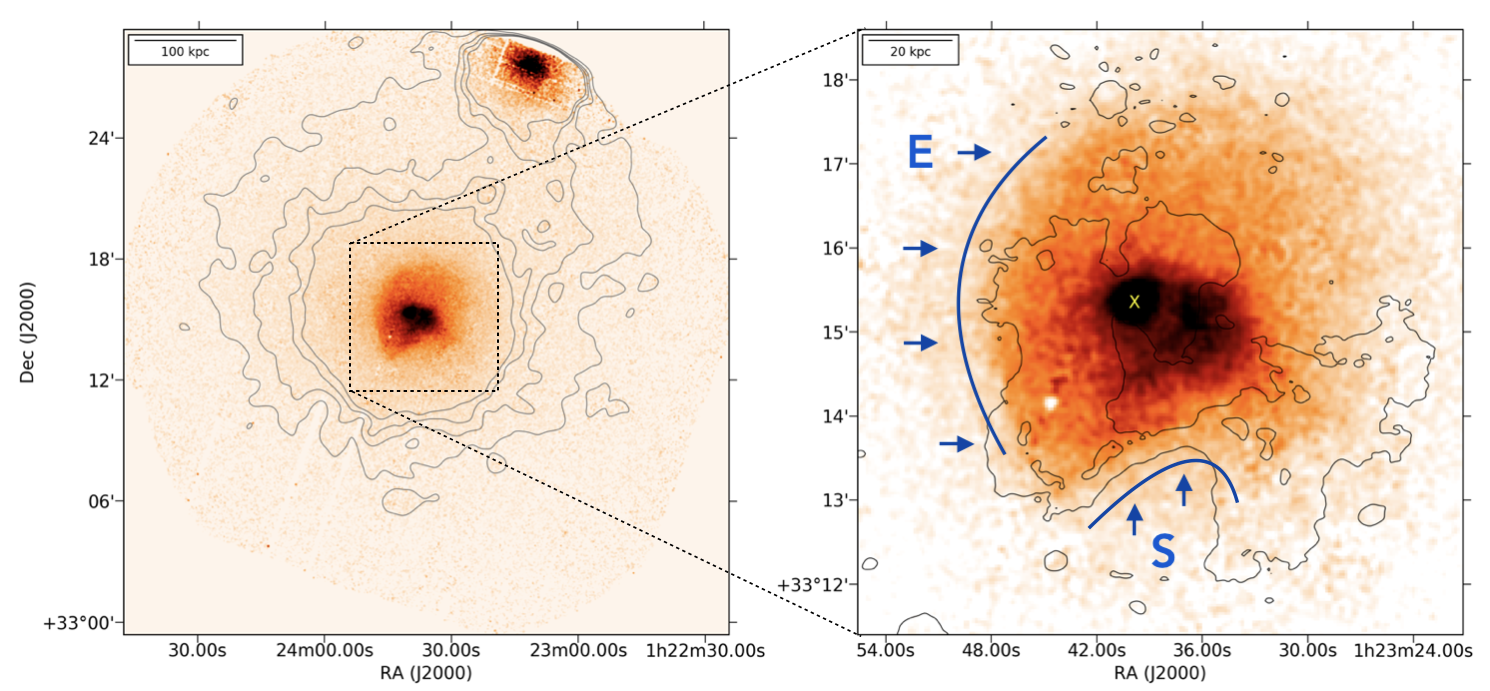}
\caption{\textit{XMM-Newton} 0.7-2\,keV-band images smoothed with a Gaussian  of 6 arcsec FWHM. Left: Full field of view where the neighbouring group NGC~499 is also detected. Grey contours represent the X-ray emission above 5$\sigma$. Right: Zoom-in on NGC~507. A 3$\sigma$ LOFAR contour at 144\,MHz with 5.57 arcsec $\times$ 8.09 arcsec resolution is overlaid in black. The X-ray emission shows a rather complex distribution with two clear discontinuities in the eastern and southern directions, which are marked in blue.}
\label{fig:507xray}
\end{figure*}

\subsection{\textit{XMM-Newton} 0.3-14 keV}
\label{sec:xray}

NGC\,507 was observed twice by {\it XMM-Newton}: once in January 2001 with a total exposure time of 35\,ks (ObsID: 0080540101)
and again in July 2013 (ObsID: 0723800301) with a total exposure time of 113\,ks. It was also observed by {\it Chandra} using the Advanced CCD Imaging Spectrometer I (ACIS-I) configuration in January 2003 for 44\,ks (ObsID 2882) and with the High Resolution Camera (HRC) configuration in November 2018 for 29\,ks (ObsID 21718).

Unfortunately, in the {\it Chandra} ACIS-I image, the southern X-ray surface-brightness jump is located exactly across the CCDs gaps (see \citealp{kraft2004}), preventing any detailed analysis of this region. On the other hand, the HRC observation lacks any spectral information.

We therefore decided to focus exclusively on the {\it XMM-Newton} observations in this work, and in particular on the deepest dataset available collected in 2013. These observations were performed in full frame mode for the MOS cameras and extended full frame mode for the pn detector, all using the medium filter. Overall, given the proximity of the target, the spatial resolution reached by the {\it XMM-Newton} image allows us to make a reliable and detailed analysis of the system in support of our new radio observations as described in Sect. \ref{sec:xray_morpho}.

Observation data files were retrieved from the {\it XMM-Newton} archive and reprocessed with the {\it XMM-Newton} Science Analysis System (SAS) v17.0.0. We used the tasks {\it emchain} and {\it epchain} to generate calibrated event files from raw data. We excluded all the events with PATTERN$>$12 for MOS data and with PATTERN$>$4 for pn data. 
In addition,  events next to CCD edges and bad pixels were excluded (i.e. FLAG==0) for all cameras. We discarded the data corresponding to the periods of high background induced by solar flares using the tasks {\it mos-filter} and {\it pn-filter}. The remaining exposure times after cleaning are 86.1\,ks for MOS1, 85.8\,ks for MOS2, and 60.4\,ks for pn.  Point-like sources were detected using the task {\it edetect-chain} and checked by eye before excluding them from the event files. All the background event files were cleaned by applying the same PATTERN selection, flare-rejection criteria, and point-source removal used for the observation events. 

The X-ray images presented in this work (see Fig. \ref{fig:507xray}) were obtained in the 0.7-2\,keV band using a binning of 40 physical pixels (corresponding to a resolution of 2 arcsec). For visualisation purposes, we refilled the point-source regions using the task {\it dmfilth} in the Chandra Interactive Analysis of Observations (CIAO). The background-subtracted and vignetting-corrected images were then used to determine the surface-brightness (SB) profiles centred on the X-ray peak.

The spectroscopic temperatures were derived by fitting the spectra in the 0.5-12\,keV energy band with an absorbed APEC thermal plasma model (\citealt{Smith2001}) with metallicities from \cite{Asplund2009}, and using C-statistics (i.e. modified Cash statistics; \citealt{Cash1979}) as implemented in XSPEC (\citealt{Arnaud1996}). The absorption was fixed at the total (neutral and molecular, see \citealt{Willingale2013}) value of $ N_H\rm=6.38\times10^{20} cm^{-2}$ estimated using the SWIFT online tool\footnote{\url{https://www.swift.ac.uk/analysis/nhtot/index.php}}. The MOS and pn spectra were fitted simultaneously allowing all normalisations to vary freely. The modelling of the background is rather complicated and we defer the interested reader to \cite{lovisari2019} for a detailed description.

In the very central regions, the spectra could be contaminated by undetected point sources (mostly low-mass X-ray binaries, LMXBs). To test whether this is the case, we included a 7.3 keV bremsstrahlung component to account for the LMXBs emission in the spectrum extracted within 30 arcsec (i.e. where their contribution should dominate). As we did not find any significant effect on the temperature estimates, we neglect this component in our study.

Another possible concern regarding our interpretation of the spectral analysis is related to the simple assumption that all the spectra can be well described by a single temperature component. In the low-mass regime where there is a strong degeneracy between temperature, abundance, normalisation, and column density, it is often necessary to include a second temperature component to avoid strong biases. However, in Appendix \ref{2kT} we show that while this is particularly relevant for the abundance determination, the pattern observed in the temperature distribution is less affected. As defining entropy and pressure (presented in Sect. \ref{sec:xray}) for a multiphase gas is not straightforward, our baseline model to derive temperature and gas densities will be a single APEC, while abundances are instead obtained with a double APEC.

\section{Results}
\label{sec:results}

\subsection{Radio morphology}
\label{sec:radiomorpho}

Figure \ref{fig:507maps} shows our new radio images of NGC\,507 in the frequency range 144-675\,MHz. To emphasise the radio emission on various spatial scales, we present two maps at each frequency, one at high and one at low resolution (see Table \ref{tab:imageparam}). For the first time, the LOFAR data allowed us to image the source at $\sim$8 arcsec at frequencies below 200\,MHz. We also note that our 400-MHz (uGMRT Band\,3) image is more than a factor ten deeper than the image presented by \cite{murgia2011} at 327\,MHz, and our 675-MHz (uGMRT Band\,4) image is about a factor of two deeper than the image presented by \cite{giacintucci2011} at 610\,MHz. 

The central double lobes known from previous observations are clearly recovered at all three frequencies. A compact core is also detected for the first time down to 144 MHz.
Most interestingly, the new data allow us to detect previously hidden features, which are crucial to improve our picture of the dynamical evolution of this system. In particular, at 144\,MHz and 400\,MHz we detect new low-surface-brightness emission with mean surface brightness $\rm SB_{144 MHz}\sim\rm 1\,mJy \,beam^{-1}$. At 675\,MHz this emission is detected only tentatively due to its steep spectral index (see Sect. \ref{sec:spec} and \ref{sec:radiospec}). 

The newly detected emission is connected with the central radio galaxy, suggesting that it might be AGN remnant plasma produced during a previous phase of jet activity. It has a very complex and filamentary morphology, which can be fully appreciated in the LOFAR image at high resolution and in the uGMRT image at 400 MHz (see Fig. \ref{fig:507maps}).

To the north of the eastern lobe, we can clearly see that the plasma is not confined within the lobe; instead it seems to be `leaking' to the northeast (see Fig. \ref{fig:507maps}, component A). This extension is also partially detected at 400\,MHz. In the southeastern region of the same lobe, we also detect  an S-shaped `filament' or `channel' of plasma at both 144\,MHz and 400\,MHz, which encompasses the  southern tip of the lobe and proceeds to the west. The filament in the south (marked as F1) is highly bent and forms a thin arc-like structure with a width of $\sim$2-5 kpc (7-15\,arcsec). Beyond the main filament, a large region of low-surface-brightness, diffuse, extended emission is detected, stretching towards the group outskirts. This 
also seems to be tentatively connected, at least in projection, to the western lobe through some thin filaments. While the filaments and all the substructures are best recovered in the high-resolution images, the full extension of this diffuse emission is best appreciated in the low-resolution images. Its largest linear size is $\sim$ 80\,kpc and its maximum distance from NGC\,507 is also $\sim$80\,kpc. From the low-resolution image at 400\,MHz it is also clear that this new emission is connected to the main double lobes, at least in projection. Unfortunately,  this cannot be appreciated in the LOFAR image at 144 MHz because of imaging artefacts.

Interestingly, we note that \cite{kraft2004} hypothesised the presence of some extra low-surface-brightness radio plasma expanding to the southeast to explain the observed X-ray surface -brightness jump. This was based on a tentative detection of radio emission at 1.4\,GHz from the NVSS and is now clearly confirmed by our new low-frequency observations.

\begin{figure*}[!htp]
\centering
\minipage{0.48\textwidth}
\centering
\includegraphics[width=1\textwidth]{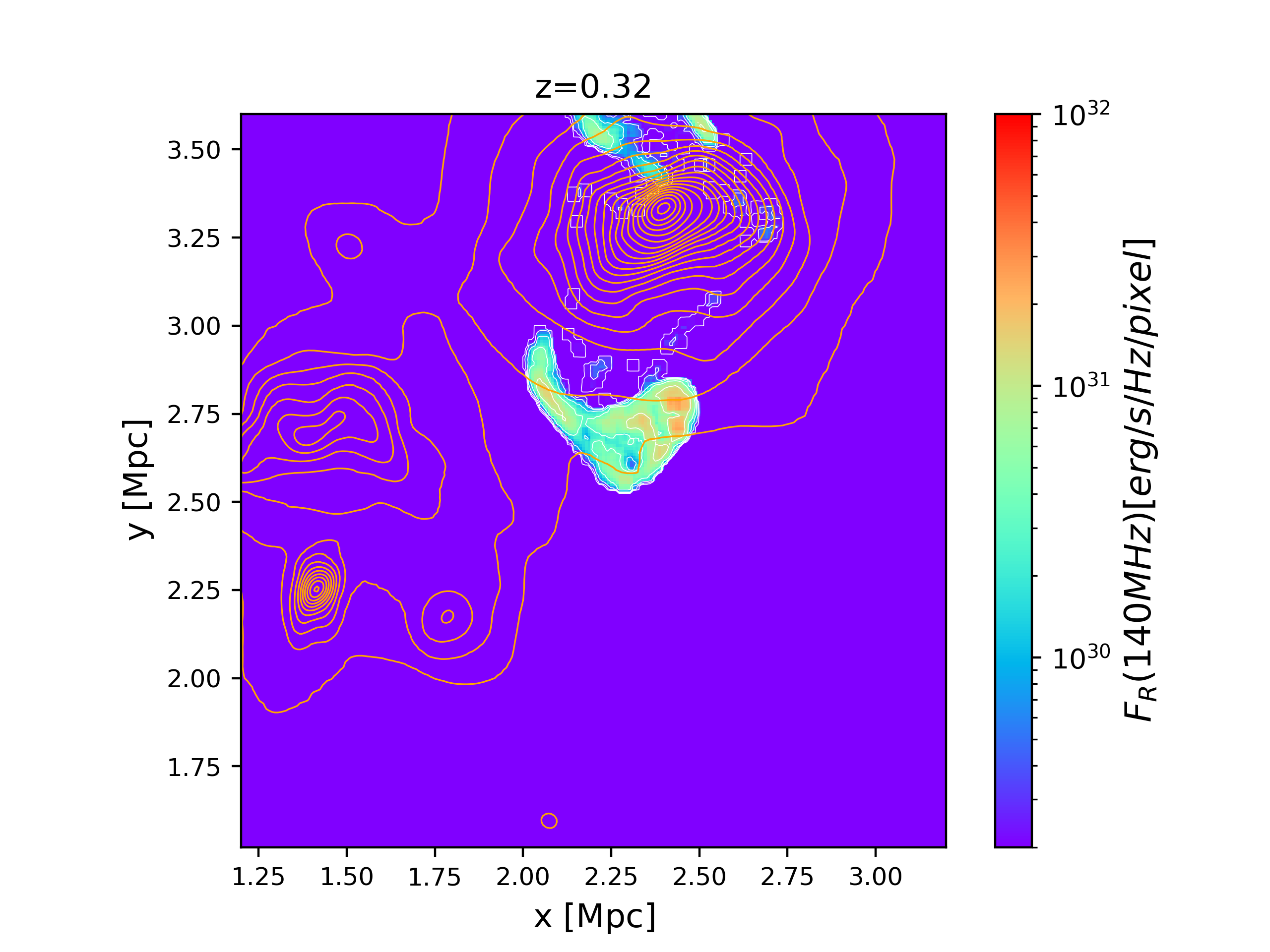}
\endminipage\hfill
\minipage{0.48\textwidth}
\centering
\includegraphics[width=1\textwidth]{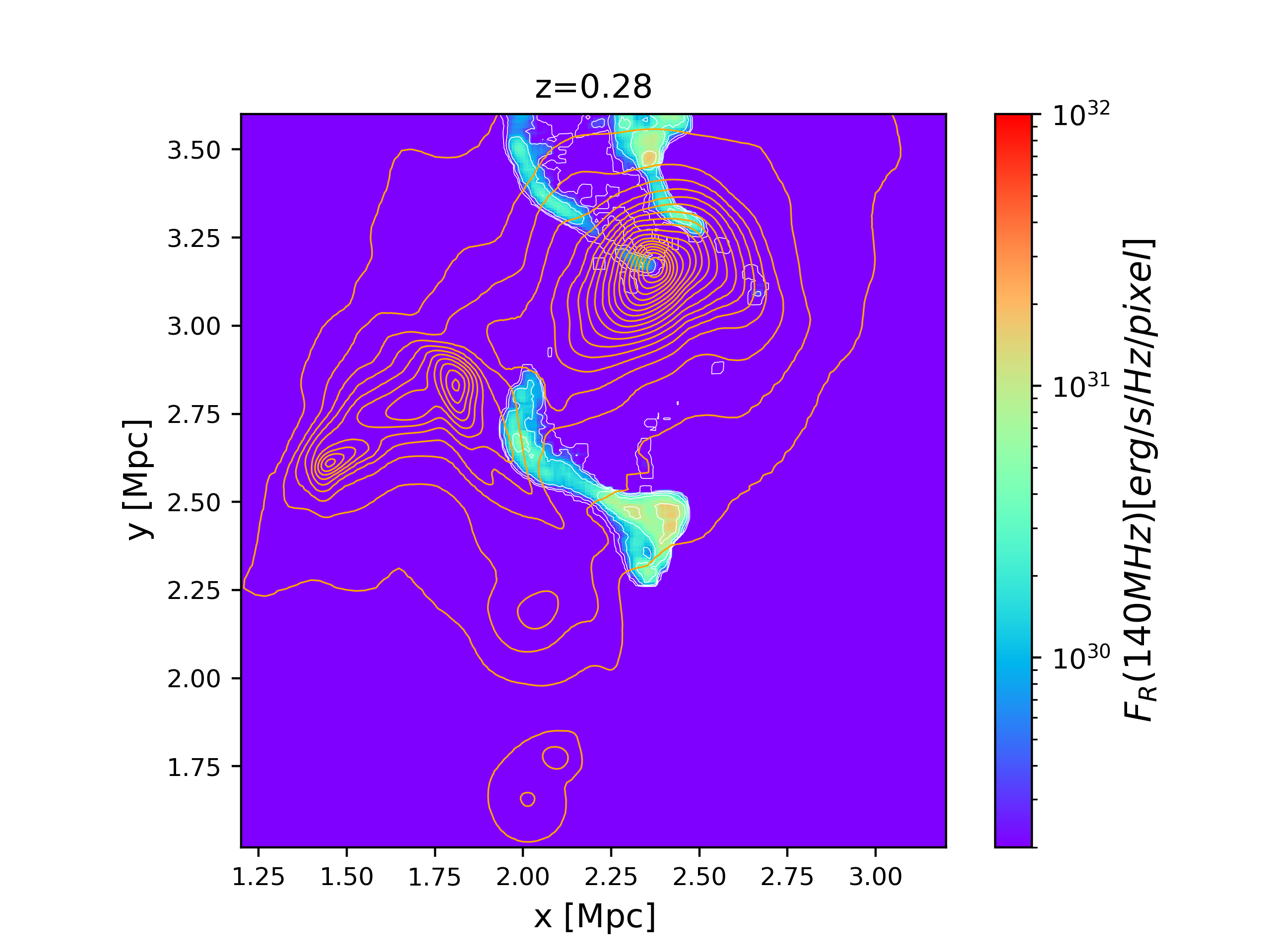}
\endminipage\hfill
\minipage{0.48\textwidth}
\centering
\includegraphics[width=1\textwidth]{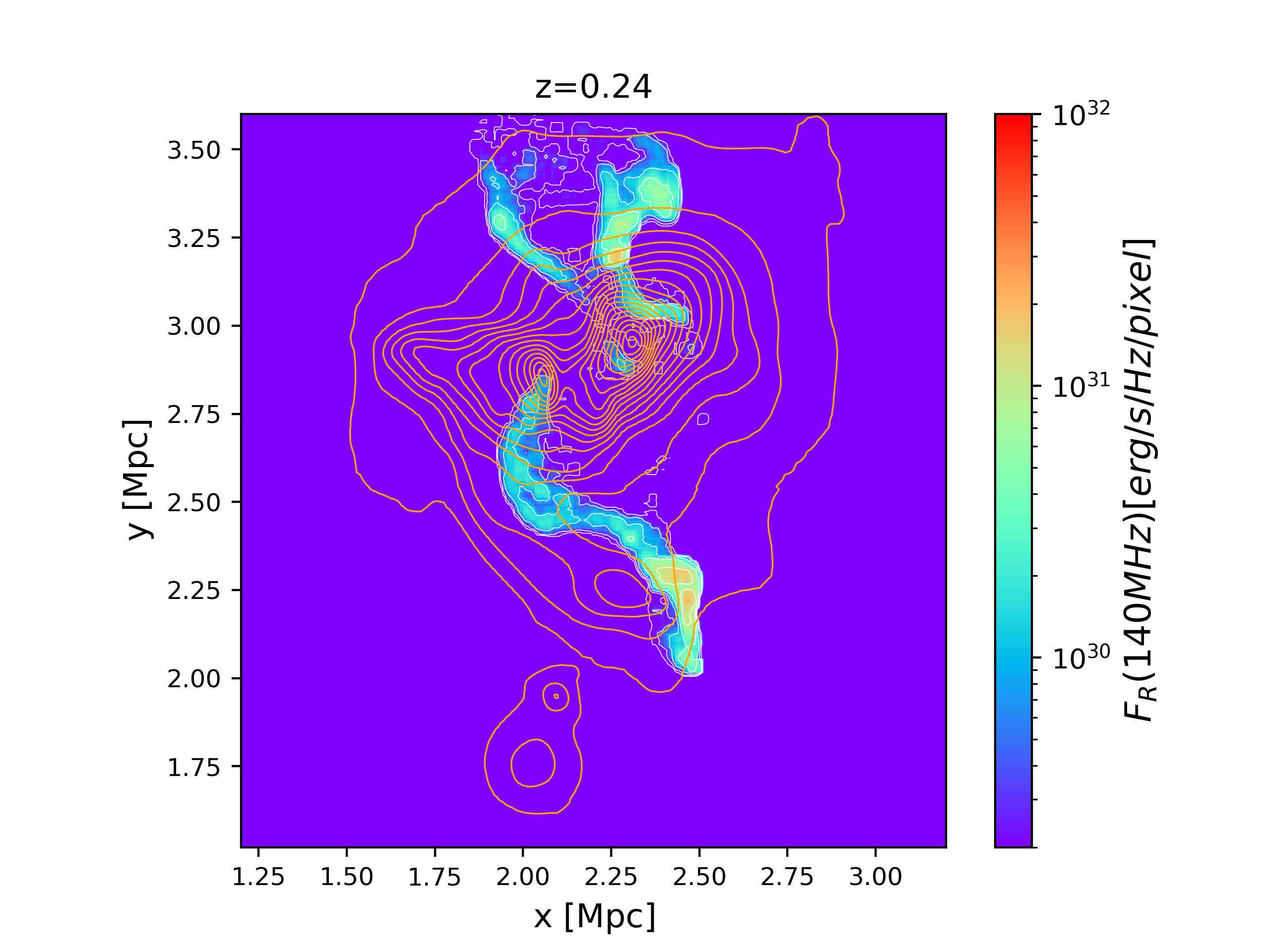}
\endminipage\hfill
\minipage{0.48\textwidth}
\centering
\includegraphics[width=1\textwidth]{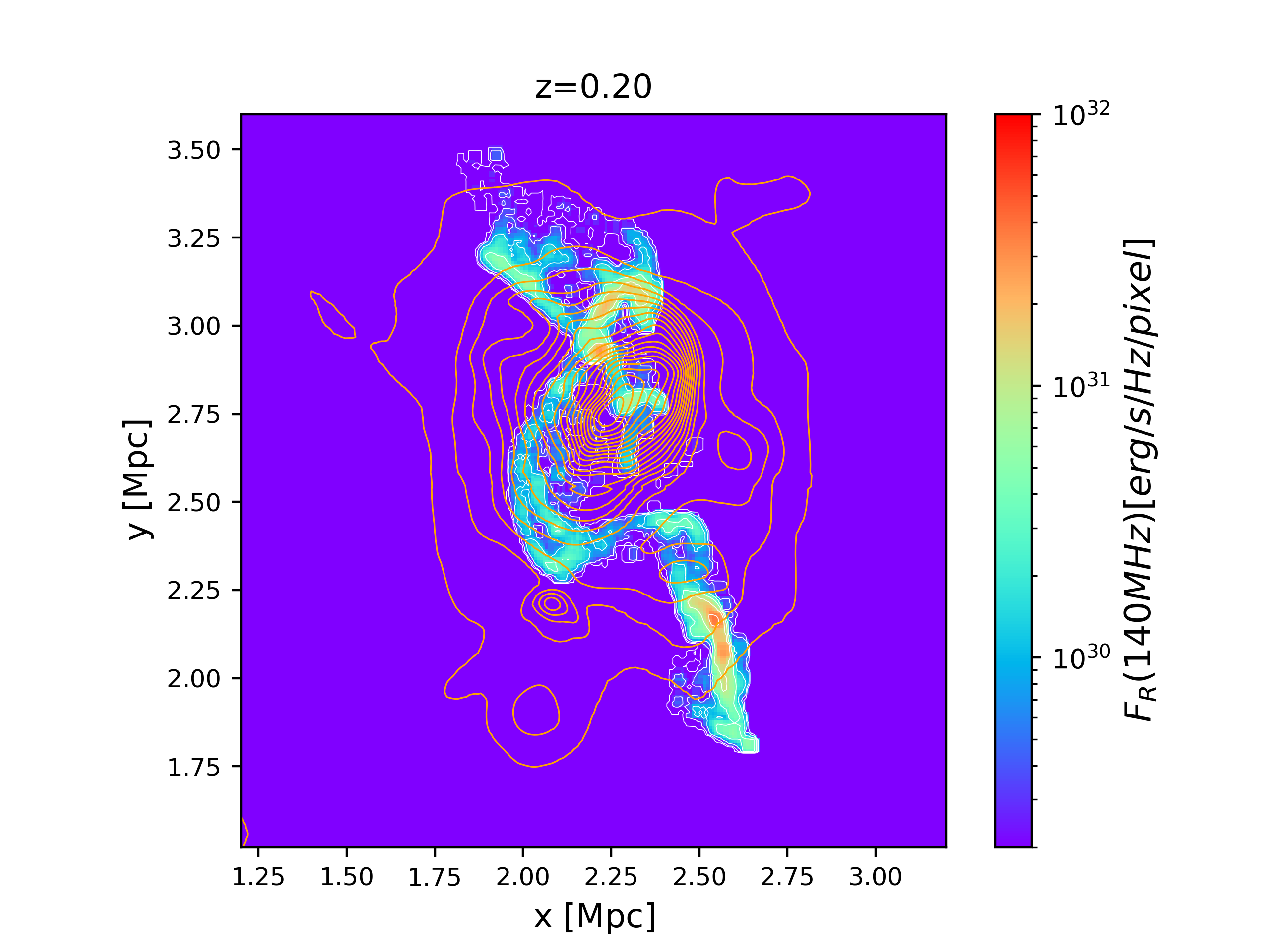}
\endminipage\hfill
\caption{\small Snapshots from a simulation by \cite{vazza2021} (not tailored to exactly reproducing the galaxy group presented in this work) showing the evolution of a cloud of AGN remnant plasma in a cluster environment: the radio emission at 140 MHz is shown overlaid with the contours of X-ray emission. The formation of a radio structure similar to the `F1' source in NGC~507 happens in the lower part of each panel. }
\label{fig:simul}
\end{figure*}

Filamentary emission within non-thermal radio sources is starting to be commonly detected when high-angular-resolution, high-sensitivity observations are available. Examples of these features are found inside the lobes and tails of radio galaxies (see e.g. NGC~1272, \citealp{gendronmarsolais2021}; NGC~326, \citealp{hardcastle2019}; Fornax A \citealp{maccagni2020}; PKS 2014-55, \citealp{cotton2020}), as well as inside cluster radio relics (e.g. \noindent Abell\,2256, \citealp{owen2014}; CIZA\,,J2242.8+5301, \citealp{digennaro2018}; RXS\, J0603.3+4214, \citealp{rajpurohit2018}), and radio phoenices \citep{manadal2020}. In the case of the lobes of radio galaxies, three-dimensional magnetohydrodynamics (MHD) simulations suggest that such filaments trace enhancements in the magnetic field distribution (e.g. \citealp{huarteespinosa2011, hardcastle2013b}), which form as a consequence of the complex internal dynamics of the lobes.

Moreover, in a few cases, collimated threads of emission have been identified outside the main body of the radio galaxy and emerging from the lobes or jets (ESO~137-006 \citealp{ramatsoku2020} and IC~4296 \citealp{condon2021}). The origin of these features is still under debate, but they could possibly be produced by the interaction of the magnetic fields of the radio lobes with the magneto-ionic surrounding medium, or by plasma escaping from jets due to Kelvin-Helmholtz instabilities.

Finally, in some cases, such as the galaxy groups Nest200047 \citep{brienza2021} and Abell 2626 \citep{ignesti2020b}, filaments of radio emission likely represent regions where old AGN remnant plasma accumulates under the influence of buoyancy combined with cluster or group weather. Recent simulations by \cite{vazza2021} indeed show how the magnetised remnant plasma from AGN can be shredded into a multitude of filamentary structures and dispersed over large spatial scales under the influence of the cluster dynamics. These filaments represent regions where particles accumulate following complex velocity fluctuations of the ICM velocity field, as well as the compression by typically weak shock waves.

In Fig. \ref{fig:simul}, for illustrative purposes, we show a few snapshots of one such simulation, where an S-shaped filamentary structure is clearly present in the synthetic radio map, which is based on the distribution of radio-emitting electrons first injected by a central radio galaxy (at $z=0.5$) and were transported by the ICM motions developed by accretion patterns in this group. This simulation is not tailored to exactly reproducing the galaxy group presented in this work but suggests a possible scenario for the formation of the arc-like filament F1 that we detect in NGC\,507 (in the lower half of each image). In the simulation, a similar structure is created as a consequence of the passage of an infalling clump moving from south to north, which creates a velocity field in the ICM that is able to stretch, stir, and compress the remnant plasma which was previously injected by the AGN (visible as a radio lobe in the first panel).

\subsection{Integrated radio spectrum}
\label{sec:spec}

To study the radio spectral properties of the source, we re-imaged all datasets with the {\tt Briggs -0.5} weighting scheme, the same pixel grid, and a final common restoring beam equal to 7 arcsec $\times$ 7 arcsec and 20 arcsec $\times$ 20 arcsec for the high and low resolution images, respectively. To make sure we recover the flux density on the same maximum spatial scales at all observed frequencies, we excluded baselines below 300$\rm \lambda$ (i.e. angular scales ${>}12$ arcmin). This cut corresponds to the shortest well-sampled baseline of the uGMRT Band-4 dataset. The size of the entire source emission is $\sim$5 arcmin, meaning that this cut is not expected to exclude useful emission from the images at any frequency.

\begin{table*}[!htp]
\caption{Flux densities of NGC\,507 and its main morphological features. The flux densities of the lobes and the filament F1 were measured using the high-resolution images described at the beginning of Sect. \ref{sec:spec}, all other values were obtained from the low-resolution images.}
\centering
\begin{tabular}{c c c c c c c}
\hline
\hline
 Region & S$\rm_{144\,MHz}$ & S$\rm_{400\,MHz}$ & S$\rm_{675\,MHz}$ & $\rm \alpha_{144\,MHz}^{400\,MHz}$& $\rm \alpha_{144\,MHz}^{675\,MHz}$ & $\rm \alpha_{400\,MHz}^{675\,MHz}$ \\
& [mJy] & [mJy] & [mJy] & & &\\
\hline
Total & 2400$\pm$350   & 630$\pm$60 & 280$\pm$30  & 1.30$\pm$0.15 & 1.40$\pm$0.10 & 1.55$\pm$0.20 \\
Lobes & 1900$\pm$300   & 500$\pm$50 & 270$\pm$30  & 1.30$\pm$0.15 & 1.30$\pm$0.10 & 1.20$\pm$0.20\\
Extended & 430$\pm$65  & 60$\pm$5   & 7$\pm$1     & 2.00$\pm$0.20 & 2.70$\pm$0.10 & 4.10$\pm$0.40\\
F1    & 95$\pm$15      & 10$\pm$1   & 4.0$\pm$0.5 & 2.15$\pm$0.15 & 2.15$\pm$0.15 & 1.80$\pm$0.30 \\
\hline
\hline  
\end{tabular}
\label{tab:flux}
\end{table*}

\begin{figure*}
\centering
\includegraphics[width=1\textwidth]{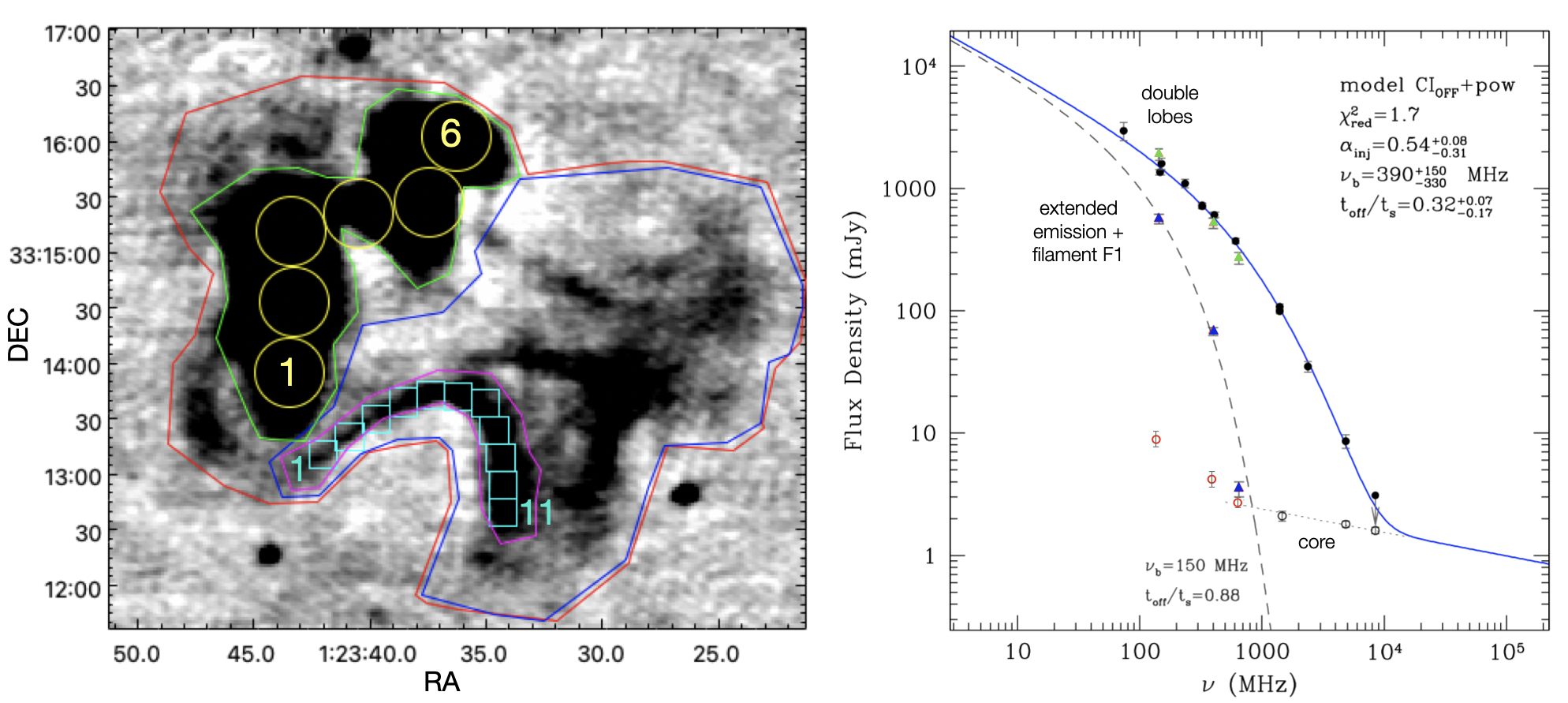}
\caption{\textit{Left panel:} LOFAR 144-MHz image with overlaid all boxes used to measure the flux densities as listed in Table \ref{tab:flux}. The red and blue boxes have been used to measure the flux density of the entire source and the newly detected extended emission, respectively, using the 20-arcsec resolution images. The green and magenta boxes were used to measure the flux density of the double lobes and the filament F1, respectively, using the 7-arcsec resolution images. Yellow circles (1-6) and cyan boxes (1-11) were used to compute the spectral profiles shown in Fig. \ref{fig:specprofile} and described in Sect. \ref{sec:radiospec}. \textit{Right panel:} Radio spectrum of the radio galaxy NGC\,507. Green filled triangles show the double-lobe flux density measurements presented in this paper (green box in the left panel and see Table \ref{tab:flux}), while black circles show the flux densities listed by \cite{murgia2011} and the value reported in the NTI catalogue at 150 MHz. The blue solid line represents the best spectral age fit of the main lobes, whose parameters are listed in the top-right corner and further described in Sect. \ref{fig:intspec}. Blue filled triangles show the flux density of the newly detected extended emission (blue box in the left panel) and the grey dashed line shows the respective best spectral age fit, whose parameters are listed in the bottom of the panel and further described in Sect. \ref{fig:intspec}. The core flux density measurements are shown as empty circles (in red is the measurement at 675 MHz presented in this work and in black are those published by \citealp{murgia2011}) and are fitted with a power-law function as shown by the dotted line.}
\label{fig:intspec}
\end{figure*}

The total flux densities of the entire source and its most significant morphological features were measured using the regions presented in Fig. \ref{fig:intspec} and are reported in Table \ref{tab:flux} together with their integrated spectral indices. The total errors on the flux densities have been computed 
by combining the flux density scale errors (assumed to be 15\% for the LOFAR data and 10\% for the uGMRT data) in quadrature with the image noise multiplied by the integration area. The errors on the spectral indices were obtained using the following formula:

\begin{equation}
\rm \alpha_{err} = \frac{1}{ln\frac{A}{B}}\sqrt{\left(\frac{\Delta S_{A}}{S_{A}}\right)^2+\left(\frac{\Delta S_{B}}{S_{B}}\right)^2}
\label{eq:err}
,\end{equation}

\noindent where $\rm {S_{A}}$ and $\rm {S_{B}}$ are the flux density values at the respective frequencies A and B, and $\rm \Delta S_{A}$ and $\rm \Delta S_{B}$ are their corresponding errors.

The integrated radio spectrum of NGC\,507 is presented in Fig.\,\ref{fig:intspec}. In particular, we show the spectrum of the main double lobes and the spectrum of the newly detected extended emission (including the filament F1), separately. Our new flux density measurements at 144\,MHz, 400\,MHz, and 675\,MHz are shown as green and blue triangles, respectively, while black circles represent all literature values presented in \cite{murgia2011} and the value reported in NTI at 150 MHz.

As visible in Fig. \ref{fig:intspec}, our new flux density measurements agree with previously published values within errors. We note that the flux density scale of \cite{scaife2012} used to calibrate our new data and the NTI catalogue is consistent within $\sim$2\% with the scale by \cite{baars1977} used as a reference for all other measurements (see \citealp{perley2017}). 

\begin{figure*}[!htp]

\centering
\minipage{0.5\textwidth}
\centering
\includegraphics[width=1\textwidth]{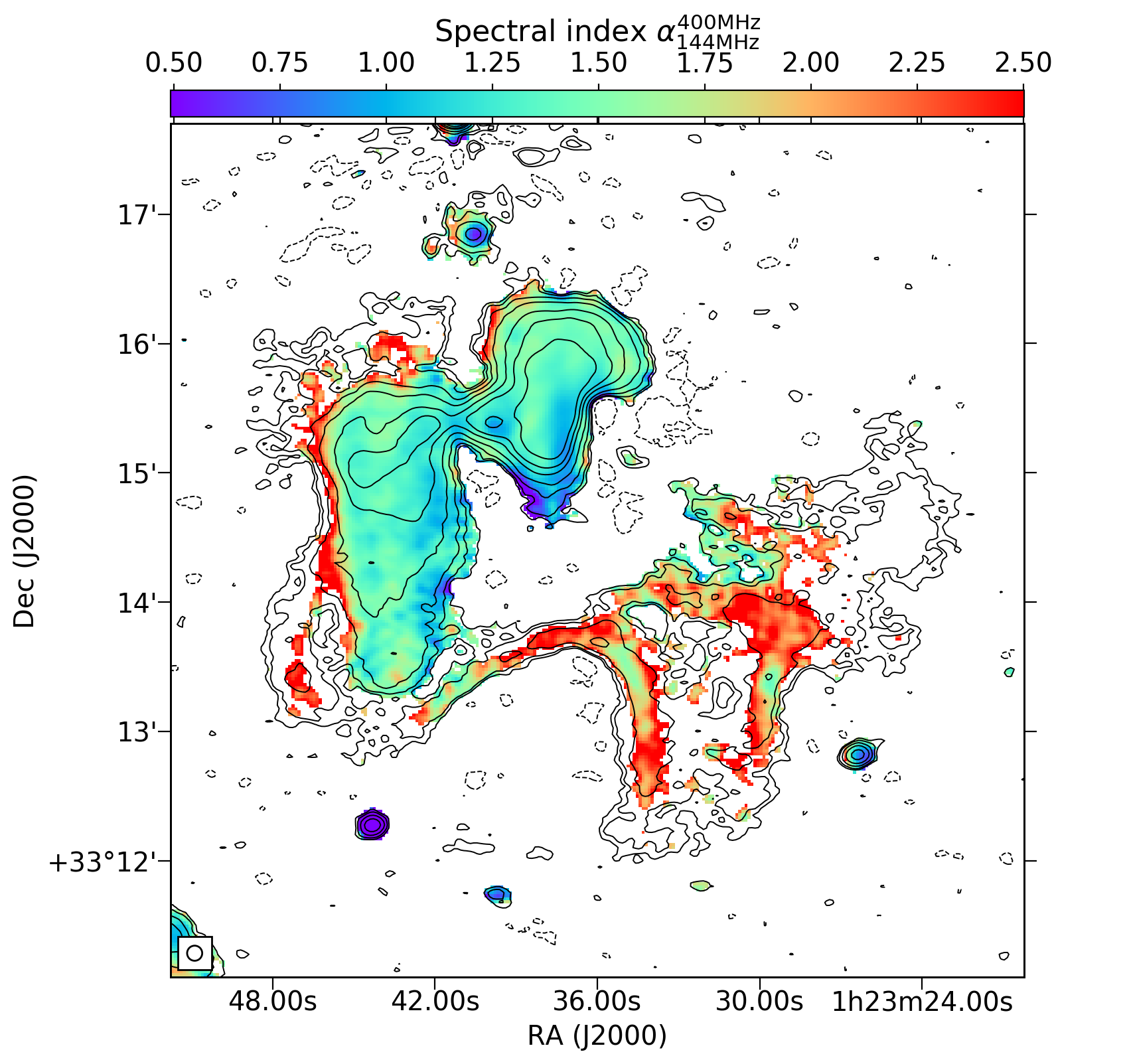}
\endminipage\hfill
\centering
\minipage{0.5\textwidth}
\centering
\includegraphics[width=1\textwidth]{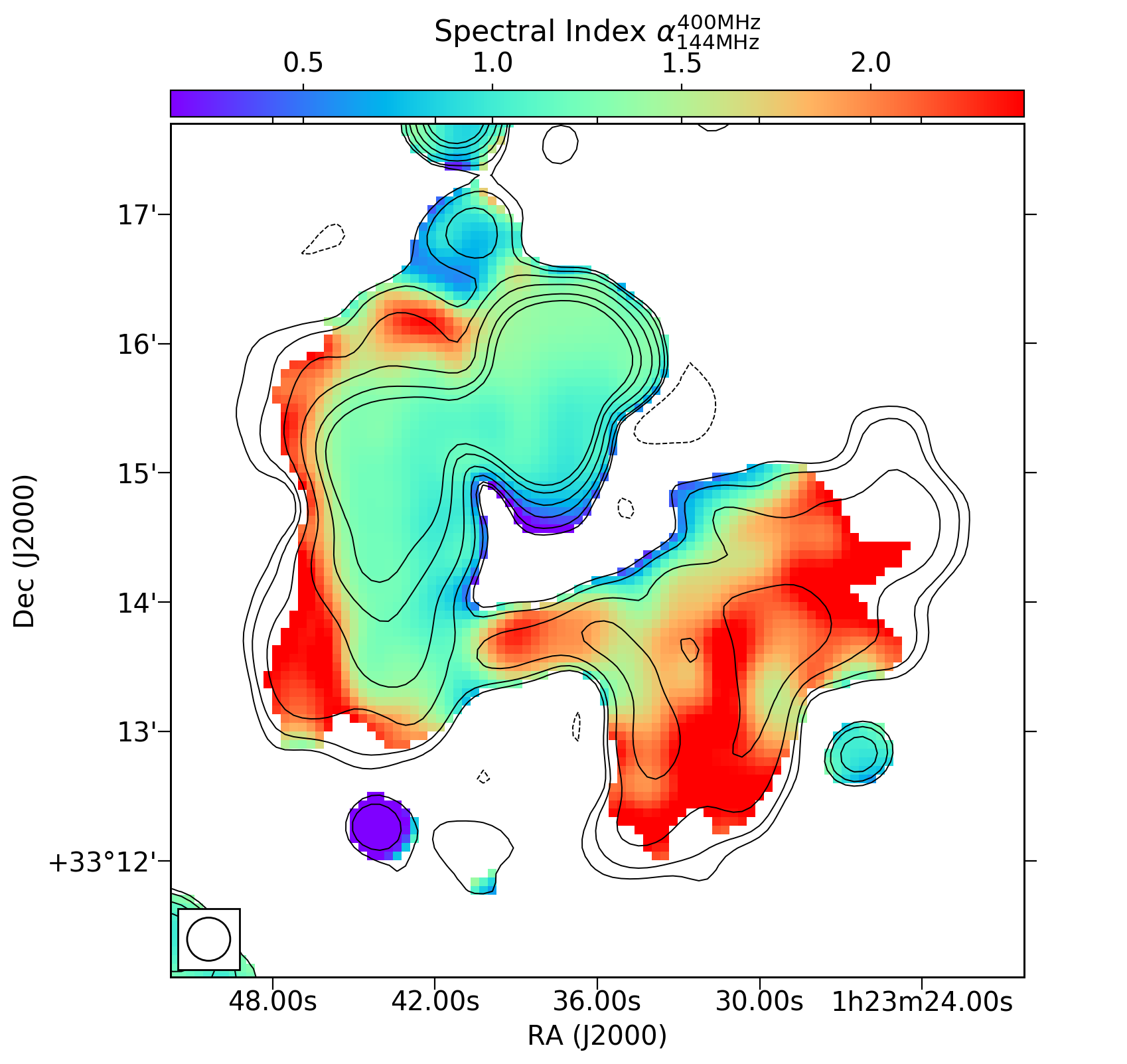}
\endminipage\hfill
\centering
\minipage{0.48\textwidth}
\endminipage\hfill

\caption{Spectral index maps of NGC\,507 between 144\,MHz and 400\,MHz at a resolution of 7 arcsec (left) and 20 arcsec (right). Only pixels with surface brightness above 3$\sigma$ in both images were used. Contours show the LOFAR emission at 144 MHz and are drawn at [-3, 3, 5, 10, 20, 40, 60, 90] $\times \sigma$. Typical errors in the brightest pixels are equal to $\sim$0.1 and in the faintest pixels to 0.15-0.2. }
\label{fig:specmaps}
 \end{figure*}
 
 \begin{figure}[htp!]
\centering
\includegraphics[width=0.47\textwidth]{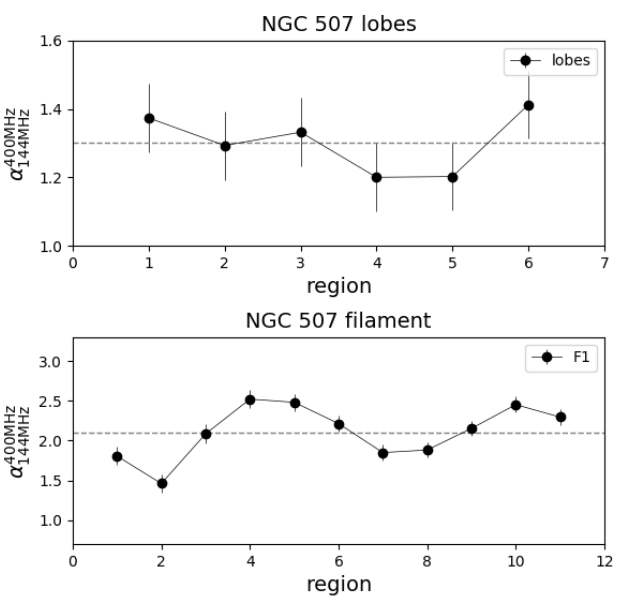}

\caption{Radio spectral index profile across the lobes of NGC\,507 (top panel) and its main filament F1 (bottom panel) in the frequency range 144-400\,MHz. The grey dashed lines represent the mean values equal to 1.3 and 2.1, respectively. The regions used to extract the profiles are shown in the left panel of Fig. \ref{fig:intspec}.}
\label{fig:specprofile}
\end{figure}

As already reported by \cite{murgia2011}, the spectrum of the double lobes is significantly curved, particularly at frequencies higher than 1400\,MHz, indicating the presence of strong radiative losses. 
The spectrum is already very steep at low frequencies with a spectral index in the frequency range 144-400\,MHz equal to $\rm \alpha_{144\,MHz}^{400\,MHz}=1.30\pm0.15$, which increases slightly in the range 400-675\,MHz (see Table \ref{tab:flux}).

The new SW diffuse emission represents about 20\% of the total flux density of the source at 144 MHz (as measured using the red box in Fig. \ref{fig:intspec}) and has a much steeper spectral index, namely $\rm \alpha_{144MHz}^{400MHz}=2.0\pm0.2$ and $\rm \alpha_{400\,MHz}^{675\,MHz}=2.7\pm0.1$. The filament F1 also shows very steep spectral indices with values $\rm \alpha_{144\,MHz}^{400\,MHz}=2.15\pm0.15$ and $\rm \alpha_{144\,MHz}^{675\,MHz}=2.15\pm0.15$.

Following \cite{murgia2011}, we used the CI$_{\rm OFF}$ model \citep{komissarov1994} to derive a spectral age for the lobes of NGC~507 and the new diffuse emission. The best fits are shown in Fig. \ref{fig:intspec} as a solid blue line and dashed grey line, together with the best-fit parameters. For the fit of the diffuse emission, the normalisation and injection index were fixed at the same values found for the lobes as a first-order approximation. 

Using an equipartition magnetic field value equal to $B_{eq}=$5.2 $\rm \mu G$ \citep{murgia2011}, we derived that the lobes have a total age of $ t_{\rm s}=150$\,Myr with an active phase of $t_{\rm on}=102$\,Myr and an inactive phase of $t_{\rm OFF}=48$\,Myr. Instead, the extended emission has a total age of $t_{\rm s}=242$\,Myr with an active phase of $t_{\rm on}=29$\,Myr and an inactive phase of $\rm t_{OFF}=213$\,Myr. We also considered a more conservative magnetic field value ---corresponding to the minimum radiative losses allowed for a plasma at a given redshift--- equal to B=$\rm B_{CMB}/\sqrt{3}=1.9 \ \mu G$, with $\rm B_{CMB}=3.25\cdot(1+z)^2 \ \mu G$. In this case, the total age increases to $t_{\rm s}$=242 Myr for the lobes and to $t_{\rm s}$=377 Myr for the extended emission. We note that these estimates should be considered as first approximations as they do not take into account the effect of any possible adiabatic expansion or compression, which are instead likely to play a role, especially in the newly detected diffuse emission.

Finally, the core has flux densities of $\rm S_{core, 675MHz}$=2.7 mJy, $\rm S_{core,400MHz}$=4.2 mJy and $\rm S_{core,144MHz}$=9 mJy, which are reported in Fig. \ref{fig:intspec} as red empty circles. From the plot, we can see that the value measured at 675 MHz is consistent with the power-law trend reported by \cite{murgia2011} based on higher frequency measurements. The values at lower frequencies instead largely exceed expectations. This excess is likely caused by the contamination of the extended emission, in which the core is embedded and becomes dominant at lower frequencies and at lower resolution. The spectral index obtained by fitting all points at frequencies $\geq$675 MHz is $\rm \alpha=0.19$, in agreement with flat-spectrum nuclear emission \citep{blandford1979}.

However, we note that the core prominence of the source at 1400 MHz, defined as $\rm CP=S_{core}/S_{tot}$, is equal to 0.02. This is more than a factor ten below the value expected for a typical radio galaxy of similar power according to \cite{deruiter1990} and more consistent with recent results on remnant radio galaxies \citep{mahatma2018, jurlin2021}. This suggests, once again, that the radio AGN is experiencing a low-power phase, and that it is not able to sustain the activity of large-scale jets.

\subsection{Resolved radio spectral analysis}
\label{sec:radiospec}

Using the matched radio images presented in Sect. \ref{sec:spec} we also created spectral index maps at low and high resolution in the frequency range 144-400\,MHz, where the new emission is detected (see Fig. \ref{fig:specmaps}). To do this, we spatially aligned the maps in the image plane using a point-like source close to the target and the tasks {\tt imhead} and {\tt imregrid} that are available in the Common Astronomy Software Applications ({\tt CASA}, \citealp{mcmullin2007}) package. This procedure is necessary to correct for any possible spatial shifts introduced by the imaging and phase self-calibration process, which would lead to unreliable spectral index values.

\begin{figure*}[htp!]
\centering
\minipage{0.5\textwidth}
\centering
\includegraphics[width=1\textwidth]{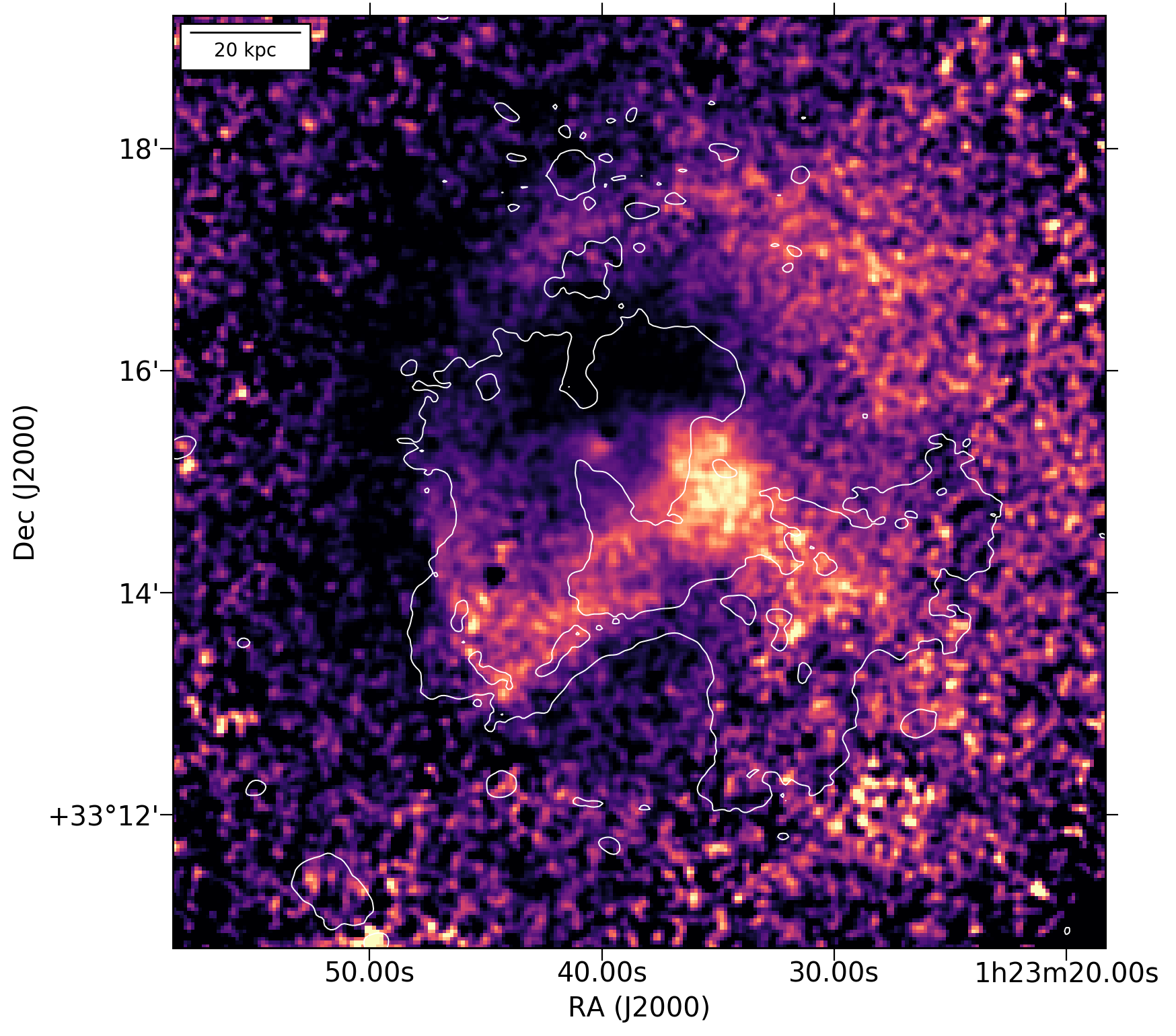}
\endminipage\hfill
\centering
\minipage{0.5\textwidth}
\centering
\includegraphics[width=1\textwidth]{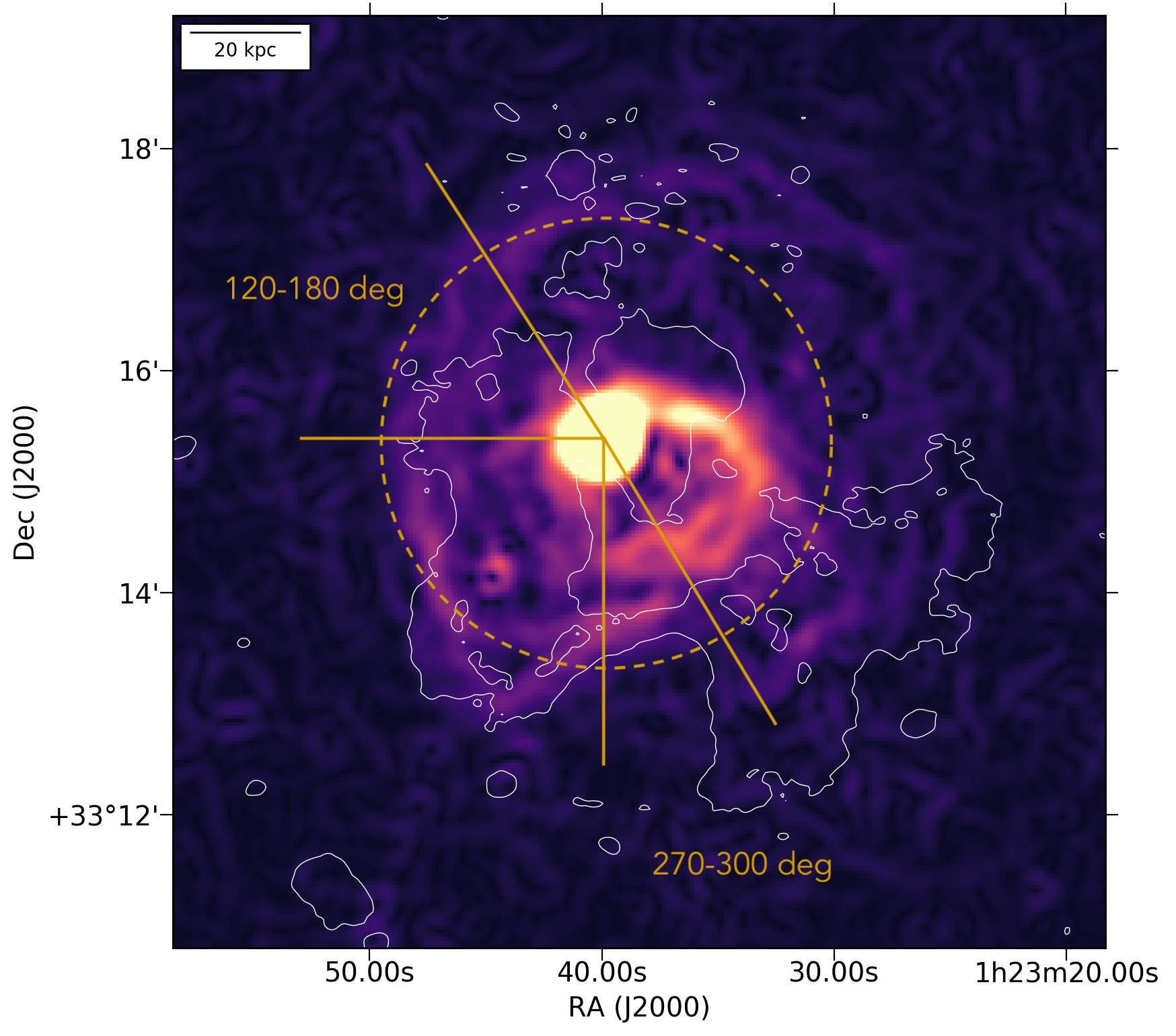}
\endminipage\hfill

\caption{Left: \textit{XMM-Newton} 0.7-2\,keV-band residual map obtained by subtracting the best-fit $\rm \beta$-model for the data  from the image shown in Fig. \ref{fig:507xray}. Right: \textit{XMM-Newton} 0.7-2\,keV-band filtered with the Gaussian Gradient Method with $\sigma$=3 pixels (1 pixel is 2 arcsec). The 3$\sigma$ LOFAR contour at 144-MHz with 5.57 arcsec $\times$ 8.09 arcsec resolution is overlaid. Yellow lines in the right panel show the sectors used to extract the profiles shown in Fig. \ref{fig:xprofile}. A dashed circle indicates the approximate distance of the E and S brightness discontinuities from the X-ray emission peak, equal to $\sim$2 arcmin.}
\label{fig:residual}
\end{figure*}

\begin{figure*}[htp!]
\centering

\includegraphics[width=1\textwidth]{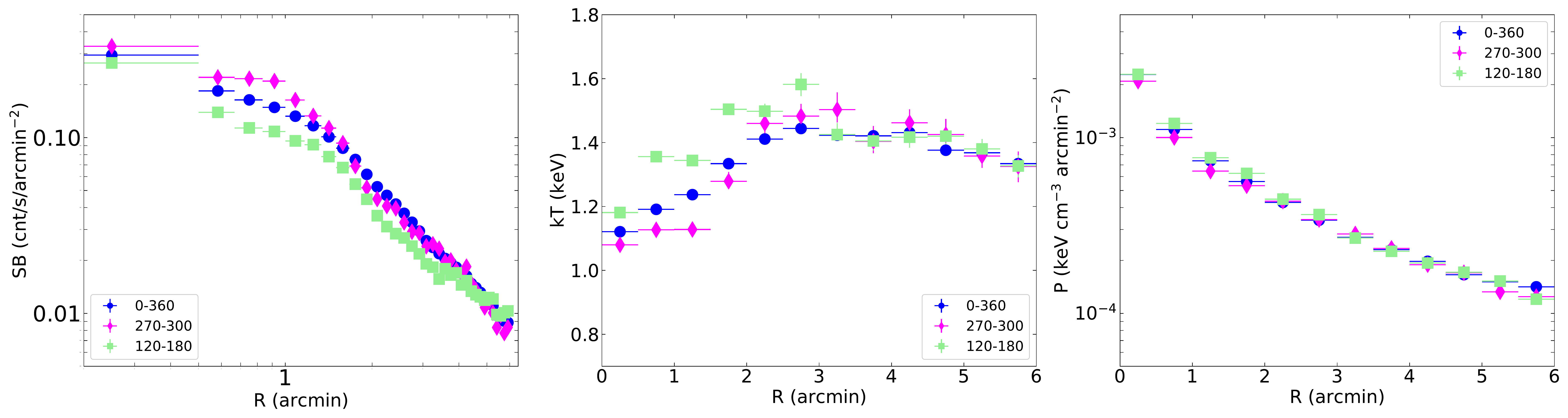}
\caption{\textit{Left:} Radial surface-brightness profile of the X-ray emission in the 0.7-2.0\,keV band extracted from three different sectors. Blue points represent the average azimuthal profile (0-360 deg), green points represent the eastern sector (120-180 deg), and magenta points represent the southern sector (270-300 deg) as shown in Fig. \ref{fig:residual}. \textit{Middle}: Temperature profiles in the same sectors as in the left panel. \textit{Right}: Pressure profiles in the same sectors as in the left panel.}

\label{fig:xprofile}
\end{figure*}

To create the spectral index maps, only pixels with surface brightness above 3$\sigma$ in all radio images were used. Spectral index errors were computed using Eq. \ref{eq:err} and have typical values of 0.1 in the brightest regions and in the range $0.15{-}0.2$ in the faintest regions.

These spectral maps allow us to study the spectral index distribution throughout the different regions of the source and search for any spatial trends. Across the double lobes, the spectral index is in the range $\rm \alpha_{144\,MHz}^{400\,MHz}=1\sim1.6$ but we do not recognise any of the clear spatial gradients typically observed in active radio galaxies, such as a steepening or a flattening moving towards the lobe edges for FRI and FRII radio galaxies, respectively (e.g. \citealp{orru2010, laing2011, heesen2015, mckean2016}). This can be appreciated further from the spectral profile shown in the top-right panel of Fig. \ref{fig:specprofile}. The regions used to extract the spectral index profile are shown in yellow in the left panel of Fig. \ref{fig:intspec}. Each region has a radius equal to 20 arcsec (7\,kpc) and they have all been chosen so as to match the regions presented in \cite{murgia2011}. Such homogeneous spectral distribution is not observed in all remnant radio galaxies (see e.g. \citealp{shulevski2017}), but was already observed at higher frequencies by \cite{murgia2011} in this source. This was interpreted as being due to the fact that the break frequency reached roughly the same value in each part of the lobes which could happen as a consequence of the jet switching off and the plasma subsequently aging. Alternatively, projection effects and consequent superposition of different particle populations might also play a role in smoothing any spectral-index trends.

As shown in Fig.\,\ref{fig:specmaps}, the distribution of the spectral index across the newly discovered diffuse emission does not show any clear gradients either. We also do not observe any clear spectral flattening in the filamentary region F1 with respect to the rest of the diffuse emission, as expected in the case where this region was created by strong compression driven by the motions of the IGrM. The effect of compression would be indeed to shift the spectral break to higher frequencies with respect to what is predicted by a simple radiative ageing of the plasma. The bottom-right
panel of Fig. \ref{fig:specprofile}  shows the spectral index profile across the main filament F1 created using the black boxes displayed in the left panel of Fig.\, \ref{fig:specprofile}  which are of 15 arcsec (5
kpc) per side. The spectral index fluctuates in the range $\rm \alpha_{144\,MHz}^{400\,MHz}=1\sim1.8$. Similar spectral index values and fluctuations have been reported in the filaments observed connecting the lobes of the radio galaxy ESO~137-006 by \cite{ramatsoku2020} and IC 4296 by \cite{condon2021} at GHz frequencies.

We note that the pixels showing spectral indices with values of $\sim0.5$ in the low-resolution map in the regions of the southern tip of the western lobe and at the northern edge of the new diffuse emission should not be considered reliable as their value is likely affected by their position at the edges of the  source  and the presence of regions of negative surface brightness in the LOFAR 144\,MHz image.

\subsection{X-ray morphology and profiles}
\label{sec:xray_morpho}

Figure \ref{fig:507xray} shows the $0.7{-}2$ keV-band \textit {XMM-Newton} image of the NGC\,507 galaxy group. In particular, the left panel shows the full \textit {XMM-Newton} field of view, where the companion group NGC~499 is  also visible, while the right panel shows a zoom onto the NGC~507 group with radio contours overlaid. 

From the large-scale map shown in the left panel, it is clear that the outer halo of NGC~507 is asymmetric and that there is an excess of emission in northwest direction towards NGC~499. There was only a hint of this latter in the ROSAT HRI map presented by \cite{paolillo2003}. Moreover, a tail of emission is observed to the south of NGC~499, in the direction of NGC~507. All this strongly supports a scenario in which the two systems are undergoing a merger.

As already discussed in previous works (see Sect. \ref{sec:ngc507}), in the central regions of NGC~507 the gas also shows a complex distribution, which is likely due to a combination of large-scale gas motions and disturbances driven by the AGN jet activity. The main peak of the X-ray emission is coincident with the optical galaxy NGC\,507 (marked with a yellow cross) and the two surface-brightness jumps detected by \cite{paolillo2003} and \cite{kraft2004} in the eastern and southern directions (using \textit{XMM-Newton} and Chandra observations, respectively) are clearly visible.

Remarkably, the filament of non-thermal radio-emitting plasma labelled F1 is perfectly aligned with the southern X-ray surface-brightness discontinuity (see Fig. \ref{fig:507xray}). This likely implies a physical connection between the motions of the non-thermal remnant plasma and the surrounding thermal medium.

To highlight the asymmetries in the X-ray surface-brightness distribution, we subtracted from the image presented in Fig. \ref{fig:507xray} the best-fit of the double $\rm \beta$-model \citep{cavaliere1976} fitted to the data and centred on the X-ray peak (core radius r$\rm_{c1}=109\pm27$ kpc, $\rm \beta_{1}=5.08\pm0.2$ and r$\rm_{c2}=91\pm3$ kpc, $\rm \beta_{2}=0.72\pm0.03$). The residual map is shown in the left panel of Fig. \ref{fig:residual}. 
We also filtered the original image using the Gaussian gradient magnitude (GGM) filter (\citealp{sanders2016}) with a three-pixel filter (1 pixel=2 arcsec), as shown in the right panel of Fig. \ref{fig:residual}. 

In both these images, the E and S discontinuities described above appear clearly enhanced. Moreover, an evident spiral pattern is revealed. Such morphology is classically interpreted as due to gas sloshing in the group potential well with a non-zero angular momentum \citep{markevitch2007, gastaldello2013, zuhone2013, ascasibar2006}. In particular, a spiral shape is expected if the interaction between the group and the perturber at the origin of the sloshing is seen face-on \citep{roediger2011}. From the residual image, an excess of emission is also visible at the southwestern side of the western lobe, which, as previously suggested, could be consistent with the presence of gas compressed by the expanding western AGN jet.

To further investigate the S and E discontinuities, we extracted radial surface-brightness, temperature, and pressure profiles using successive annular regions extending from the X-ray surface brightness peak up to $\sim$0.2R$_{500}$ (i.e. $\sim$5.9 arcmin), where all relevant features are located. Two angular sectors were chosen to match the eastern and southern surface-brightness features and spanning the range 120-180 deg and 270-300 deg, respectively (measured from the west; see Fig. \ref{fig:residual}, left panel). An azimuthally averaged radial profile was also created for comparison. For the temperature and pressure profiles, the annuli were determined by requiring a minimum width of 30 arcsec to minimise the flux redistribution due to the \textit{XMM-Newton} point spread function (PSF).  
We note that, because of the complex geometry of the X-ray surface-brightness distribution, the profiles were not deprojected.

The final profiles are shown in Fig. \ref{fig:xprofile}. From this we can see that the surface-brightness jump reported by \cite{kraft2004} (using \textit{Chandra} data) at $\sim$2 arcmin distance from the group centre in the eastern direction is recovered. A drop in the surface-brightness distribution is also observed at approximately the same distance ($\lesssim$2 arcmin) in the southern direction. We stress that the southern discontinuity is asymmetrical with respect to the surface-brightness distribution centred on the X-ray peak and this likely contributes to smoothing the observed surface-brightness jump, in addition to the natural blurring by the \textit{XMM-Newton} PSF.

\begin{figure*}[!htp]

\centering
\minipage{0.5\textwidth}
\centering
\includegraphics[width=0.9\textwidth]{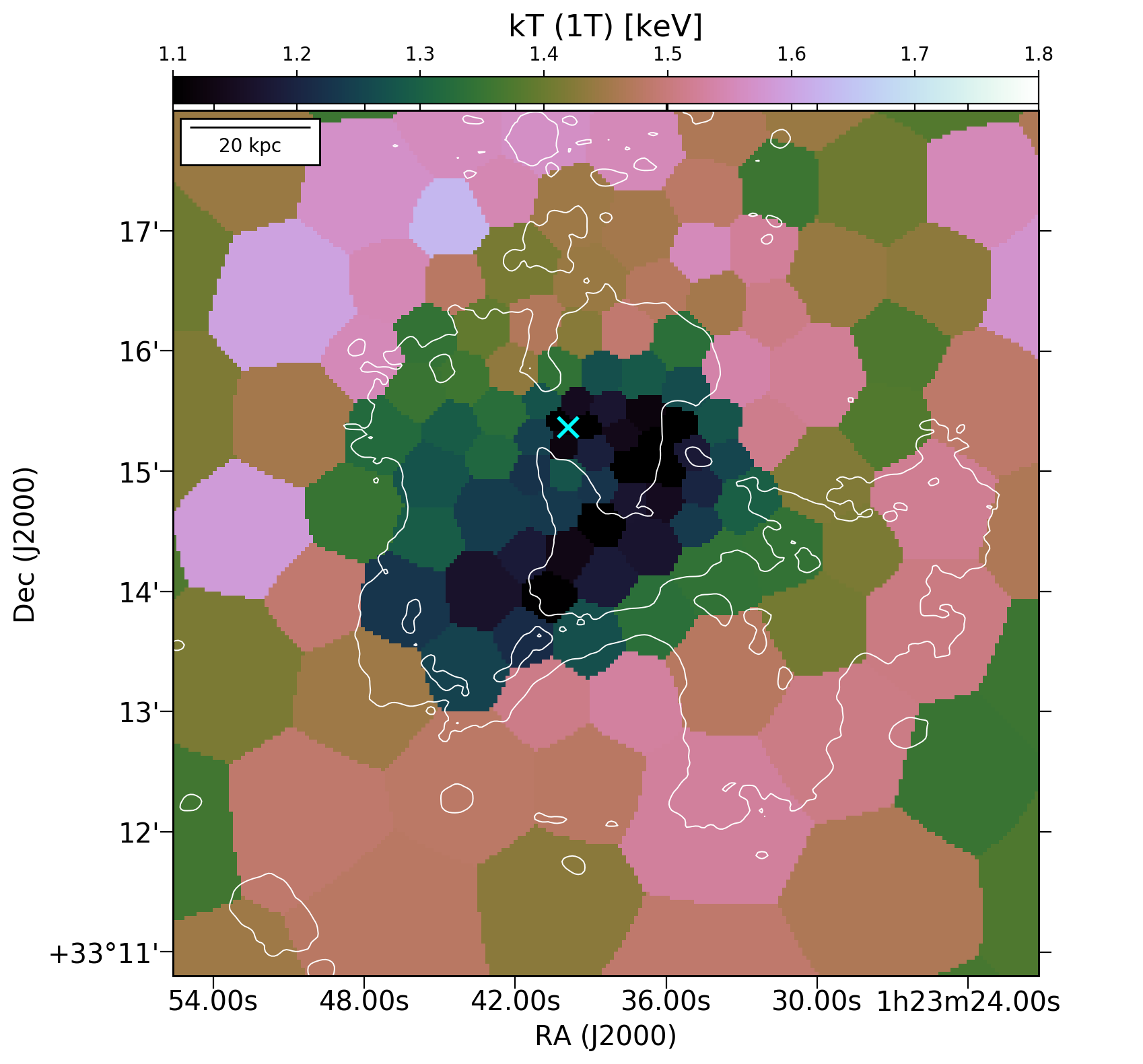}
\endminipage\hfill
\centering
\minipage{0.5\textwidth}
\centering
\includegraphics[width=0.9\textwidth]{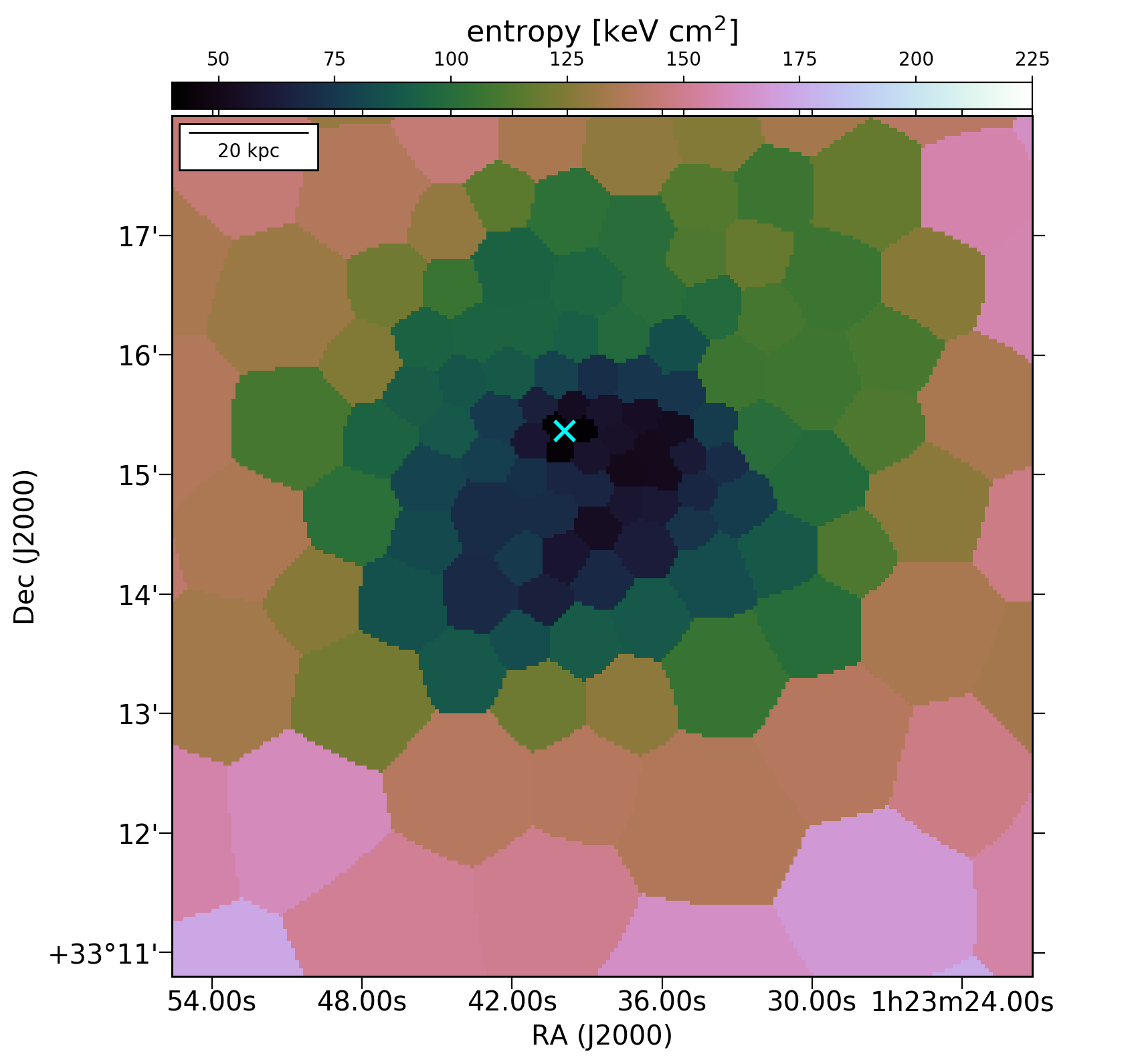}
\endminipage\hfill
\centering
\minipage{0.5\textwidth}
\centering
\includegraphics[width=0.9\textwidth]{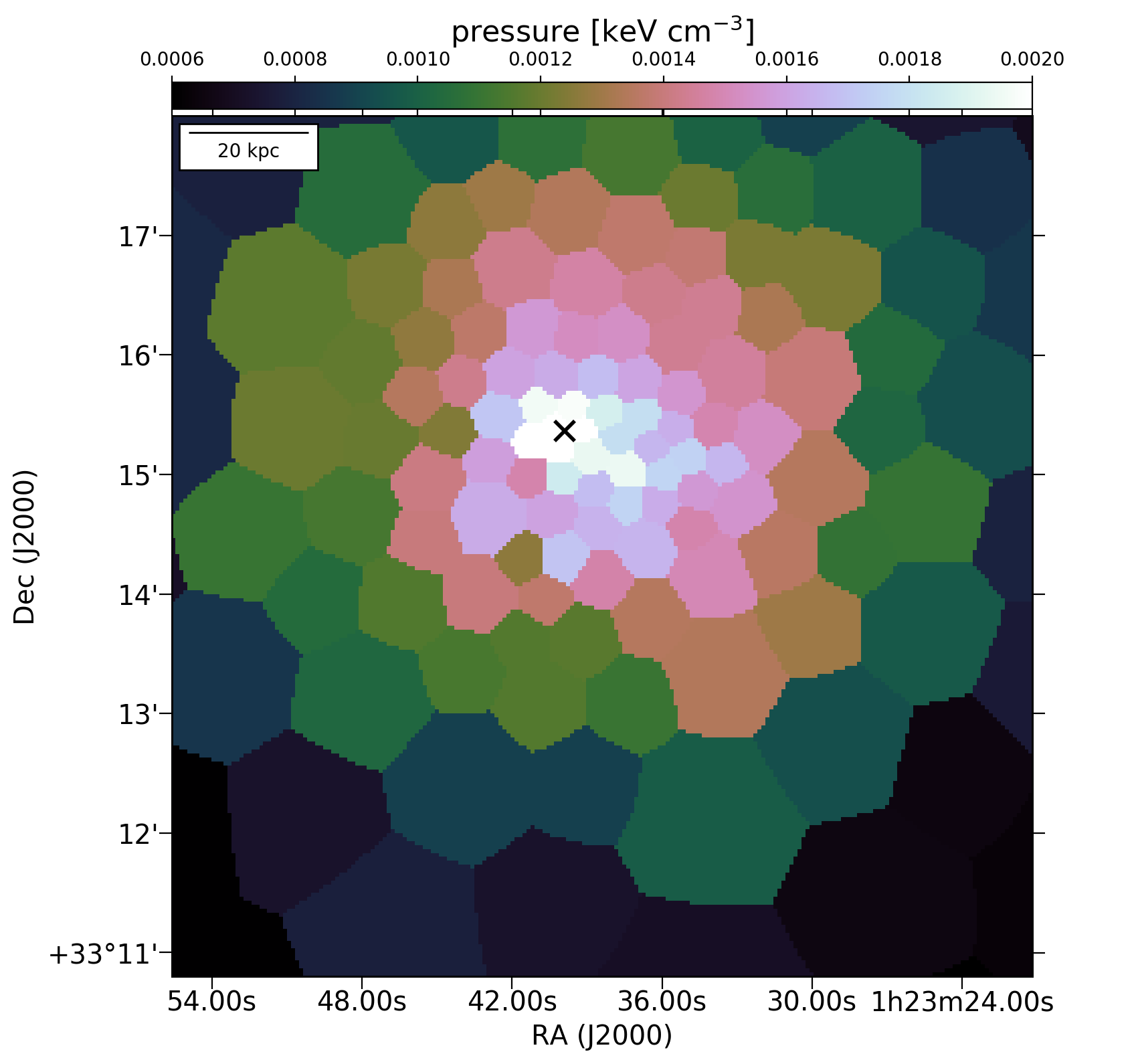}
\endminipage\hfill
\centering
\minipage{0.5\textwidth}
\centering
\includegraphics[width=0.9\textwidth]{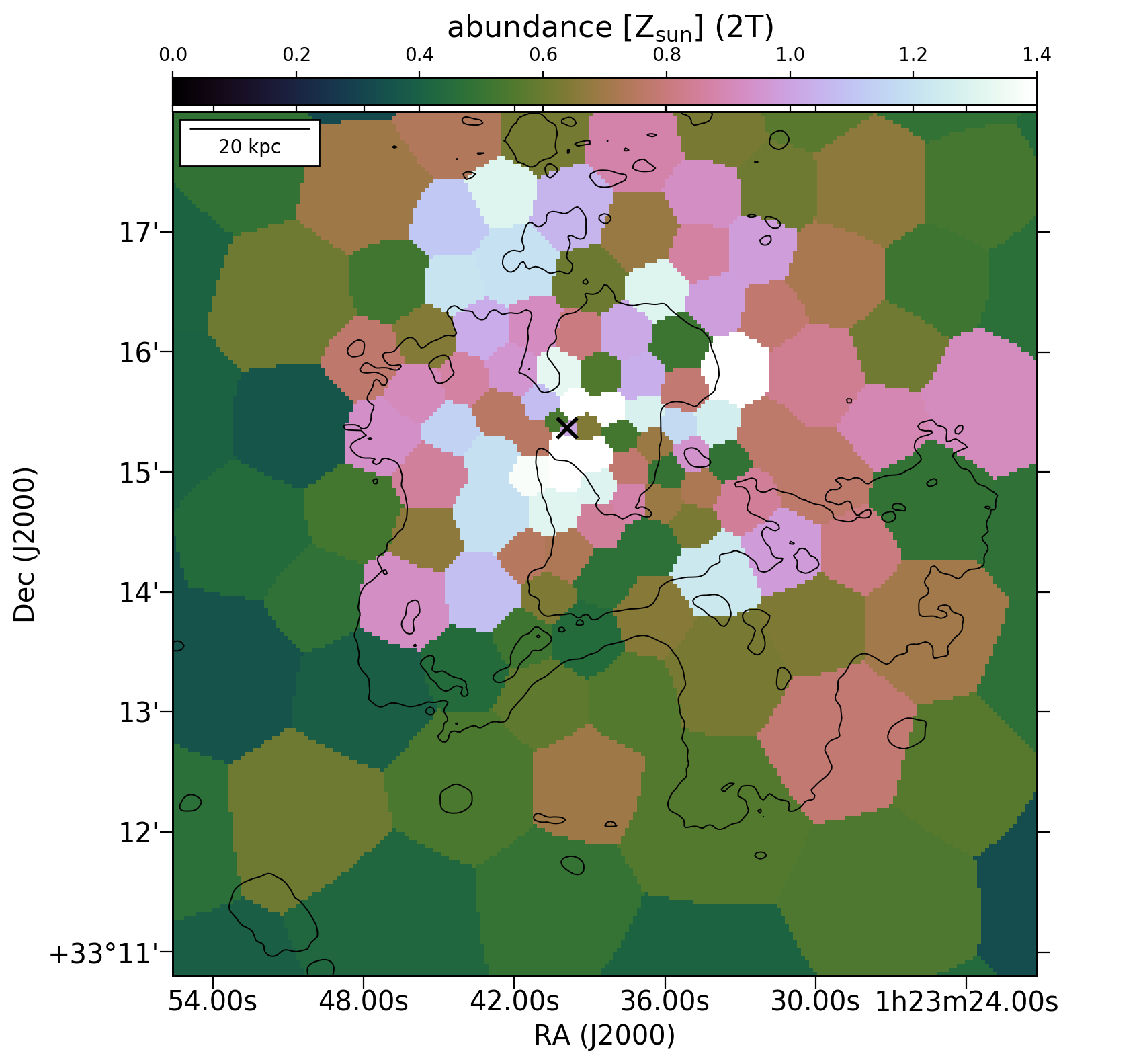} 
\endminipage\hfill
\caption{Projected temperature (top-left), entropy (top-right), pressure (bottom-left) maps created using a 1T model fit and abundance map created using a 2T model (bottom-right) as described in Sect. \ref{sec:xray}. A cross marks the centre of the galaxy NGC~507. $\rm 3\sigma$ LOFAR contours at 144-MHz with 5.57 arcsec $\times$ 8.09 arcsec resolution are overlaid in the top-left panel. }
\label{fig:tmap1}
 \end{figure*}

To understand the nature of these surface-brightness edges, we looked at the temperature and pressure profiles across the edges as shown in the middle and right panels of Fig.\,\ref{fig:xprofile}. As we can see in the middle plot, a temperature jump across the two edges is also present, with increasing values moving from the brighter (and therefore denser) side of the edge to the fainter. In particular, we measure a temperature increase from kT=1.28$\pm$0.03 keV to kT=1.46$\pm$0.03 keV in the southern sector and from kT=1.35$\pm$0.01 keV to kT=1.50$\pm$0.02 keV in the eastern sector. No jump is detected in the pressure profile. This behaviour is classically observed in the so-called `cold fronts' (e.g. \citealp{markevitch2007}).

We note that the observed jumps are more subtle than those observed in merging systems, where they trace discontinuities between gas from different subclusters (e.g. \citealp{markevitch2000}). Instead, here they appear more consistent with a simple redistribution of the gas of the group itself, as observed in more relaxed systems. Projection effects  may also likely contribute to smoothing any observed gradients.

\subsection{X-ray spectral maps}
\label{sec:xrayspec}

In order to get a more detailed view of the two-dimensional spatial distribution of the thermal gas properties in the system (temperature, entropy, pressure, and metal abundance) we created the spectral maps shown in Fig. \ref{fig:tmap1}.
To this end, we subdivided the data into small regions from which spectra were extracted. The sizes of the spatial bins were obtained with the Weighted Voronoi Tessellation (WVT) binning algorithm by \cite{Diehl2006}, which is a generalisation of the \cite{Cappellari2003} Voronoi binning algorithm. We ran WVT on the background-subtracted and exposure-corrected image and created maps with a S/N=50 per bin. As we are interested only in the central regions, the maps were obtained for $r<$0.3R$_{500}$.

The entropy (K) and pressure (P) maps can be obtained by combining temperature and electron number density as K=Tn$_e^{-2/3}$ and P=Tn$_e$, respectively.
As the normalisation in XSPEC is defined as 
\begin{equation}
N=\frac{10^{-14}}{4\pi D_A^2 (1+z)^2} \int{n_e n_H dV} \ \rm cm^{-5},
\end{equation}
if we assume the volume of the emitting region, it is straightforward to derive the electron density. Under the assumption that there is no material projected onto the region of interest, the volume can be easily computed as V=2A$\sqrt{R_{500}^2-X^2- Y^2}$, where A is the area of the region, while X and Y are the projected distances in the east--west and north--south directions, respectively.

From the maps, we can clearly see that the spatial distribution of all four quantities is highly asymmetrical with respect to the centre, consistent with previous results by \cite{kim2019} and \cite{islam2021}, and again suggesting a dynamically disturbed environment. In particular, the temperature and entropy maps show a spiral pattern, similar to that already observed in surface brightness. The temperature discontinuities corresponding to the jumps described in Sect. \ref{sec:xray_morpho} are also clearly visible. 

The metallicity map shows a very scattered distribution with the highest values observed at the group centre, as expected for relaxed systems. This could be due to a combination of intrinsic variations in the gas properties and errors in the fit. Indeed, because of the multi-phase properties of the gas, the error bars associated with each measurement are relatively large, making it difficult to interpret the map. For completeness, the abundance 1-$\sigma$ error map is included in Appendix A (Fig. \ref{fig:2apecerror}).
We see a mild abundance drop, from values of 0.8-0.9 to values of 0.4-0.5, corresponding to the eastern discontinuity. However, this gradient is not as unusual as previously suggested by \cite{kraft2004}; indeed it is expected at regular cold fronts \citep{simionescu2010}. Overall, it seems more plausible that on larger scales the abundance distribution is mainly determined by sloshing rather than by the AGN activity.

\begin{figure*}[!htp]

\centering
\minipage{0.48\textwidth}
\centering
\includegraphics[width=0.85\textwidth]{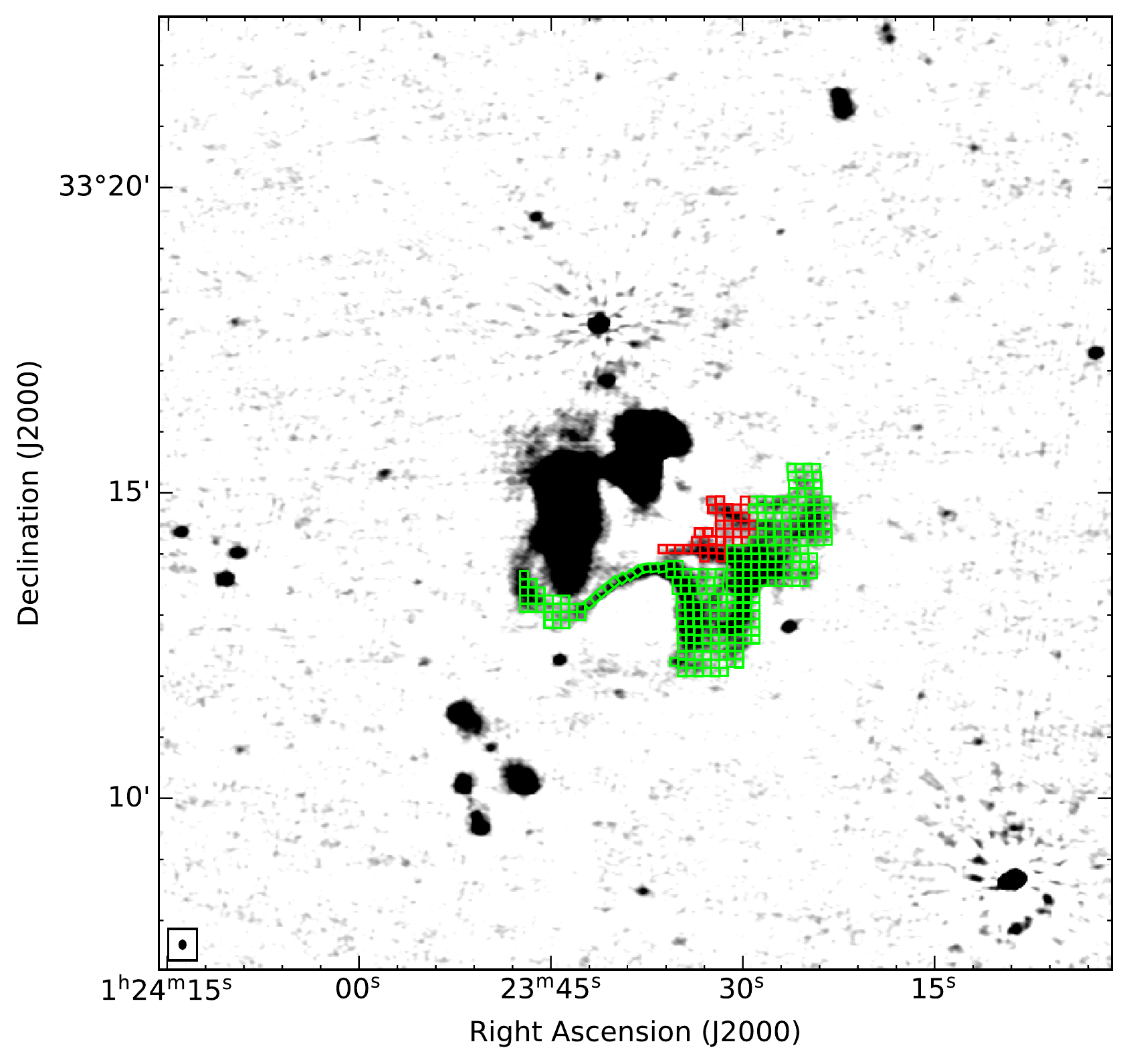}
\endminipage\hfill
\centering
\minipage{0.49\textwidth}
\centering
\includegraphics[width=1\textwidth]{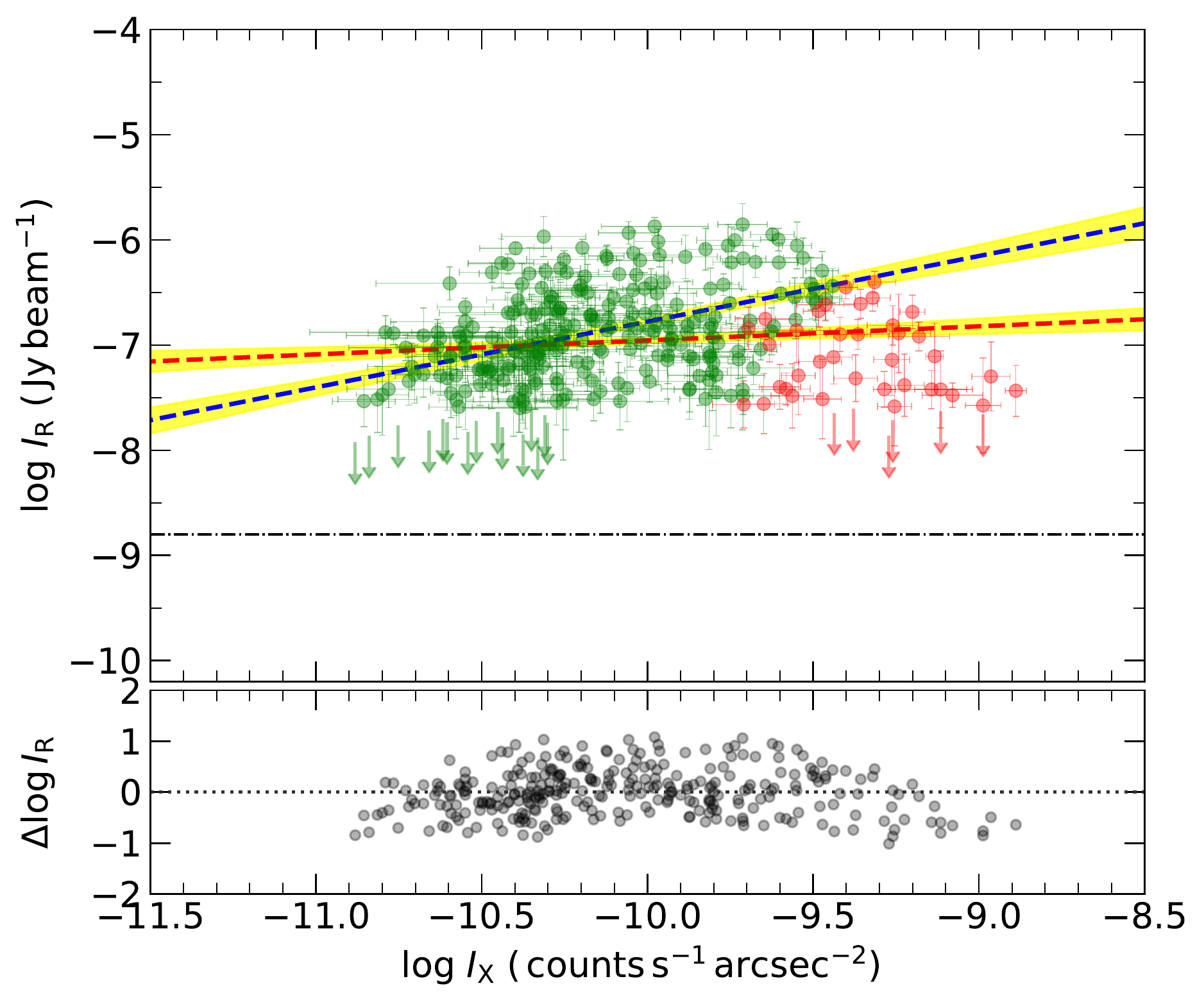}
\endminipage\hfill
\caption{$\rm I_{R}-I_{X}$ relation of NGC\,507. The high-resolution LOFAR image at 144\,MHz is used for the radio. \textit{Left:} Squared boxes, with 8-arcsec ($\simeq$3\,kpc) side, used to extract surface-brightness values. \textit{Right:} $\rm I_{R}-I_{X}$ relation obtained using all regions. Points are coloured following the colour scheme shown in the left panel. Circles depict cells where the radio and X-ray surface brightness is above 3$\sigma$. The upper limits (arrows) represent cells with radio surface brightness in the range 2$\sigma$-3$\sigma$. The horizontal black dash-dotted line indicates the $1\sigma$ in the LOFAR 144\,MHz map. The red and blue lines correspond to the best-fit obtained for the entire emission and for green regions only, respectively. The lower panel shows the residuals of log $\rm I_R$ and log $\rm I_X$ with respect to the {\tt Linmix} best fit line obtained using all regions (green+red). A negligible correlation is found with  a Spearman coefficient of $0.13\pm0.06$ in the case where all boxes are included (red+green). A mild positive correlation is found instead with a Spearman coefficient equal to $r_s=0.50\pm0.06$ and a slope equal to $b_{\rm green}=0.63\pm0.09$, if the red boxes are excluded from the fit. }
 
\label{fig:507xradiocorr}
 \end{figure*}

\subsection{X-ray versus radio correlation}

In recent years, an increasing number of studies have reported evidence of a point-to-point correlation between the radio surface brightness, $I_R$, and the X-ray surface brightness, $I_X$, of radio halos (e.g. \citealp{govoni2001, giacintucci2005, rajpurohit2018, botteon2020B, rajpurohit2021,  duchesne2021}) and mini-halos \citep{giacintucci2019, ignesti2020, biava2021b} harbouring massive galaxy clusters, as well as radio bridges between galaxy clusters \citep{botteon2020A, bonafede2021}. This is interpreted as a signature of the dynamical coupling of the thermal and non-thermal components of the ICM. In particular, according to current studies, it seems that halos show preferentially sublinear slopes and mini-halos super-linear slopes.

While it is not obvious whether or not such a relation should be expected for the newly detected diffuse radio emission in NGC\,507, we investigate it here to further understand the plasma origin. The relation is generally described by a power law of the form

\begin{equation}
I{\rm_R}\propto I_{\rm X}^{b},
\end{equation}

\noindent where $b$ is the correlation slope. For this, we used the LOFAR 144-MHz high-resolution image with a beam size of $8.0\arcsec\times5.5\arcsec$. The \textit {XMM-Newton} X-ray image was smoothed with a Gaussian of $2\arcsec$ FWHM. The grid used for the point-to-point analysis overlaid on the LOFAR image is shown in the left panel of Fig. \ref{fig:507xradiocorr}. Boxes are 8 arcsec ($\sim$3kpc) on one side. To perform the fit, we use the {\tt Linmix} method \citep{kelly2007} following \cite{ rajpurohit2021}. Also, 2$\sigma$ limits on the radio surface brightness are included in the fit.

As shown in the right panel of Fig. \ref{fig:507xradiocorr}, if all regions are considered (green and red boxes), the Spearman coefficient is $0.13\pm0.06$. There is clearly no evidence of any correlation between the two quantities, suggesting that the non-thermal plasma is still not mixed up with the ICM on large scales. This behaviour is expected if the radio emission of the newly detected plasma is still mostly driven by the initial particle population accelerated by the AGN.

However, it should be mentioned that if the red boxes are excluded from the fit, a mild positive correlation is found with Spearman correlation coefficient equal to $r_s=0.50\pm0.06$ and slope equal to $b_{\rm green}=0.63\pm0.09$. These red boxes coincide with some filamentary emission in both the radio and X-ray images (see Fig. \ref{fig:residual}) and might be suggestive of different physical conditions with respect to the rest of the medium, for example different levels of compression. Current data do not allow us to explore this possibility further but future observations at lower frequencies (e.g. using the LOFAR Low Band Antennas) may help us to study this in more detail.

\subsection{Optical analysis}
\label{sec:optical}

In order to further investigate the dynamical state of NGC~507 and to identify the presence of any substructures in the member galaxy distribution, we performed an optical analysis of the system. Here, we present a summary of the results obtained and we describe the methods used in Appendix \ref{opt_app}  in more detail.

NGC~507 is part of the Pisces supercluster, one of the largest structures in the nearby Universe traced both optically \citep[e.g.][]{tully2015} and in X-rays \citep{bohringer2021}. It was first recognised as a galaxy group in the Lyon catalogue \citep{garcia1993}, who found it to be a separate structure from the nearby NGC~499 group (the latter with only five members). However, more recent analyses, such as those based on the Two-Micron All-Sky Survey (2MASS, \citealp{skrutskie2006}), report only one group \citep{crook2007,lavaux2011,tully2015}. 

We investigated the membership of the group using the "velocity gap" method \citep{depropris2002} and we clearly detect the NGC~507 group as a peak at the redshift of the galaxy NGC~507, populated by 74 galaxies. This result is also confirmed using the `shifting gapper' method \citep{fadda1996, owers2009}. The value of the bi-weight location estimator of the mean group velocity is $5030\pm74$ \kms, which corresponds to $z_{\rm{cos}}=0.01678\pm0.00025$. 

\begin{figure*}[htp!]
\centering
\minipage{0.5\textwidth}
\centering
\includegraphics[width=0.9\textwidth]{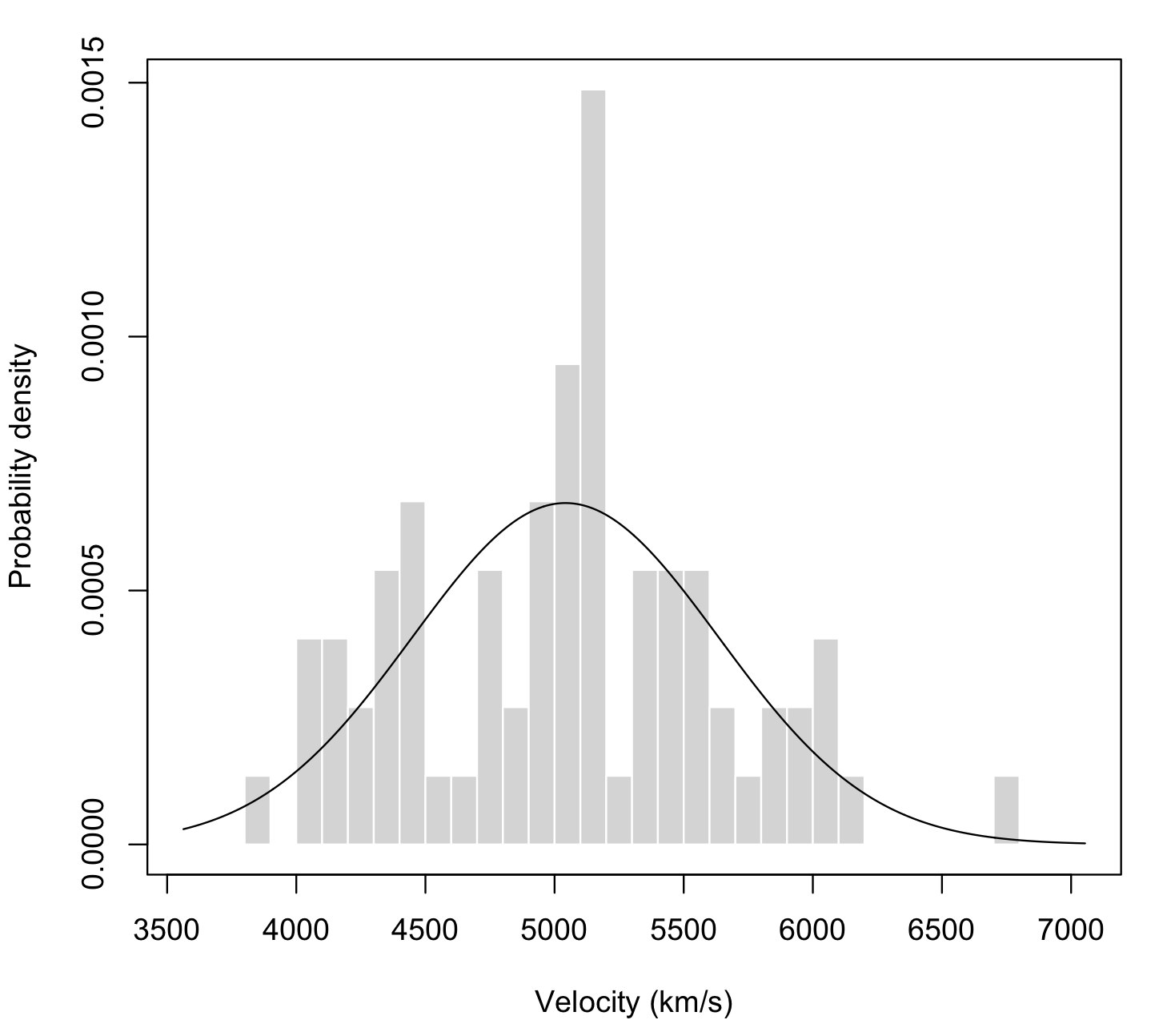}
\endminipage\hfill
\centering
\minipage{0.5\textwidth}
\centering
\includegraphics[width=0.9\textwidth]{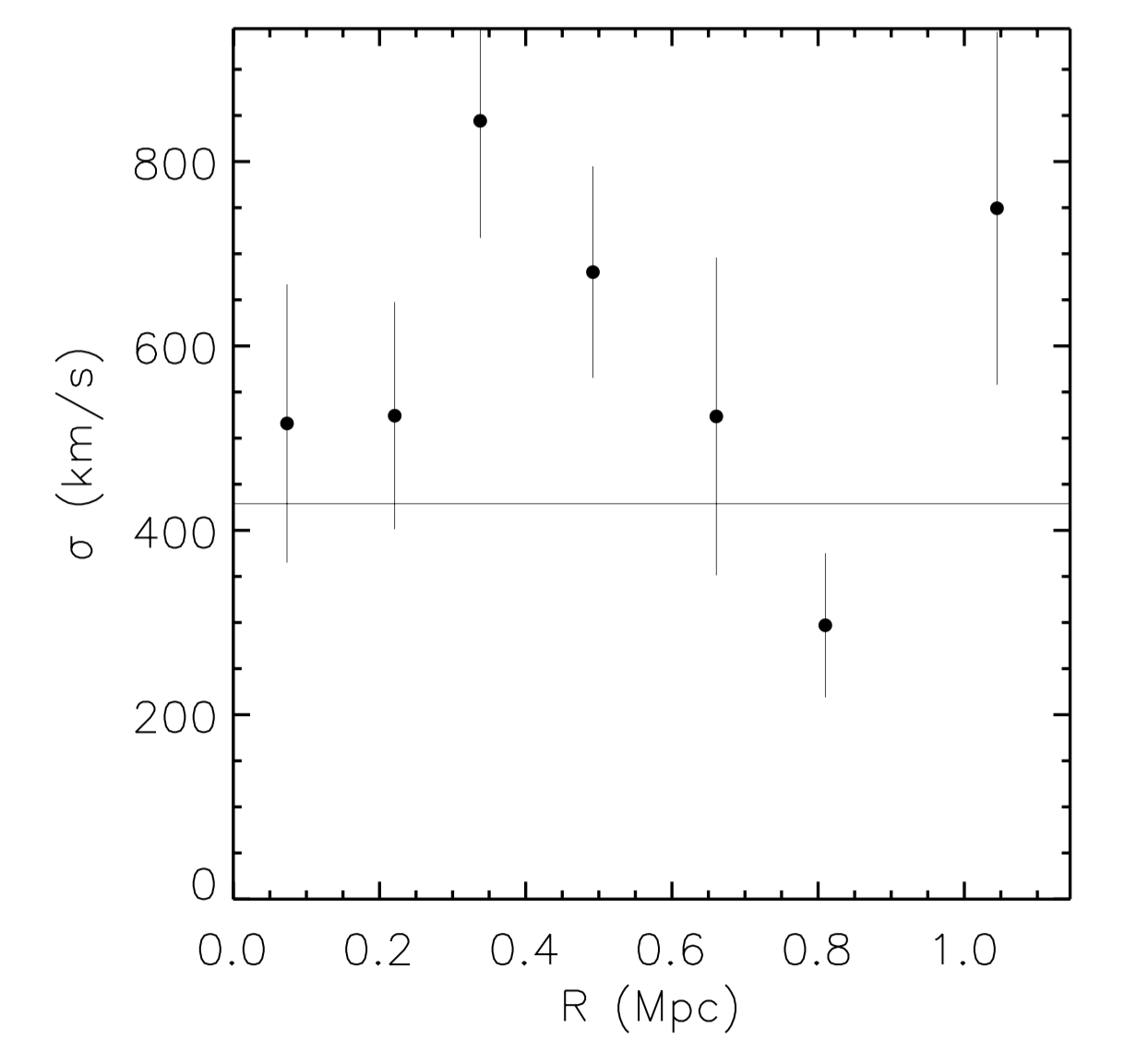}
\endminipage\hfill
\caption{\textit{Left:} Histogram of the velocities of the galaxy members of the NGC~507 group. The best-fit Gaussian distribution of the Gaussian mixture model \textsc{mclust} is also shown. \textit{Right}: Velocity dispersion profile of the NGC~507 group, where the galaxy NGC~507 is assumed to be the centre of the group. Bins contain ten galaxies each and the 1$\sigma$ `jacknife' errors are shown. The solid line shows the expected value of dispersion velocity assuming density--energy equipartition between IGrM and galaxies, i.e. $\beta_{spec} =1$, assuming a temperature for the IGrM of $kT =1.19$ keV.}
 
\label{fig:507optical}
 \end{figure*}
 
We note that the membership allocation process groups together the galaxies NGC~507 (with a velocity of 4934 \kms\ reporting the NED  velocity after heliocentric correction and a $Ks$ magnitude of 8.28 as reported by \citealp{tully2015}) and NGC~499 (4399 \kms and $Ks$ magnitude of 8.71). These two are separated in velocity by 535 \kms\ and are the brightest members of the group with a difference in magnitude of $\Delta\,m_{12}=0.43$. A low value of $\Delta\,m_{12}$ for such an evolved X-ray-bright system is an indication that the system is undergoing a merger.

A further indication that the system is dynamically disturbed comes from the global value of its dispersion velocity, which is equal to $611\pm50$ \kms \ , and from the velocity dispersion radial profile (see Fig. \ref{fig:507optical}). In the case of a dynamically complex, merging system, the profile is generally observed to deviate from the declining shape of dynamically relaxed systems \citep[e.g.][]{girardi1996,menci1996,hou2009,costa2018}.

The analysis of the scaling relations of the X-ray properties with the dispersion velocity and with the profile of the dispersion velocity \citep{lovisari2021} provides further support to the disturbed nature of the group. If we take the temperature of the system as measured in the annulus 0.15-0.5 $R_{500}$ equal to 1.19 keV as a reference value, the NGC~507 group is a clear outlier in the $\sigma - T$ relation \citep[see e.g.][]{farahi2018} as it is located well above the expectation of a constant specific energy ratio, $\beta_{spec}=1$.

We finally applied a series of substructure tests from one-dimensional (1D) in the velocity space to three dimensional (3D) involving both spatial and velocity spaces, because the existence of a correlation between the positions and velocities of cluster galaxies is a signature of true substructure. However, all these tests provide a null result, highlighting again the difficult task of finding statistically significant evidence for a merger with an optical analysis.

Overall, while our optical analysis of the group galaxy members was not very effective in confirming the merger scenario, clear indications that the NGC~507 group is a dynamically disturbed system are present. These include a globally high value of the dispersion velocity of the system, its non-declining velocity radial profile, and a low magnitude difference between the two most massive galaxies NGC 507 and NGC 499.

\section{Summary and Discussion}
\label{sec:concl}

Since its discovery, NGC~507 has always been shown to be a complex system with peculiar and interesting features. Our analysis of new low-frequency  radio observations combined with X-ray and optical data confirms this view and helps us to make further steps forward in our understanding of its history.

Based on the X-ray spectral profiles and maps shown in Sect. \ref{sec:xray_morpho} and \ref{sec:xray}, and supported by the extensive literature built up in the last decade (see e.g. \citealp{markevitch2007, zuhone2016} for reviews), we suggest that the eastern discontinuity observed in the X-ray surface brightness distribution is not as unusual as previously suggested by \cite{kraft2004}. The authors attributed it to an `abundance front' created by the jet expansion of the central AGN. Instead, we think that both the eastern and southern discontinuities can reasonably be interpreted as typical cold fronts defined as contact edges between regions of gas with different entropies.

The central AGN is likely contributing to shaping the thermal gas distribution in the inner regions of the group. However, the spiral morphology of the X-ray emission (best detected in our X-ray residual image and GGM-filtered image shown in Fig. \ref{fig:residual}) and of the temperature and entropy maps suggests that sloshing is inducing gas motions on larger scales. The spiral morphology is indeed characteristic of a gas that has gained angular momentum, and therefore is unlikely to be initiated by the AGN-jet activity \citep{ascasibar2006}.

We note that, sloshing cold fronts have been observed in a large number of massive, mostly relaxed  clusters, but they remain more rarely detected and studied in galaxy groups, possibly also due to observational limitations. Among the best such cases known so far are NGC 5098 \citep{randall2009}, NGC 5044 (e.g. \citealp{gastaldello2013, osullivan2014}), NGC 5846 \citep{machacek2011, gastaldello2013}, IC 1860 \citep{gastaldello2013}, NGC 7618 (\citealp{roediger2012}, but see \citealp{sheardown2019} for a slingshot interpretation), and NGC 1550 \citep{kolokythas2020}.

In some of these systems, the perturber initiating the sloshing is clearly identified with a nearby galaxy or galaxy group (see e.g. IC~1860, \citealp{gastaldello2013}). In the case of NGC~507, it remains difficult for us to identify  the perturber from an optical analysis only, as discussed in Sect. \ref{sec:optical}. However, the nearby galaxy group NGC~499, located at a projected distance of 270\,kpc, remains a very good candidate. The excess X-ray emission between NGC~507 and NGC~499, the tail of X-ray emission present in the southern extent of NGC~499 (see Fig. \ref{fig:507xray}), and the high velocity dispersion measured clearly suggest that the two systems are interacting. The small magnitude difference between the two systems further implies a sufficiently high mass ratio to give rise to the observed perturbations.

The spiral pattern observed in the X-ray emission of NGC~507 would then imply that the orbit of NGC~499 is nearly in the plane of the sky. According to simulations, the morphology of the X-ray distribution in sloshing systems is indeed a strong indicator of the inclination between the orbital plane of the pertuber and our line of sight \citep{roediger2011}. Spiral patterns are expected in the case of face-on interactions, while arc-like features at the sides of the group or cluster core are expected in the case of edge-on interactions. The velocity difference between NGC~507 and NGC~499 is equal to $\rm\Delta v \sim$500 km/s. This is a factor of about two lower than the velocity difference observed between NGC~5044 and IC1860 and their respective perturbers, which are also claimed to be experiencing face-on interactions. This therefore seems to be consistent with the above scenario.

The origin of the arc-like (concave) X-ray discontinuity observed to the south and coincident with the radio filament F1 (see Fig. \ref{fig:rgb}) merits further discussion. One possibility is that it may be created by a second interaction between NGC~507 and another system, in addition to that initiating the sloshing motion. Simulations by \cite{vazza2021} as shown in Sect. \ref{sec:radiomorpho} suggest for example that such a structure might originate from the passage of a small clump moving from south towards north.

As the galaxy group NGC~507 is located in a larger gravitationally bound structure, which is the Pisces supercluster, assuming that its dynamical history is dictated by interactions with multiple systems is probably not unrealistic. The nearby (30-kpc projected distance) galaxy NGC~508 could for example play a role in this. NGC~508 and NGC~507 show a significant magnitude difference equal to $\Delta m$=1.42, and so it is unlikely that their interaction is responsible of initiating the sloshing motions of NGC~507. However, \cite{jeltema2008} report that NGC~508 shows small tails in its X-ray emission extending for $\sim$6\,kpc. These features are very rarely detected in the coronae of early-type galaxies \citep{sun2007}, and might support the idea of an interaction with the nearby group NGC~507.

An alternative interpretation for the arc-like X-ray discontinuity could be that it represents a giant roll resulting from Kelvin-Helmholtz instabilities developing at the interface of the  cold front. According to simulations, these instabilities are expected to develop on long timescales (of the order of a gigayear) due to the presence of shear motions (e.g. \citealp{roediger2011}). This argument has already been used in the literature to explain the unusual concave `bay'-like features observed in some relaxed clusters (e.g. Perseus, Centaurus and Abell 1795, \citealp{walker2017}; Abell 2319, \citealp{ichinohe2021}). Nevertheless, this possibility remains difficult to assess and only dedicated hydro-dynamical simulations will be able to shed further light on the overall dynamics of the system.

Regardless of the precise dynamical history of NGC~507, the most likely interpretation of the newly detected extended radio emission is that it is associated with remnant plasma produced by a past phase of activity of the central AGN, which is being transported by the large-scale sloshing motions during its buoyant rise in the group atmosphere. The lack of a strong correlation between the radio and X-ray surface brightness and the lack of strong spectral index gradients overall suggests that the newly detected plasma is still in the process of mixing up with the external medium. This picture is consistent with simulations showing that interactions between old AGN plasma and sloshing cold fronts can deflect the original trajectory of  plasma and transport it up to very large distances from its original position \citep{zuhone2021}. 

Following \cite{osullivan2014}, we computed a first-order estimate of the age of the sloshing structures by assuming that the eastern front has travelled from the group core at no more than half the sound speed $c_s$. The sound speed can be derived as

\begin{equation}
c_s=\sqrt{\Gamma \frac{ k  T}{\mu m_p}},  
\end{equation}

\noindent where $kT$ is the average IGrM temperature as obtained from the X-ray analysis equal to $kT_{X}=$1.19\,keV, $\mu$=0.62 is the mean molecular weight, $\rm \Gamma=5/3$ is the adiabatic index, and $m_p$ is the proton mass. The derived sound speed is equal to $c_s=550$\,km/s. If we assume a (projected) distance between the eastern cold front and the group core of equal to $d=40$\,kpc (2 arcmin), we find that the sloshing motion has taken approximately 145 Myr to reach the current position.

Considering that the radiative age of the newly detected AGN remnant emission as estimated from its radio spectrum is in the range 250-350 Myr, this implies that the AGN remnant plasma was already there at the time when the sloshing motion was initiated. This is consistent with the observed co-spatiality between the radio and X-ray features. Again, a more robust estimate of the age of the sloshing structures will only be obtained in the future with ad hoc simulations.

To the best of our knowledge, the only other clear case in the literature of a galaxy group showing AGN remnant plasma interacting with the sloshing IGrM is NGC~5044 \citep{osullivan2014}. Here an old AGN lobe was suggested to be `trapped' at a distance of $\sim$50 kpc in the arm of hot thermal gas that was pushed towards the core by sloshing motions. Interestingly, this system also features a filament of plasma originating from the central AGN and twisting around a cold front but never connects with the more distant older lobe, contrary to what is observed in NGC~507. 

Hints of AGN remnant emission contained within a cold front and possibly redistributed by sloshing motions are also found in IC 1860 \citep{gastaldello2013}. However, radio observations at higher spatial resolution and sensitivity would be required to confirm this and to constrain the possible interaction between the thermal and non-thermal plasma in more detail.

Overall, NGC~507 represents the clearest observational example to date of how old AGN plasma can be transported away from the host galaxy and diffuse into the surrounding medium in a galaxy group. On longer timescales, this non-thermal plasma will presumably be further disrupted and transported even further in the surrounding IGrM until its particle population and associated magnetic fields completely merge and mix with it. 

This will ultimately represent a pool of mildly energised particles, which, in the case of subsequent sufficiently strong shock passages or turbulence injection, might be re-energised. This seed particle population permeating the ambient medium is considered relevant for the formation of diffuse radio sources in more massive systems, especially mini-halos in cool-core clusters \citep{richardlaferrire2020}. However, in a galaxy group such as NGC~507, mergers and external dynamical interactions might never be energetic enough to initiate such processes. Indeed, classical mini-halos have not so far been detected in such low-mass systems.

Future highly sensitive observations at low frequencies, such as those that can be obtained with the LOFAR Low Band Antennas and in the future with LOFAR 2.0, will be ideal to investigate the presence and distribution of further, older diffuse radio emission in the system.

\begin{acknowledgements} 

M. Brienza acknowledges S. Ettori, G. Migliori and P. Parma for their help and useful discussions. M. Brienza, A. Bonafede, E. Bonnaisseux and C. Riseley acknowledge financial support from the ERC-Stg 674 DRANOEL, no 714245. K. Rajpurohit, M. Brienza and F. Vazza acknowledge financial support from the ERC Starting Grant ``MAGCOW", no. 714196. F. Vazza acknowledges the use of computing resources on JUWELS at JSC, under project Id. 22552 (“RADGALICM”). L. Lovisari, F. Gastaldello acknowledge financial contribution from the contracts ASI-INAF Athena 2019-27-HH.0, "Attività di Studio per la comunità scientifica di Astrofisica delle Alte Energie e Fisica Astroparticellare" (Accordo Attuativo ASI-INAF n. 2017-14-H.0), and from INAF "Call per interventi aggiuntivi a sostegno della ricerca di main stream di INAF".

M.J. Hardcastle acknowledges support from STFC [ST/V000624/1]. A. Drabent acknowledges support by the BMBF Verbundforschung under the grant 05A20STA. R.J. van Weeren acknowledges support from the ERC Starting Grant ClusterWeb 804208. A. Botteon acknowledges support from the VIDI research programme with project number 639.042.729, which is financed by the Netherlands Organisation for Scientific Research (NWO). TP is supported by the BMBF Verbundforschung under grant number 50OR1906. A.S. is supported by the Women In Science Excel (WISE) programme of the Netherlands Organisation for Scientific Research (NWO), and acknowledges the World Premier Research Center Initiative (WPI) and the Kavli IPMU for the continued hospitality. SRON Netherlands Institute for Space Research is supported financially by NWO.

LOFAR, the Low Frequency Array designed and constructed by ASTRON, has facilities in several countries, which are owned by various parties (each with their own funding sources), and are collectively operated by the International LOFAR Telescope (ILT) foundation under a joint scientific policy. The ILT resources have benefited from the following recent major funding sources: CNRS-INSU, Observatoire de Paris and Universit\'e d'Orl\'eans, France; BMBF, MIWF-NRW, MPG, Germany; Science Foundation Ireland (SFI), Department of Business, Enterprise and Innovation (DBEI), Ireland; NWO, The Netherlands; the Science and Technology Facilities Council, UK; Ministry of Science and Higher Education, Poland; The Istituto Nazionale di Astrofisica (INAF), Italy.

Part of this work was carried out on the Dutch national e-infrastructure with the support of the SURF Cooperative through grant e-infra 160022 \& 160152. The LOFAR software and dedicated reduction packages on \url{https://github.com/apmechev/GRID_LRT} were deployed on the e-infrastructure by the LOFAR e-infragrop, consisting of J.\ B.\ R.\ Oonk (ASTRON \& Leiden Observatory), A.\ P.\ Mechev (Leiden Observatory) and T. Shimwell (ASTRON) with support from N.\ Danezi
(SURFsara) and C.\ Schrijvers (SURFsara). The J\"ulich LOFAR Long Term Archive and the German LOFAR network are both coordinated and operated by the J\"ulich Supercomputing Centre (JSC), and computing resources on the supercomputer JUWELS at JSC were provided by the Gauss Centre for supercomputing e.V. (grant CHTB00) through the John von Neumann Institute for Computing (NIC).

This research made use of the University of Hertfordshire
high-performance computing facility and the LOFAR-UK computing facility located at the University of Hertfordshire and supported by STFC (ST/P000096/1), and of the Italian LOFAR IT computing infrastructure supported and operated by INAF, and by the Physics Department of Turin University (under an agreement with Consorzio Interuniversitario per la Fisica Spaziale) at the C3S Supercomputing Centre, Italy.

We thank the staff of the GMRT who have made these observations possible. The GMRT is run by the National Centre for Radio Astrophysics of the Tata Institute of Fundamental Research.

This research has made use of the NASA/IPAC Extragalactic Database (NED), which is operated by the Jet Propulsion Laboratory, California Institute of Technology,under contract with the National Aeronautics and Space Administration. This research made use of APLpy, an open-source plotting package for Python hosted at \url{http://aplpy.github.com}.

\end{acknowledgements}

\bibliographystyle{aa}
\bibliography{ngc507.bib}

\begin{thebibliography}{10}

\bibitem{mcnamara2007}
B.~R. {McNamara} and P.~E.~J. {Nulsen}.
\newblock {Heating Hot Atmospheres with Active Galactic Nuclei}.
\newblock {\em A\&A}, 45:117--175, September 2007.

\bibitem{fabian2012}
A.~C. {Fabian}.
\newblock {Observational Evidence of Active Galactic Nuclei Feedback}.
\newblock {\em \araa}, 50:455--489, September 2012.

\bibitem{gull1973}
S.~F. {Gull} and K.~J.~E. {Northover}.
\newblock {Bubble Model of Extragalactic Radio Sources}.
\newblock {\em \nat}, 244:80--83, July 1973.

\bibitem{werner2019}
N.~{Werner}, B.~R. {McNamara}, E.~{Churazov}, and E.~{Scannapieco}.
\newblock {Hot Atmospheres, Cold Gas, AGN Feedback and the Evolution of Early
  Type Galaxies: A Topical Perspective}.
\newblock {\em \ssr}, 215(1):5, January 2019.

\bibitem{dunn2005}
R.~J.~H. {Dunn}, A.~C. {Fabian}, and G.~B. {Taylor}.
\newblock {Radio bubbles in clusters of galaxies}.
\newblock {\em \mnras}, 364(4):1343--1353, December 2005.

\bibitem{churazov2000}
E.~{Churazov}, W.~{Forman}, C.~{Jones}, and H.~{B{\"o}hringer}.
\newblock {Asymmetric, arc minute scale structures around NGC 1275}.
\newblock {\em \aap}, 356:788--794, April 2000.

\bibitem{bruggen2003}
M.~{Br{\"u}ggen}.
\newblock {Simulations of Buoyant Bubbles in Galaxy Clusters}.
\newblock {\em \apj}, 592(2):839--845, August 2003.

\bibitem{reynolds2005}
Christopher~S. {Reynolds}, Barry {McKernan}, Andrew~C. {Fabian}, James~M.
  {Stone}, and John~C. {Vernaleo}.
\newblock {Buoyant radio lobes in a viscous intracluster medium}.
\newblock {\em \mnras}, 357(1):242--250, February 2005.

\bibitem{ruszkowski2008}
M.~{Ruszkowski}, T.~A. {En{\ss}lin}, M.~{Br{\"u}ggen}, M.~C. {Begelman}, and
  E.~{Churazov}.
\newblock {Cosmic ray confinement in fossil cluster bubbles}.
\newblock {\em \mnras}, 383(4):1359--1365, February 2008.

\bibitem{tully2015}
R.~Brent {Tully}.
\newblock {Galaxy Groups: A 2MASS Catalog}.
\newblock {\em \aj}, 149(5):171, May 2015.

\bibitem{huchra2012}
John~P. {Huchra}, Lucas~M. {Macri}, Karen~L. {Masters}, Thomas~H. {Jarrett},
  Perry {Berlind}, Michael {Calkins}, Aidan~C. {Crook}, Roc {Cutri}, Pirin
  {Erdo{\v{g}}du}, Emilio {Falco}, Teddy {George}, Conrad~M. {Hutcheson}, Ofer
  {Lahav}, Jeff {Mader}, Jessica~D. {Mink}, Nathalie {Martimbeau}, Stephen
  {Schneider}, Michael {Skrutskie}, Susan {Tokarz}, and Michael {Westover}.
\newblock {The 2MASS Redshift Survey{\textemdash}Description and Data Release}.
\newblock {\em \apjs}, 199(2):26, April 2012.

\bibitem{vanhaarlem2013}
M.~P. {van Haarlem}, M.~W. {Wise}, A.~W. {Gunst}, G.~{Heald}, J.~P. {McKean},
  J.~W.~T. {Hessels}, A.~G. {de Bruyn}, R.~{Nijboer}, J.~{Swinbank},
  R.~{Fallows}, M.~{Brentjens}, A.~{Nelles}, R.~{Beck}, H.~{Falcke},
  R.~{Fender}, J.~{H{\"o}randel}, L.~V.~E. {Koopmans}, G.~{Mann}, G.~{Miley},
  H.~{R{\"o}ttgering}, B.~W. {Stappers}, R.~A.~M.~J. {Wijers}, S.~{Zaroubi},
  M.~{van den Akker}, A.~{Alexov}, J.~{Anderson}, K.~{Anderson}, A.~{van
  Ardenne}, M.~{Arts}, A.~{Asgekar}, I.~M. {Avruch}, F.~{Batejat},
  L.~{B{\"a}hren}, M.~E. {Bell}, M.~R. {Bell}, I.~{van Bemmel}, P.~{Bennema},
  M.~J. {Bentum}, G.~{Bernardi}, P.~{Best}, L.~{B{\^i}rzan}, A.~{Bonafede},
  A.-J. {Boonstra}, R.~{Braun}, J.~{Bregman}, F.~{Breitling}, R.~H. {van de
  Brink}, J.~{Broderick}, P.~C. {Broekema}, W.~N. {Brouw}, M.~{Br{\"u}ggen},
  H.~R. {Butcher}, W.~{van Cappellen}, B.~{Ciardi}, T.~{Coenen}, J.~{Conway},
  A.~{Coolen}, A.~{Corstanje}, S.~{Damstra}, O.~{Davies}, A.~T. {Deller}, R.-J.
  {Dettmar}, G.~{van Diepen}, K.~{Dijkstra}, P.~{Donker}, A.~{Doorduin},
  J.~{Dromer}, M.~{Drost}, A.~{van Duin}, J.~{Eisl{\"o}ffel}, J.~{van Enst},
  C.~{Ferrari}, W.~{Frieswijk}, H.~{Gankema}, M.~A. {Garrett}, F.~{de
  Gasperin}, M.~{Gerbers}, E.~{de Geus}, J.-M. {Grie{\ss}meier}, T.~{Grit},
  P.~{Gruppen}, J.~P. {Hamaker}, T.~{Hassall}, M.~{Hoeft}, H.~A. {Holties},
  A.~{Horneffer}, A.~{van der Horst}, A.~{van Houwelingen}, A.~{Huijgen},
  M.~{Iacobelli}, H.~{Intema}, N.~{Jackson}, V.~{Jelic}, A.~{de Jong},
  E.~{Juette}, D.~{Kant}, A.~{Karastergiou}, A.~{Koers}, H.~{Kollen}, V.~I.
  {Kondratiev}, E.~{Kooistra}, Y.~{Koopman}, A.~{Koster}, M.~{Kuniyoshi},
  M.~{Kramer}, G.~{Kuper}, P.~{Lambropoulos}, C.~{Law}, J.~{van Leeuwen},
  J.~{Lemaitre}, M.~{Loose}, P.~{Maat}, G.~{Macario}, S.~{Markoff},
  J.~{Masters}, R.~A. {McFadden}, D.~{McKay-Bukowski}, H.~{Meijering},
  H.~{Meulman}, M.~{Mevius}, E.~{Middelberg}, R.~{Millenaar}, J.~C.~A.
  {Miller-Jones}, R.~N. {Mohan}, J.~D. {Mol}, J.~{Morawietz}, R.~{Morganti},
  D.~D. {Mulcahy}, E.~{Mulder}, H.~{Munk}, L.~{Nieuwenhuis}, R.~{van
  Nieuwpoort}, J.~E. {Noordam}, M.~{Norden}, A.~{Noutsos}, A.~R. {Offringa},
  H.~{Olofsson}, A.~{Omar}, E.~{Orr{\'u}}, R.~{Overeem}, H.~{Paas},
  M.~{Pandey-Pommier}, V.~N. {Pandey}, R.~{Pizzo}, A.~{Polatidis},
  D.~{Rafferty}, S.~{Rawlings}, W.~{Reich}, J.-P. {de Reijer}, J.~{Reitsma},
  G.~A. {Renting}, P.~{Riemers}, E.~{Rol}, J.~W. {Romein}, J.~{Roosjen},
  M.~{Ruiter}, A.~{Scaife}, K.~{van der Schaaf}, B.~{Scheers}, P.~{Schellart},
  A.~{Schoenmakers}, G.~{Schoonderbeek}, M.~{Serylak}, A.~{Shulevski},
  J.~{Sluman}, O.~{Smirnov}, C.~{Sobey}, H.~{Spreeuw}, M.~{Steinmetz}, C.~G.~M.
  {Sterks}, H.-J. {Stiepel}, K.~{Stuurwold}, M.~{Tagger}, Y.~{Tang},
  C.~{Tasse}, I.~{Thomas}, S.~{Thoudam}, M.~C. {Toribio}, B.~{van der Tol},
  O.~{Usov}, M.~{van Veelen}, A.-J. {van der Veen}, S.~{ter Veen}, J.~P.~W.
  {Verbiest}, R.~{Vermeulen}, N.~{Vermaas}, C.~{Vocks}, C.~{Vogt}, M.~{de Vos},
  E.~{van der Wal}, R.~{van Weeren}, H.~{Weggemans}, P.~{Weltevrede},
  S.~{White}, S.~J. {Wijnholds}, T.~{Wilhelmsson}, O.~{Wucknitz},
  S.~{Yatawatta}, P.~{Zarka}, A.~{Zensus}, and J.~{van Zwieten}.
\newblock {LOFAR: The LOw-Frequency ARray}.
\newblock {\em \aap}, 556:A2, August 2013.

\bibitem{shimwell2019}
T.~W. {Shimwell}, C.~{Tasse}, M.~J. {Hardcastle}, A.~P. {Mechev}, W.~L.
  {Williams}, P.~N. {Best}, H.~J.~A. {R{\"o}ttgering}, J.~R. {Callingham},
  T.~J. {Dijkema}, F.~{de Gasperin}, D.~N. {Hoang}, B.~{Hugo}, M.~{Mirmont},
  J.~B.~R. {Oonk}, I.~{Prandoni}, D.~{Rafferty}, J.~{Sabater}, O.~{Smirnov},
  R.~J. {van Weeren}, G.~J. {White}, M.~{Atemkeng}, L.~{Bester},
  E.~{Bonnassieux}, M.~{Br{\"u}ggen}, G.~{Brunetti}, K.~T. {Chy{\.z}y},
  R.~{Cochrane}, J.~E. {Conway}, J.~H. {Croston}, A.~{Danezi}, K.~{Duncan},
  M.~{Haverkorn}, G.~H. {Heald}, M.~{Iacobelli}, H.~T. {Intema}, N.~{Jackson},
  M.~{Jamrozy}, M.~J. {Jarvis}, R.~{Lakhoo}, M.~{Mevius}, G.~K. {Miley},
  L.~{Morabito}, R.~{Morganti}, D.~{Nisbet}, E.~{Orr{\'u}}, S.~{Perkins}, R.~F.
  {Pizzo}, C.~{Schrijvers}, D.~J.~B. {Smith}, R.~{Vermeulen}, M.~W. {Wise},
  L.~{Alegre}, D.~J. {Bacon}, I.~M. {van Bemmel}, R.~J. {Beswick},
  A.~{Bonafede}, A.~{Botteon}, S.~{Bourke}, M.~{Brienza}, G.~{Calistro Rivera},
  R.~{Cassano}, A.~O. {Clarke}, C.~J. {Conselice}, R.~J. {Dettmar},
  A.~{Drabent}, C.~{Dumba}, K.~L. {Emig}, T.~A. {En{\ss}lin}, C.~{Ferrari},
  M.~A. {Garrett}, R.~T. {G{\'e}nova-Santos}, A.~{Goyal}, G.~{G{\"u}rkan},
  C.~{Hale}, J.~J. {Harwood}, V.~{Heesen}, M.~{Hoeft}, C.~{Horellou},
  C.~{Jackson}, G.~{Kokotanekov}, R.~{Kondapally}, M.~{Kunert-Bajraszewska},
  V.~{Mahatma}, E.~K. {Mahony}, S.~{Mandal}, J.~P. {McKean}, A.~{Merloni},
  B.~{Mingo}, A.~{Miskolczi}, S.~{Mooney}, B.~{Nikiel-Wroczy{\'n}ski}, S.~P.
  {O'Sullivan}, J.~{Quinn}, W.~{Reich}, C.~{Roskowi{\'n}ski}, A.~{Rowlinson},
  F.~{Savini}, A.~{Saxena}, D.~J. {Schwarz}, A.~{Shulevski}, S.~S. {Sridhar},
  H.~R. {Stacey}, S.~{Urquhart}, M.~H.~D. {van der Wiel}, E.~{Varenius},
  B.~{Webster}, and A.~{Wilber}.
\newblock {The LOFAR Two-metre Sky Survey. II. First data release}.
\newblock {\em \aap}, 622:A1, February 2019.

\bibitem{predehl2021}
P.~{Predehl}, R.~{Andritschke}, V.~{Arefiev}, V.~{Babyshkin}, O.~{Batanov},
  W.~{Becker}, H.~{B{\"o}hringer}, A.~{Bogomolov}, T.~{Boller}, K.~{Borm},
  W.~{Bornemann}, H.~{Br{\"a}uninger}, M.~{Br{\"u}ggen}, H.~{Brunner},
  M.~{Brusa}, E.~{Bulbul}, M.~{Buntov}, V.~{Burwitz}, W.~{Burkert}, N.~{Clerc},
  E.~{Churazov}, D.~{Coutinho}, T.~{Dauser}, K.~{Dennerl}, V.~{Doroshenko},
  J.~{Eder}, V.~{Emberger}, T.~{Eraerds}, A.~{Finoguenov}, M.~{Freyberg},
  P.~{Friedrich}, S.~{Friedrich}, M.~{F{\"u}rmetz}, A.~{Georgakakis},
  M.~{Gilfanov}, S.~{Granato}, C.~{Grossberger}, A.~{Gueguen}, P.~{Gureev},
  F.~{Haberl}, O.~{H{\"a}lker}, G.~{Hartner}, G.~{Hasinger}, H.~{Huber},
  L.~{Ji}, A.~v. {Kienlin}, W.~{Kink}, F.~{Korotkov}, I.~{Kreykenbohm},
  G.~{Lamer}, I.~{Lomakin}, I.~{Lapshov}, T.~{Liu}, C.~{Maitra},
  N.~{Meidinger}, B.~{Menz}, A.~{Merloni}, T.~{Mernik}, B.~{Mican}, J.~{Mohr},
  S.~{M{\"u}ller}, K.~{Nandra}, V.~{Nazarov}, F.~{Pacaud}, M.~{Pavlinsky},
  E.~{Perinati}, E.~{Pfeffermann}, D.~{Pietschner}, M.~E. {Ramos-Ceja},
  A.~{Rau}, J.~{Reiffers}, T.~H. {Reiprich}, J.~{Robrade}, M.~{Salvato},
  J.~{Sanders}, A.~{Santangelo}, M.~{Sasaki}, H.~{Scheuerle}, C.~{Schmid},
  J.~{Schmitt}, A.~{Schwope}, A.~{Shirshakov}, M.~{Steinmetz}, I.~{Stewart},
  L.~{Str{\"u}der}, R.~{Sunyaev}, C.~{Tenzer}, L.~{Tiedemann},
  J.~{Tr{\"u}mper}, V.~{Voron}, P.~{Weber}, J.~{Wilms}, and V.~{Yaroshenko}.
\newblock {The eROSITA X-ray telescope on SRG}.
\newblock {\em \aap}, 647:A1, March 2021.

\bibitem{schoenmakers2000}
A.~P. {Schoenmakers}, A.~G. {de Bruyn}, H.~J.~A. {R{\"o}ttgering}, H.~{van der
  Laan}, and C.~R. {Kaiser}.
\newblock {Radio galaxies with a `double-double morphology' - I. Analysis of
  the radio properties and evidence for interrupted activity in active galactic
  nuclei}.
\newblock {\em \mnras}, 315:371--380, June 2000.

\bibitem{finoguenov2001}
A.~{Finoguenov} and C.~{Jones}.
\newblock {Chandra Observation of M84, a Radio Lobe Elliptical Galaxy in the
  Virgo Cluster}.
\newblock {\em \apjl}, 547(2):L107--L110, February 2001.

\bibitem{owen2000}
Frazer~N. {Owen}, Jean~A. {Eilek}, and Namir~E. {Kassim}.
\newblock {M87 at 90 Centimeters: A Different Picture}.
\newblock {\em \apj}, 543(2):611--619, November 2000.

\bibitem{churazov2001}
E.~{Churazov}, M.~{Br{\"u}ggen}, C.~R. {Kaiser}, H.~{B{\"o}hringer}, and
  W.~{Forman}.
\newblock {Evolution of Buoyant Bubbles in M87}.
\newblock {\em \apj}, 554(1):261--273, June 2001.

\bibitem{yang2019}
H.~Y.~Karen {Yang}, Massimo {Gaspari}, and Carl {Marlow}.
\newblock {The Impact of Radio AGN Bubble Composition on the Dynamics and
  Thermal Balance of the Intracluster Medium}.
\newblock {\em \apj}, 871(1):6, January 2019.

\bibitem{markevitch2005}
M.~{Markevitch}, F.~{Govoni}, G.~{Brunetti}, and D.~{Jerius}.
\newblock {Bow Shock and Radio Halo in the Merging Cluster A520}.
\newblock {\em \apj}, 627(2):733--738, July 2005.

\bibitem{ensslin2002}
T.~A. {En{\ss}lin} and M.~{Br{\"u}ggen}.
\newblock {On the formation of cluster radio relics}.
\newblock {\em \mnras}, 331:1011--1019, April 2002.

\bibitem{rajpurohit2018}
K.~{Rajpurohit}, M.~{Hoeft}, R.~J. {van Weeren}, L.~{Rudnick}, H.~J.~A.
  {R{\"o}ttgering}, W.~R. {Forman}, M.~{Br{\"u}ggen}, J.~H. {Croston},
  F.~{Andrade-Santos}, W.~A. {Dawson}, H.~T. {Intema}, R.~P. {Kraft},
  C.~{Jones}, and M.~James {Jee}.
\newblock {Deep VLA Observations of the Cluster 1RXS J0603.3+4214 in the
  Frequency Range of 1-2 GHz}.
\newblock {\em \apj}, 852(2):65, January 2018.

\bibitem{degasperin2014b}
F.~{de Gasperin}, R.~J. {van Weeren}, M.~{Br{\"u}ggen}, F.~{Vazza},
  A.~{Bonafede}, and H.~T. {Intema}.
\newblock {A new double radio relic in PSZ1 G096.89+24.17 and a radio relic
  mass-luminosity relation}.
\newblock {\em \mnras}, 444(4):3130--3138, November 2014.

\bibitem{wise2007}
M.~W. {Wise}, B.~R. {McNamara}, P.~E.~J. {Nulsen}, J.~C. {Houck}, and L.~P.
  {David}.
\newblock {X-Ray Supercavities in the Hydra A Cluster and the Outburst History
  of the Central Galaxy's Active Nucleus}.
\newblock {\em \apj}, 659(2):1153--1158, April 2007.

\bibitem{birzan2004}
L.~{B{\^\i}rzan}, D.~A. {Rafferty}, B.~R. {McNamara}, M.~W. {Wise}, and
  P.~E.~J. {Nulsen}.
\newblock {A Systematic Study of Radio-induced X-Ray Cavities in Clusters,
  Groups, and Galaxies}.
\newblock {\em \apj}, 607(2):800--809, June 2004.

\bibitem{oneill2009}
S.~M. {O'Neill}, D.~S. {De Young}, and T.~W. {Jones}.
\newblock {Three-Dimensional Magnetohydrodynamic Simulations of Buoyant Bubbles
  in Galaxy Clusters}.
\newblock {\em \apj}, 694(2):1317--1330, April 2009.

\bibitem{sunyaev2003}
R.~A. {Sunyaev}, M.~L. {Norman}, and G.~L. {Bryan}.
\newblock {On the Detectability of Turbulence and Bulk Flows in X-ray
  Clusters}.
\newblock {\em Astronomy Letters}, 29:783--790, December 2003.

\bibitem{porter2015}
David~H. {Porter}, T.~W. {Jones}, and Dongsu {Ryu}.
\newblock {Vorticity, Shocks, and Magnetic Fields in Subsonic, ICM-like
  Turbulence}.
\newblock {\em \apj}, 810(2):93, September 2015.

\bibitem{xu2019}
Siyao {Xu}, Suoqing {Ji}, and A.~{Lazarian}.
\newblock {On the Formation of Density Filaments in the Turbulent Interstellar
  Medium}.
\newblock {\em \apj}, 878(2):157, June 2019.

\bibitem{zhuravleva2014}
I.~{Zhuravleva}, E.~{Churazov}, A.~A. {Schekochihin}, S.~W. {Allen},
  P.~{Ar{\'e}valo}, A.~C. {Fabian}, W.~R. {Forman}, J.~S. {Sanders},
  A.~{Simionescu}, R.~{Sunyaev}, A.~{Vikhlinin}, and N.~{Werner}.
\newblock {Turbulent heating in galaxy clusters brightest in X-rays}.
\newblock {\em \nat}, 515(7525):85--87, November 2014.

\bibitem{ehlert2018}
K.~{Ehlert}, R.~{Weinberger}, C.~{Pfrommer}, R.~{Pakmor}, and V.~{Springel}.
\newblock {Simulations of the dynamics of magnetized jets and cosmic rays in
  galaxy clusters}.
\newblock {\em \mnras}, 481(3):2878--2900, December 2018.

\bibitem{degasperin2019}
F.~{de Gasperin}, T.~J. {Dijkema}, A.~{Drabent}, M.~{Mevius}, D.~{Rafferty},
  R.~{van Weeren}, M.~{Br{\"u}ggen}, J.~R. {Callingham}, K.~L. {Emig},
  G.~{Heald}, H.~T. {Intema}, L.~K. {Morabito}, A.~R. {Offringa}, R.~{Oonk},
  E.~{Orr{\`u}}, H.~{R{\"o}ttgering}, J.~{Sabater}, T.~{Shimwell},
  A.~{Shulevski}, and W.~{Williams}.
\newblock {Systematic effects in LOFAR data: A unified calibration strategy}.
\newblock {\em \aap}, 622:A5, February 2019.

\bibitem{vanweeren2016}
R.~J. {van Weeren}, W.~L. {Williams}, M.~J. {Hardcastle}, T.~W. {Shimwell},
  D.~A. {Rafferty}, J.~{Sabater}, G.~{Heald}, S.~S. {Sridhar}, T.~J. {Dijkema},
  G.~{Brunetti}, M.~{Br{\"u}ggen}, F.~{Andrade-Santos}, G.~A. {Ogrean},
  H.~J.~A. {R{\"o}ttgering}, W.~A. {Dawson}, W.~R. {Forman}, F.~{de Gasperin},
  C.~{Jones}, G.~K. {Miley}, L.~{Rudnick}, C.~L. {Sarazin}, A.~{Bonafede},
  P.~N. {Best}, L.~{B{\^i}rzan}, R.~{Cassano}, K.~T. {Chy{\.z}y}, J.~H.
  {Croston}, T.~{Ensslin}, C.~{Ferrari}, M.~{Hoeft}, C.~{Horellou}, M.~J.
  {Jarvis}, R.~P. {Kraft}, M.~{Mevius}, H.~T. {Intema}, S.~S. {Murray},
  E.~{Orr{\'u}}, R.~{Pizzo}, A.~{Simionescu}, A.~{Stroe}, S.~{van der Tol}, and
  G.~J. {White}.
\newblock {LOFAR Facet Calibration}.
\newblock {\em \apjs}, 223:2, March 2016.

\bibitem{williams2016}
W.~L. {Williams}, R.~J. {van Weeren}, H.~J.~A. {R{\"o}ttgering}, P.~{Best},
  T.~J. {Dijkema}, F.~{de Gasperin}, M.~J. {Hardcastle}, G.~{Heald},
  I.~{Prandoni}, J.~{Sabater}, T.~W. {Shimwell}, C.~{Tasse}, I.~M. {van
  Bemmel}, M.~{Br{\"u}ggen}, G.~{Brunetti}, J.~E. {Conway}, T.~{En{\ss}lin},
  D.~{Engels}, H.~{Falcke}, C.~{Ferrari}, M.~{Haverkorn}, N.~{Jackson}, M.~J.
  {Jarvis}, A.~D. {Kapi{\'n}ska}, E.~K. {Mahony}, G.~K. {Miley}, L.~K.
  {Morabito}, R.~{Morganti}, E.~{Orr{\'u}}, E.~{Retana-Montenegro}, S.~S.
  {Sridhar}, M.~C. {Toribio}, G.~J. {White}, M.~W. {Wise}, and J.~T.~L.
  {Zwart}.
\newblock {LOFAR 150-MHz observations of the Bo{\"o}tes field: catalogue and
  source counts}.
\newblock {\em \mnras}, 460:2385--2412, August 2016.

\bibitem{tasse2021}
C.~{Tasse}, T.~{Shimwell}, M.~J. {Hardcastle}, S.~P. {O'Sullivan}, R.~{van
  Weeren}, P.~N. {Best}, L.~{Bester}, B.~{Hugo}, O.~{Smirnov}, J.~{Sabater},
  G.~{Calistro-Rivera}, F.~{de Gasperin}, L.~K. {Morabito},
  H.~{R{\"o}ttgering}, W.~L. {Williams}, M.~{Bonato}, M.~{Bondi}, A.~{Botteon},
  M.~{Br{\"u}ggen}, G.~{Brunetti}, K.~T. {Chy{\.z}y}, M.~A. {Garrett},
  G.~{G{\"u}rkan}, M.~J. {Jarvis}, R.~{Kondapally}, S.~{Mandal}, I.~{Prandoni},
  A.~{Repetti}, E.~{Retana-Montenegro}, D.~J. {Schwarz}, A.~{Shulevski}, and
  Y.~{Wiaux}.
\newblock {The LOFAR Two-meter Sky Survey: Deep Fields Data Release 1. I.
  Direction-dependent calibration and imaging}.
\newblock {\em \aap}, 648:A1, April 2021.

\bibitem{tasse2014}
C.~{Tasse}.
\newblock {Nonlinear Kalman filters for calibration in radio interferometry}.
\newblock {\em \aap}, 566:A127, June 2014.

\bibitem{smirnov2015}
O.~M. {Smirnov} and C.~{Tasse}.
\newblock {Radio interferometric gain calibration as a complex optimization
  problem}.
\newblock {\em \mnras}, 449(3):2668--2684, May 2015.

\bibitem{tasse2018}
C.~{Tasse}, B.~{Hugo}, M.~{Mirmont}, O.~{Smirnov}, M.~{Atemkeng}, L.~{Bester},
  M.~J. {Hardcastle}, R.~{Lakhoo}, S.~{Perkins}, and T.~{Shimwell}.
\newblock {Faceting for direction-dependent spectral deconvolution}.
\newblock {\em \aap}, 611:A87, April 2018.

\bibitem{vanweeren2020}
R.~J. {van Weeren}, T.~W. {Shimwell}, A.~{Botteon}, G.~{Brunetti},
  M.~{Br{\"u}ggen}, J.~M. {Boxelaar}, R.~{Cassano}, G.~{Di Gennaro},
  F.~{Andrade-Santos}, E.~{Bonnassieux}, A.~{Bonafede}, V.~{Cuciti},
  D.~{Dallacasa}, F.~{de Gasperin}, F.~{Gastaldello}, M.~J. {Hardcastle},
  M.~{Hoeft}, R~.~P. {Kraft}, S.~{Mandal}, M.~{Rossetti}, H.~J.~A.
  {R{\"o}ttgering}, C.~{Tasse}, and A.~G. {Wilber}.
\newblock {LOFAR observations of galaxy clusters in HETDEX}.
\newblock {\em arXiv e-prints}, page arXiv:2011.02387, November 2020.

\bibitem{offringa2014}
A.~R. {Offringa}, B.~{McKinley}, N.~{Hurley-Walker}, F.~H. {Briggs}, R.~B.
  {Wayth}, D.~L. {Kaplan}, M.~E. {Bell}, L.~{Feng}, A.~R. {Neben}, J.~D.
  {Hughes}, J.~{Rhee}, T.~{Murphy}, N.~D.~R. {Bhat}, G.~{Bernardi}, J.~D.
  {Bowman}, R.~J. {Cappallo}, B.~E. {Corey}, A.~A. {Deshpande}, D.~{Emrich},
  A.~{Ewall-Wice}, B.~M. {Gaensler}, R.~{Goeke}, L.~J. {Greenhill}, B.~J.
  {Hazelton}, L.~{Hindson}, M.~{Johnston-Hollitt}, D.~C. {Jacobs}, J.~C.
  {Kasper}, E.~{Kratzenberg}, E.~{Lenc}, C.~J. {Lonsdale}, M.~J. {Lynch}, S.~R.
  {McWhirter}, D.~A. {Mitchell}, M.~F. {Morales}, E.~{Morgan},
  N.~{Kudryavtseva}, D.~{Oberoi}, S.~M. {Ord}, B.~{Pindor}, P.~{Procopio},
  T.~{Prabu}, J.~{Riding}, D.~A. {Roshi}, N.~U. {Shankar}, K.~S. {Srivani},
  R.~{Subrahmanyan}, S.~J. {Tingay}, M.~{Waterson}, R.~L. {Webster}, A.~R.
  {Whitney}, A.~{Williams}, and C.~L. {Williams}.
\newblock {WSCLEAN: an implementation of a fast, generic wide-field imager for
  radio astronomy}.
\newblock {\em \mnras}, 444:606--619, October 2014.

\bibitem{offringa2012}
A.~R. {Offringa}, J.~J. {van de Gronde}, and J.~B.~T.~M. {Roerdink}.
\newblock {A morphological algorithm for improving radio-frequency interference
  detection}.
\newblock {\em \aap}, 539:A95, March 2012.

\bibitem{degasperin2020}
F.~{de Gasperin}, T.~J.~W. {Lazio}, and M.~{Knapp}.
\newblock {Radio Observations of HD80606 Near Planetary Periastron: II. LOFAR
  Low Band Antenna Observations at 30-78 MHz}.
\newblock {\em arXiv e-prints}, page arXiv:2011.05696, November 2020.

\bibitem{mevius2016}
M.~{Mevius}, S.~{van der Tol}, V.~N. {Pandey}, H.~K. {Vedantham}, M.~A.
  {Brentjens}, A.~G. {de Bruyn}, F.~B. {Abdalla}, K.~M.~B. {Asad}, J.~D.
  {Bregman}, W.~N. {Brouw}, S.~{Bus}, E.~{Chapman}, B.~{Ciardi}, E.~R.
  {Fernandez}, A.~{Ghosh}, G.~{Harker}, I.~T. {Iliev}, V.~{Jeli{\'c}},
  S.~{Kazemi}, L.~V.~E. {Koopmans}, J.~E. {Noordam}, A.~R. {Offringa}, A.~H.
  {Patil}, R.~J. {van Weeren}, S.~{Wijnholds}, S.~{Yatawatta}, and
  S.~{Zaroubi}.
\newblock {Probing ionospheric structures using the LOFAR radio telescope}.
\newblock {\em Radio Science}, 51(7):927--941, July 2016.

\bibitem{degasperin2018}
F.~{de Gasperin}, M.~{Mevius}, D.~A. {Rafferty}, H.~T. {Intema}, and R.~A.
  {Fallows}.
\newblock {The effect of the ionosphere on ultra-low-frequency
  radio-interferometric observations}.
\newblock {\em \aap}, 615:A179, August 2018.

\bibitem{degasperin2021}
F.~{de Gasperin}, W.~L. {Williams}, P.~{Best}, M.~{Bruggen}, G.~{Brunetti},
  V.~{Cuciti}, T.~J. {Dijkema}, M.~J. {Hardcastle}, M.~J. {Norden},
  A.~{Offringa}, T.~{Shimwell}, R.~{van Weeren}, D.~{Bomans}, A.~{Bonafede},
  A.~{Botteon}, J.~R. {Callingham}, R.~{Cassano}, K.~T. {Chyzy}, K.~L. {Emig},
  H.~{Edler}, M.~{Haverkorn}, G.~{Heald}, V.~{Heesen}, M.~{Iacobelli}, H.~T.
  {Intema}, M.~{Kadler}, K.~{Malek}, M.~{Mevius}, G.~{Miley}, B.~{Mingo}, L.~K.
  {Morabito}, J.~{Sabater}, R.~{Morganti}, E.~{Orru}, R.~{Pizzo},
  I.~{Prandoni}, A.~{Shulevski}, C.~{Tasse}, M.~{Vaccari}, P.~{Zarka}, and
  H.~{Rottgering}.
\newblock {The LOFAR LBA Sky Survey I. survey description and preliminary data
  release}.
\newblock {\em arXiv e-prints}, page arXiv:2102.09238, February 2021.

\bibitem{rossetti2016}
M.~{Rossetti}, F.~{Gastaldello}, G.~{Ferioli}, M.~{Bersanelli}, S.~{De Grandi},
  D.~{Eckert}, S.~{Ghizzardi}, D.~{Maino}, and S.~{Molendi}.
\newblock {Measuring the dynamical state of Planck SZ-selected clusters: X-ray
  peak - BCG offset}.
\newblock {\em \mnras}, 457(4):4515--4524, April 2016.

\bibitem{hi4pi2016}
{HI4PI Collaboration}, N.~{Ben Bekhti}, L.~{Fl{\"o}er}, R.~{Keller}, J.~{Kerp},
  D.~{Lenz}, B.~{Winkel}, J.~{Bailin}, M.~R. {Calabretta}, L.~{Dedes}, H.~A.
  {Ford}, B.~K. {Gibson}, U.~{Haud}, S.~{Janowiecki}, P.~M.~W. {Kalberla},
  F.~J. {Lockman}, N.~M. {McClure-Griffiths}, T.~{Murphy}, H.~{Nakanishi},
  D.~J. {Pisano}, and L.~{Staveley-Smith}.
\newblock {HI4PI: A full-sky H I survey based on EBHIS and GASS}.
\newblock {\em \aap}, 594:A116, October 2016.

\bibitem{planck2014}
{Planck Collaboration}, A.~{Abergel}, P.~A.~R. {Ade}, N.~{Aghanim}, M.~I.~R.
  {Alves}, G.~{Aniano}, C.~{Armitage-Caplan}, M.~{Arnaud}, M.~{Ashdown},
  F.~{Atrio-Barandela}, J.~{Aumont}, C.~{Baccigalupi}, A.~J. {Banday}, R.~B.
  {Barreiro}, J.~G. {Bartlett}, E.~{Battaner}, K.~{Benabed}, A.~{Beno{\^\i}t},
  A.~{Benoit-L{\'e}vy}, J.~P. {Bernard}, M.~{Bersanelli}, P.~{Bielewicz},
  J.~{Bobin}, J.~J. {Bock}, A.~{Bonaldi}, J.~R. {Bond}, J.~{Borrill}, F.~R.
  {Bouchet}, F.~{Boulanger}, M.~{Bridges}, M.~{Bucher}, C.~{Burigana}, R.~C.
  {Butler}, J.~F. {Cardoso}, A.~{Catalano}, A.~{Chamballu}, R.~R. {Chary},
  H.~C. {Chiang}, L.~Y. {Chiang}, P.~R. {Christensen}, S.~{Church},
  M.~{Clemens}, D.~L. {Clements}, S.~{Colombi}, L.~P.~L. {Colombo},
  C.~{Combet}, F.~{Couchot}, A.~{Coulais}, B.~P. {Crill}, A.~{Curto},
  F.~{Cuttaia}, L.~{Danese}, R.~D. {Davies}, R.~J. {Davis}, P.~{de Bernardis},
  A.~{de Rosa}, G.~{de Zotti}, J.~{Delabrouille}, J.~M. {Delouis}, F.~X.
  {D{\'e}sert}, C.~{Dickinson}, J.~M. {Diego}, H.~{Dole}, S.~{Donzelli},
  O.~{Dor{\'e}}, M.~{Douspis}, B.~T. {Draine}, X.~{Dupac}, G.~{Efstathiou},
  T.~A. {En{\ss}lin}, H.~K. {Eriksen}, E.~{Falgarone}, F.~{Finelli},
  O.~{Forni}, M.~{Frailis}, A.~A. {Fraisse}, E.~{Franceschi}, S.~{Galeotta},
  K.~{Ganga}, T.~{Ghosh}, M.~{Giard}, G.~{Giardino}, Y.~{Giraud-H{\'e}raud},
  J.~{Gonz{\'a}lez-Nuevo}, K.~M. {G{\'o}rski}, S.~{Gratton}, A.~{Gregorio},
  I.~A. {Grenier}, A.~{Gruppuso}, V.~{Guillet}, F.~K. {Hansen}, D.~{Hanson},
  D.~L. {Harrison}, G.~{Helou}, S.~{Henrot-Versill{\'e}},
  C.~{Hern{\'a}ndez-Monteagudo}, D.~{Herranz}, S.~R. {Hildebrandt}, E.~{Hivon},
  M.~{Hobson}, W.~A. {Holmes}, A.~{Hornstrup}, W.~{Hovest}, K.~M.
  {Huffenberger}, A.~H. {Jaffe}, T.~R. {Jaffe}, J.~{Jewell}, G.~{Joncas}, W.~C.
  {Jones}, M.~{Juvela}, E.~{Keih{\"a}nen}, R.~{Keskitalo}, T.~S. {Kisner},
  J.~{Knoche}, L.~{Knox}, M.~{Kunz}, H.~{Kurki-Suonio}, G.~{Lagache},
  A.~{L{\"a}hteenm{\"a}ki}, J.~M. {Lamarre}, A.~{Lasenby}, R.~J. {Laureijs},
  C.~R. {Lawrence}, R.~{Leonardi}, J.~{Le{\'o}n-Tavares}, J.~{Lesgourgues},
  F.~{Levrier}, M.~{Liguori}, P.~B. {Lilje}, M.~{Linden-V{\o}rnle},
  M.~{L{\'o}pez-Caniego}, P.~M. {Lubin}, J.~F. {Mac{\'\i}as-P{\'e}rez},
  B.~{Maffei}, D.~{Maino}, N.~{Mandolesi}, M.~{Maris}, D.~J. {Marshall}, P.~G.
  {Martin}, E.~{Mart{\'\i}nez-Gonz{\'a}lez}, S.~{Masi}, M.~{Massardi},
  S.~{Matarrese}, F.~{Matthai}, P.~{Mazzotta}, P.~{McGehee}, A.~{Melchiorri},
  L.~{Mendes}, A.~{Mennella}, M.~{Migliaccio}, S.~{Mitra}, M.~A.
  {Miville-Desch{\^e}nes}, A.~{Moneti}, L.~{Montier}, G.~{Morgante},
  D.~{Mortlock}, D.~{Munshi}, J.~A. {Murphy}, P.~{Naselsky}, F.~{Nati},
  P.~{Natoli}, C.~B. {Netterfield}, H.~U. {N{\o}rgaard-Nielsen}, F.~{Noviello},
  D.~{Novikov}, I.~{Novikov}, S.~{Osborne}, C.~A. {Oxborrow}, F.~{Paci},
  L.~{Pagano}, F.~{Pajot}, R.~{Paladini}, D.~{Paoletti}, F.~{Pasian},
  G.~{Patanchon}, O.~{Perdereau}, L.~{Perotto}, F.~{Perrotta}, F.~{Piacentini},
  M.~{Piat}, E.~{Pierpaoli}, D.~{Pietrobon}, S.~{Plaszczynski},
  E.~{Pointecouteau}, G.~{Polenta}, N.~{Ponthieu}, L.~{Popa}, T.~{Poutanen},
  G.~W. {Pratt}, G.~{Pr{\'e}zeau}, S.~{Prunet}, J.~L. {Puget}, J.~P. {Rachen},
  W.~T. {Reach}, R.~{Rebolo}, M.~{Reinecke}, M.~{Remazeilles}, C.~{Renault},
  S.~{Ricciardi}, T.~{Riller}, I.~{Ristorcelli}, G.~{Rocha}, C.~{Rosset},
  G.~{Roudier}, M.~{Rowan-Robinson}, J.~A. {Rubi{\~n}o-Mart{\'\i}n},
  B.~{Rusholme}, M.~{Sandri}, D.~{Santos}, G.~{Savini}, D.~{Scott}, M.~D.
  {Seiffert}, E.~P.~S. {Shellard}, L.~D. {Spencer}, J.~L. {Starck},
  V.~{Stolyarov}, R.~{Stompor}, R.~{Sudiwala}, R.~{Sunyaev}, F.~{Sureau},
  D.~{Sutton}, A.~S. {Suur-Uski}, J.~F. {Sygnet}, J.~A. {Tauber},
  D.~{Tavagnacco}, L.~{Terenzi}, L.~{Toffolatti}, M.~{Tomasi}, M.~{Tristram},
  M.~{Tucci}, J.~{Tuovinen}, M.~{T{\"u}rler}, G.~{Umana}, L.~{Valenziano},
  J.~{Valiviita}, B.~{Van Tent}, L.~{Verstraete}, P.~{Vielva}, F.~{Villa},
  N.~{Vittorio}, L.~A. {Wade}, B.~D. {Wandelt}, N.~{Welikala}, N.~{Ysard},
  D.~{Yvon}, A.~{Zacchei}, and A.~{Zonca}.
\newblock {Planck 2013 results. XI. All-sky model of thermal dust emission}.
\newblock {\em \aap}, 571:A11, November 2014.

\bibitem{schlegel1998}
David~J. {Schlegel}, Douglas~P. {Finkbeiner}, and Marc {Davis}.
\newblock {Maps of Dust Infrared Emission for Use in Estimation of Reddening
  and Cosmic Microwave Background Radiation Foregrounds}.
\newblock {\em \apj}, 500(2):525--553, June 1998.

\bibitem{foster2012}
A.~R. {Foster}, L.~{Ji}, R.~K. {Smith}, and N.~S. {Brickhouse}.
\newblock {Updated Atomic Data and Calculations for X-Ray Spectroscopy}.
\newblock {\em \apj}, 756(2):128, September 2012.

\bibitem{lovisari2015}
L.~{Lovisari}, T.~H. {Reiprich}, and G.~{Schellenberger}.
\newblock {Scaling properties of a complete X-ray selected galaxy group
  sample}.
\newblock {\em \aap}, 573:A118, January 2015.

\bibitem{hardcastle2013}
M.~J. {Hardcastle} and M.~G.~H. {Krause}.
\newblock {Numerical modelling of the lobes of radio galaxies in cluster
  environments}.
\newblock {\em \mnras}, 430:174--196, March 2013.

\bibitem{hardcastle1998}
M.~J. {Hardcastle}, M.~{Birkinshaw}, and D.~M. {Worrall}.
\newblock {Magnetic field strengths in the hotspots of 3C 33 and 111.}
\newblock {\em Monthly Notices of the Royal Astronomical Society},
  294:615--621, Mar 1998.

\bibitem{jaffe1973}
W.~J. {Jaffe} and G.~C. {Perola}.
\newblock {Dynamical Models of Tailed Radio Sources in Clusters of Galaxies}.
\newblock {\em \aap}, 26:423, August 1973.

\bibitem{carilli1991}
C.~L. {Carilli}, R.~A. {Perley}, J.~W. {Dreher}, and J.~P. {Leahy}.
\newblock {Multifrequency radio observations of Cygnus A - Spectral aging in
  powerful radio galaxies}.
\newblock {\em \apj}, 383:554--573, December 1991.

\bibitem{croston2018}
J.~H. {Croston}, J.~{Ineson}, and M.~J. {Hardcastle}.
\newblock {Particle content, radio-galaxy morphology, and jet power: all
  radio-loud AGN are not equal}.
\newblock {\em \mnras}, 476(2):1614--1623, May 2018.

\bibitem{binney1987}
James {Binney} and Scott {Tremaine}.
\newblock {\em {Galactic dynamics}}.
\newblock 1987.

\bibitem{zhang2018}
Congyao {Zhang}, Eugene {Churazov}, and Alexander~A. {Schekochihin}.
\newblock {Generation of internal waves by buoyant bubbles in galaxy clusters
  and heating of intracluster medium}.
\newblock {\em \mnras}, 478(4):4785--4798, August 2018.

\bibitem{harwood2013}
J.~J. {Harwood}, M.~J. {Hardcastle}, J.~H. {Croston}, and J.~L. {Goodger}.
\newblock {Spectral ageing in the lobes of FR-II radio galaxies: new methods of
  analysis for broad-band radio data}.
\newblock {\em \mnras}, 435:3353--3375, November 2013.

\bibitem{vandenbosch2015}
Remco C.~E. {van den Bosch}, Karl {Gebhardt}, Kayhan {G{\"u}ltekin}, Akin
  {Y{\i}ld{\i}r{\i}m}, and Jonelle~L. {Walsh}.
\newblock {Hunting for Supermassive Black Holes in Nearby Galaxies With the
  Hobby-Eberly Telescope}.
\newblock {\em \apjs}, 218(1):10, May 2015.

\bibitem{macchetto1997}
F.~{Macchetto}, A.~{Marconi}, D.~J. {Axon}, A.~{Capetti}, W.~{Sparks}, and
  P.~{Crane}.
\newblock {The Supermassive Black Hole of M87 and the Kinematics of Its
  Associated Gaseous Disk}.
\newblock {\em \apj}, 489(2):579--600, November 1997.

\bibitem{anderson2018}
Michael~E. {Anderson} and Rashid {Sunyaev}.
\newblock {FUV line emission, gas kinematics, and discovery of [Fe XXI]
  {\ensuremath{\lambda}}1354.1 in the sightline toward a filament in M87}.
\newblock {\em \aap}, 617:A123, October 2018.

\bibitem{wright2010}
Edward~L. {Wright}, Peter R.~M. {Eisenhardt}, Amy~K. {Mainzer}, Michael~E.
  {Ressler}, Roc~M. {Cutri}, Thomas {Jarrett}, J.~Davy {Kirkpatrick}, Deborah
  {Padgett}, Robert~S. {McMillan}, Michael {Skrutskie}, S.~A. {Stanford},
  Martin {Cohen}, Russell~G. {Walker}, John~C. {Mather}, David {Leisawitz}, III
  {Gautier}, Thomas~N., Ian {McLean}, Dominic {Benford}, Carol~J. {Lonsdale},
  Andrew {Blain}, Bryan {Mendez}, William~R. {Irace}, Valerie {Duval},
  Fengchuan {Liu}, Don {Royer}, Ingolf {Heinrichsen}, Joan {Howard}, Mark
  {Shannon}, Martha {Kendall}, Amy~L. {Walsh}, Mark {Larsen}, Joel~G. {Cardon},
  Scott {Schick}, Mark {Schwalm}, Mohamed {Abid}, Beth {Fabinsky}, Larry
  {Naes}, and Chao-Wei {Tsai}.
\newblock {The Wide-field Infrared Survey Explorer (WISE): Mission Description
  and Initial On-orbit Performance}.
\newblock {\em \aj}, 140(6):1868--1881, December 2010.

\bibitem{gurkan2014}
G.~{G{\"u}rkan}, M.~J. {Hardcastle}, and M.~J. {Jarvis}.
\newblock {The Wide-field Infrared Survey Explorer properties of complete
  samples of radio-loud active galactic nucleus}.
\newblock {\em \mnras}, 438(2):1149--1161, February 2014.

\bibitem{cohen2007}
A.~S. {Cohen}, W.~M. {Lane}, W.~D. {Cotton}, N.~E. {Kassim}, T.~J.~W. {Lazio},
  R.~A. {Perley}, J.~J. {Condon}, and W.~C. {Erickson}.
\newblock {The VLA Low-Frequency Sky Survey}.
\newblock {\em \aj}, 134:1245--1262, September 2007.

\bibitem{lane2012}
W.~M. {Lane}, W.~D. {Cotton}, J.~F. {Helmboldt}, and N.~E. {Kassim}.
\newblock {VLSS redux: Software improvements applied to the Very Large Array
  Low-Frequency Sky Survey}.
\newblock {\em Radio Science}, 47(3):RS0K04, Jan 2012.

\bibitem{rengelink1997}
R.~B. {Rengelink}, Y.~{Tang}, A.~G. {de Bruyn}, G.~K. {Miley}, M.~N. {Bremer},
  H.~J.~A. {Roettgering}, and M.~A.~R. {Bremer}.
\newblock {The Westerbork Northern Sky Survey (WENSS), I. A 570 square degree
  Mini-Survey around the North Ecliptic Pole}.
\newblock {\em \aaps}, 124:259--280, August 1997.

\bibitem{wayth2015}
R.~B. {Wayth}, E.~{Lenc}, M.~E. {Bell}, J.~R. {Callingham}, K.~S.
  {Dwarakanath}, T.~M.~O. {Franzen}, B.-Q. {For}, B.~{Gaensler}, P.~{Hancock},
  L.~{Hindson}, N.~{Hurley-Walker}, C.~A. {Jackson}, M.~{Johnston-Hollitt},
  A.~D. {Kapi{\'n}ska}, B.~{McKinley}, J.~{Morgan}, A.~R. {Offringa},
  P.~{Procopio}, L.~{Staveley-Smith}, C.~{Wu}, Q.~{Zheng}, C.~M. {Trott},
  G.~{Bernardi}, J.~D. {Bowman}, F.~{Briggs}, R.~J. {Cappallo}, B.~E. {Corey},
  A.~A. {Deshpande}, D.~{Emrich}, R.~{Goeke}, L.~J. {Greenhill}, B.~J.
  {Hazelton}, D.~L. {Kaplan}, J.~C. {Kasper}, E.~{Kratzenberg}, C.~J.
  {Lonsdale}, M.~J. {Lynch}, S.~R. {McWhirter}, D.~A. {Mitchell}, M.~F.
  {Morales}, E.~{Morgan}, D.~{Oberoi}, S.~M. {Ord}, T.~{Prabu}, A.~E.~E.
  {Rogers}, A.~{Roshi}, N.~U. {Shankar}, K.~S. {Srivani}, R.~{Subrahmanyan},
  S.~J. {Tingay}, M.~{Waterson}, R.~L. {Webster}, A.~R. {Whitney},
  A.~{Williams}, and C.~L. {Williams}.
\newblock {GLEAM: The GaLactic and Extragalactic All-Sky MWA Survey}.
\newblock {\em \pasa}, 32:e025, June 2015.

\bibitem{meisner2015}
Aaron~M. {Meisner} and Douglas~P. {Finkbeiner}.
\newblock {Modeling Thermal Dust Emission with Two Components: Application to
  the Planck High Frequency Instrument Maps}.
\newblock {\em \apj}, 798(2):88, January 2015.

\end{thebibliography}


\begin{thebibliography}{156}
\expandafter\ifx\csname natexlab\endcsname\relax\def\natexlab#1{#1}\fi

\bibitem[{{Allen} {et~al.}(2006){Allen}, {Dunn}, {Fabian}, {Taylor}, \&
  {Reynolds}}]{allen2006}
{Allen}, S.~W., {Dunn}, R.~J.~H., {Fabian}, A.~C., {Taylor}, G.~B., \&
  {Reynolds}, C.~S. 2006, \mnras, 372, 21

\bibitem[{{Arnaud}(1996)}]{Arnaud1996}
{Arnaud}, K.~A. 1996, in Astronomical Society of the Pacific Conference Series,
  Vol. 101, Astronomical Data Analysis Software and Systems V, ed. G.~H.
  {Jacoby} \& J.~{Barnes}, 17

\bibitem[{{Ascasibar} \& {Markevitch}(2006)}]{ascasibar2006}
{Ascasibar}, Y. \& {Markevitch}, M. 2006, \apj, 650, 102

\bibitem[{{Asplund} {et~al.}(2009){Asplund}, {Grevesse}, {Sauval}, \&
  {Scott}}]{Asplund2009}
{Asplund}, M., {Grevesse}, N., {Sauval}, A.~J., \& {Scott}, P. 2009, \araa, 47,
  481

\bibitem[{{Baars} {et~al.}(1977){Baars}, {Genzel}, {Pauliny-Toth}, \&
  {Witzel}}]{baars1977}
{Baars}, J.~W.~M., {Genzel}, R., {Pauliny-Toth}, I.~I.~K., \& {Witzel}, A.
  1977, \aap, 500, 135

\bibitem[{{Barton} {et~al.}(1998){Barton}, {Bromley}, \& {Geller}}]{barton1998}
{Barton}, E.~J., {Bromley}, B.~C., \& {Geller}, M.~J. 1998, in American
  Astronomical Society Meeting Abstracts, Vol. 193, American Astronomical
  Society Meeting Abstracts, 105.01

\bibitem[{{Beers} {et~al.}(1990){Beers}, {Flynn}, \& {Gebhardt}}]{beers1990}
{Beers}, T.~C., {Flynn}, K., \& {Gebhardt}, K. 1990, \aj, 100, 32

\bibitem[{{Biava} {et~al.}(2021{\natexlab{a}}){Biava}, {Brienza}, {Bonafede},
  {Gitti}, {Bonnassieux}, {Harwood}, {Edge}, {Riseley}, \&
  {Vantyghem}}]{biava2021a}
{Biava}, N., {Brienza}, M., {Bonafede}, A., {et~al.} 2021{\natexlab{a}}, \aap,
  650, A170

\bibitem[{{Biava} {et~al.}(2021{\natexlab{b}}){Biava}, {de Gasperin},
  {Bonafede}, {Edler}, {Giacintucci}, {Mazzotta}, {Brunetti}, {Botteon},
  {Br{\"u}ggen}, {Cassano}, {Drabent}, {Edge}, {En{\ss}lin}, {Gastaldello},
  {Riseley}, {Rossetti}, {Rottgering}, {Shimwell}, {Tasse}, \& {van
  Weeren}}]{biava2021b}
{Biava}, N., {de Gasperin}, F., {Bonafede}, A., {et~al.} 2021{\natexlab{b}},
  \mnras, 508, 3995

\bibitem[{{Bird} \& {Beers}(1993)}]{bird1993}
{Bird}, C.~M. \& {Beers}, T.~C. 1993, \aj, 105, 1596

\bibitem[{{Blandford} \& {K{\"o}nigl}(1979)}]{blandford1979}
{Blandford}, R.~D. \& {K{\"o}nigl}, A. 1979, \apj, 232, 34

\bibitem[{{B{\"o}hringer} {et~al.}(2021){B{\"o}hringer}, {Chon}, \&
  {Tr{\"u}mper}}]{bohringer2021}
{B{\"o}hringer}, H., {Chon}, G., \& {Tr{\"u}mper}, J. 2021, \aap, 651, A16

\bibitem[{{Bonafede} {et~al.}(2021){Bonafede}, {Brunetti}, {Vazza},
  {Simionescu}, {Giovannini}, {Bonnassieux}, {Shimwell}, {Br{\"u}ggen}, {van
  Weeren}, {Botteon}, {Brienza}, {Cassano}, {Drabent}, {Feretti}, {de
  Gasperin}, {Gastaldello}, {di Gennaro}, {Rossetti}, {Rottgering}, {Stuardi},
  \& {Venturi}}]{bonafede2021}
{Bonafede}, A., {Brunetti}, G., {Vazza}, F., {et~al.} 2021, \apj, 907, 32

\bibitem[{{Botteon} {et~al.}(2020{\natexlab{a}}){Botteon}, {Brunetti}, {Ryu},
  \& {Roh}}]{botteon2020C}
{Botteon}, A., {Brunetti}, G., {Ryu}, D., \& {Roh}, S. 2020{\natexlab{a}},
  \aap, 634, A64

\bibitem[{{Botteon} {et~al.}(2020{\natexlab{b}}){Botteon}, {Brunetti}, {van
  Weeren}, {Shimwell}, {Pizzo}, {Cassano}, {Iacobelli}, {Gastaldello},
  {B{\^\i}rzan}, {Bonafede}, {Br{\"u}ggen}, {Cuciti}, {Dallacasa}, {de
  Gasperin}, {Di Gennaro}, {Drabent}, {Hardcastle}, {Hoeft}, {Mandal},
  {R{\"o}ttgering}, \& {Simionescu}}]{botteon2020B}
{Botteon}, A., {Brunetti}, G., {van Weeren}, R.~J., {et~al.}
  2020{\natexlab{b}}, \apj, 897, 93

\bibitem[{{Botteon} {et~al.}(2020{\natexlab{c}}){Botteon}, {van Weeren},
  {Brunetti}, {de Gasperin}, {Intema}, {Osinga}, {Di Gennaro}, {Shimwell},
  {Bonafede}, {Br{\"u}ggen}, {Cassano}, {Cuciti}, {Dallacasa}, {Gastaldello},
  {Mandal}, {Rossetti}, \& {R{\"o}ttgering}}]{botteon2020A}
{Botteon}, A., {van Weeren}, R.~J., {Brunetti}, G., {et~al.}
  2020{\natexlab{c}}, \mnras, 499, L11

\bibitem[{{Brienza} {et~al.}(2017){Brienza}, {Godfrey}, {Morganti}, {Prandoni},
  {Harwood}, {Mahony}, {Hardcastle}, {Murgia}, {R{\"o}ttgering}, {Shimwell}, \&
  {Shulevski}}]{brienza2017}
{Brienza}, M., {Godfrey}, L., {Morganti}, R., {et~al.} 2017, \aap, 606, A98

\bibitem[{{Brienza} {et~al.}(2021){Brienza}, {Shimwell}, {de Gasperin},
  {Bikmaev}, {Bonafede}, {Botteon}, {Br{\"u}ggen}, {Brunetti}, {Burenin},
  {Capetti}, {Churazov}, {Hardcastle}, {Khabibullin}, {Lyskova},
  {R{\"o}ttgering}, {Sunyaev}, {van Weeren}, {Gastaldello}, {Mandal}, {Purser},
  {Simionescu}, \& {Tasse}}]{brienza2021}
{Brienza}, M., {Shimwell}, T.~W., {de Gasperin}, F., {et~al.} 2021, Nature
  Astronomy, 5, 1261

\bibitem[{{Brunetti} \& {Jones}(2014)}]{brunetti2014}
{Brunetti}, G. \& {Jones}, T.~W. 2014, International Journal of Modern Physics
  D, 23, 1430007

\bibitem[{{Cappellari} \& {Copin}(2003)}]{Cappellari2003}
{Cappellari}, M. \& {Copin}, Y. 2003, \mnras, 342, 345

\bibitem[{{Cash}(1979)}]{Cash1979}
{Cash}, W. 1979, \apj, 228, 939

\bibitem[{{Cavaliere} \& {Fusco-Femiano}(1976)}]{cavaliere1976}
{Cavaliere}, A. \& {Fusco-Femiano}, R. 1976, \aap, 500, 95

\bibitem[{{Cohen} {et~al.}(2007){Cohen}, {Lane}, {Cotton}, {Kassim}, {Lazio},
  {Perley}, {Condon}, \& {Erickson}}]{cohen2007}
{Cohen}, A.~S., {Lane}, W.~M., {Cotton}, W.~D., {et~al.} 2007, \aj, 134, 1245

\bibitem[{{Colless} \& {Dunn}(1996)}]{colless1996}
{Colless}, M. \& {Dunn}, A.~M. 1996, \apj, 458, 435

\bibitem[{{Condon} {et~al.}(1998){Condon}, {Cotton}, {Greisen}, {Yin},
  {Perley}, {Taylor}, \& {Broderick}}]{condon1998}
{Condon}, J.~J., {Cotton}, W.~D., {Greisen}, E.~W., {et~al.} 1998, \aj, 115,
  1693

\bibitem[{{Condon} {et~al.}(2021){Condon}, {Cotton}, {White}, {Legodi},
  {Goedhart}, {McAlpine}, {Ratcliffe}, \& {Camilo}}]{condon2021}
{Condon}, J.~J., {Cotton}, W.~D., {White}, S.~V., {et~al.} 2021, \apj, 917, 18

\bibitem[{{Costa} {et~al.}(2018){Costa}, {Ribeiro}, \& {de
  Carvalho}}]{costa2018}
{Costa}, A.~P., {Ribeiro}, A.~L.~B., \& {de Carvalho}, R.~R. 2018, \mnras, 473,
  L31

\bibitem[{{Cotton} {et~al.}(2020){Cotton}, {Thorat}, {Condon}, {Frank},
  {J{\'o}zsa}, {White}, {Deane}, {Oozeer}, {Atemkeng}, {Bester}, {Fanaroff},
  {Kupa}, {Smirnov}, {Mauch}, {Krishnan}, \& {Camilo}}]{cotton2020}
{Cotton}, W.~D., {Thorat}, K., {Condon}, J.~J., {et~al.} 2020, \mnras, 495,
  1271

\bibitem[{{Crook} {et~al.}(2007){Crook}, {Huchra}, {Martimbeau}, {Masters},
  {Jarrett}, \& {Macri}}]{crook2007}
{Crook}, A.~C., {Huchra}, J.~P., {Martimbeau}, N., {et~al.} 2007, \apj, 655,
  790

\bibitem[{{de Gasperin} {et~al.}(2019){de Gasperin}, {Dijkema}, {Drabent},
  {Mevius}, {Rafferty}, {van Weeren}, {Br{\"u}ggen}, {Callingham}, {Emig},
  {Heald}, {Intema}, {Morabito}, {Offringa}, {Oonk}, {Orr{\`u}},
  {R{\"o}ttgering}, {Sabater}, {Shimwell}, {Shulevski}, \&
  {Williams}}]{degasperin2019}
{de Gasperin}, F., {Dijkema}, T.~J., {Drabent}, A., {et~al.} 2019, \aap, 622,
  A5

\bibitem[{{de Gasperin} {et~al.}(2018){de Gasperin}, {Mevius}, {Rafferty},
  {Intema}, \& {Fallows}}]{degasperin2018}
{de Gasperin}, F., {Mevius}, M., {Rafferty}, D.~A., {Intema}, H.~T., \&
  {Fallows}, R.~A. 2018, \aap, 615, A179

\bibitem[{{De Propris} {et~al.}(2002){De Propris}, {Couch}, {Colless},
  {Dalton}, {Collins}, {Baugh}, {Bland-Hawthorn}, {Bridges}, {Cannon}, {Cole},
  {Cross}, {Deeley}, {Driver}, {Efstathiou}, {Ellis}, {Frenk}, {Glazebrook},
  {Jackson}, {Lahav}, {Lewis}, {Lumsden}, {Maddox}, {Madgwick}, {Moody},
  {Norberg}, {Peacock}, {Percival}, {Peterson}, {Sutherland}, \&
  {Taylor}}]{depropris2002}
{De Propris}, R., {Couch}, W.~J., {Colless}, M., {et~al.} 2002, \mnras, 329, 87

\bibitem[{{de Ruiter} {et~al.}(1990){de Ruiter}, {Parma}, {Fanti}, \&
  {Fanti}}]{deruiter1990}
{de Ruiter}, H.~R., {Parma}, P., {Fanti}, C., \& {Fanti}, R. 1990, \aap, 227,
  351

\bibitem[{{Di Gennaro} {et~al.}(2018){Di Gennaro}, {van Weeren}, {Hoeft},
  {Kang}, {Ryu}, {Rudnick}, {Forman}, {R{\"o}ttgering}, {Br{\"u}ggen},
  {Dawson}, {Golovich}, {Hoang}, {Intema}, {Jones}, {Kraft}, {Shimwell}, \&
  {Stroe}}]{digennaro2018}
{Di Gennaro}, G., {van Weeren}, R.~J., {Hoeft}, M., {et~al.} 2018, \apj, 865,
  24

\bibitem[{{Diehl} \& {Statler}(2006)}]{Diehl2006}
{Diehl}, S. \& {Statler}, T.~S. 2006, \mnras, 368, 497

\bibitem[{{Dong} {et~al.}(2010){Dong}, {Rasmussen}, \& {Mulchaey}}]{dong2010}
{Dong}, R., {Rasmussen}, J., \& {Mulchaey}, J.~S. 2010, \apj, 712, 883

\bibitem[{{Dressler} \& {Shectman}(1988)}]{dressler1988}
{Dressler}, A. \& {Shectman}, S.~A. 1988, \aj, 95, 985

\bibitem[{{Duchesne} {et~al.}(2021){Duchesne}, {Johnston-Hollitt}, \&
  {Wilber}}]{duchesne2021}
{Duchesne}, S.~W., {Johnston-Hollitt}, M., \& {Wilber}, A.~G. 2021, \pasa, 38,
  e031

\bibitem[{{Duchesne} {et~al.}(2020){Duchesne}, {Johnston-Hollitt}, {Zhu},
  {Wayth}, \& {Line}}]{duchesne2020}
{Duchesne}, S.~W., {Johnston-Hollitt}, M., {Zhu}, Z., {Wayth}, R.~B., \&
  {Line}, J.~L.~B. 2020, \pasa, 37, e037

\bibitem[{{Eckert} {et~al.}(2021){Eckert}, {Gaspari}, {Gastaldello}, {Le Brun},
  \& {O'Sullivan}}]{eckert2021}
{Eckert}, D., {Gaspari}, M., {Gastaldello}, F., {Le Brun}, A. M.~C., \&
  {O'Sullivan}, E. 2021, Universe, 7, 142

\bibitem[{{Einasto} {et~al.}(2012){Einasto}, {Vennik}, {Nurmi}, {Tempel},
  {Ahvensalmi}, {Tago}, {Liivam{\"a}gi}, {Saar}, {Hein{\"a}m{\"a}ki},
  {Einasto}, \& {Mart{\'\i}nez}}]{einasto2012}
{Einasto}, M., {Vennik}, J., {Nurmi}, P., {et~al.} 2012, \aap, 540, A123

\bibitem[{{English} {et~al.}(2019){English}, {Hardcastle}, \&
  {Krause}}]{english2019}
{English}, W., {Hardcastle}, M.~J., \& {Krause}, M.~G.~H. 2019, \mnras, 490,
  5807

\bibitem[{{En{\ss}lin} \& {Br{\"u}ggen}(2002)}]{ensslin2002}
{En{\ss}lin}, T.~A. \& {Br{\"u}ggen}, M. 2002, \mnras, 331, 1011

\bibitem[{{En{\ss}lin} \& {Gopal-Krishna}(2001)}]{ensslin2001}
{En{\ss}lin}, T.~A. \& {Gopal-Krishna}. 2001, \aap, 366, 26

\bibitem[{{Fabian}(2012)}]{fabian2012}
{Fabian}, A.~C. 2012, \araa, 50, 455

\bibitem[{{Fabian} {et~al.}(2021){Fabian}, {ZuHone}, \& {Walker}}]{fabian2021}
{Fabian}, A.~C., {ZuHone}, J.~A., \& {Walker}, S.~A. 2021, \mnras

\bibitem[{{Fadda} {et~al.}(1996){Fadda}, {Girardi}, {Giuricin}, {Mardirossian},
  \& {Mezzetti}}]{fadda1996}
{Fadda}, D., {Girardi}, M., {Giuricin}, G., {Mardirossian}, F., \& {Mezzetti},
  M. 1996, \apj, 473, 670

\bibitem[{{Fanaroff} \& {Riley}(1974)}]{fanaroff1974}
{Fanaroff}, B.~L. \& {Riley}, J.~M. 1974, \mnras, 167, 31P

\bibitem[{{Farahi} {et~al.}(2018){Farahi}, {Guglielmo}, {Evrard}, {Poggianti},
  {Adami}, {Ettori}, {Gastaldello}, {Giles}, {Maughan}, {Rapetti}, {Sereno},
  {Altieri}, {Baldry}, {Birkinshaw}, {Bolzonella}, {Bongiorno}, {Brown},
  {Chiappetti}, {Driver}, {Elyiv}, {Garilli}, {Guennou}, {Hopkins}, {Iovino},
  {Koulouridis}, {Liske}, {Maurogordato}, {Owers}, {Pacaud}, {Pierre},
  {Plionis}, {Ponman}, {Robotham}, {Sadibekova}, {Scodeggio}, {Tuffs}, \&
  {Valtchanov}}]{farahi2018}
{Farahi}, A., {Guglielmo}, V., {Evrard}, A.~E., {et~al.} 2018, \aap, 620, A8

\bibitem[{{Fraley} \& {Raftery}(2002)}]{mclust1}
{Fraley}, C. \& {Raftery}, A. 2002, Journal of the American Statistical
  Association, 97, 611

\bibitem[{Fraley {et~al.}(2012)Fraley, Raftery, Murphy, \& Scrucca}]{mclust2}
Fraley, C., Raftery, A.~E., Murphy, T.~B., \& Scrucca, L. 2012, mclust Version
  4 for R: Normal Mixture Modeling for Model-Based Clustering, Classification,
  and Density Estimation

\bibitem[{{Garcia}(1993)}]{garcia1993}
{Garcia}, A.~M. 1993, \aaps, 100, 47

\bibitem[{{Gastaldello} {et~al.}(2013){Gastaldello}, {Di Gesu}, {Ghizzardi},
  {Giacintucci}, {Girardi}, {Roediger}, {Rossetti}, {Brighenti}, {Buote},
  {Eckert}, {Ettori}, {Humphrey}, \& {Mathews}}]{gastaldello2013}
{Gastaldello}, F., {Di Gesu}, L., {Ghizzardi}, S., {et~al.} 2013, \apj, 770, 56

\bibitem[{{Gebhardt} \& {Beers}(1991)}]{gebhardt1991}
{Gebhardt}, K. \& {Beers}, T.~C. 1991, \apj, 383, 72

\bibitem[{{Gendron-Marsolais} {et~al.}(2021){Gendron-Marsolais}, {Hull},
  {Perley}, {Rudnick}, {Kraft}, {Hlavacek-Larrondo}, {Fabian}, {Roediger}, {van
  Weeren}, {Richard-Laferri{\`e}re}, {Golden-Marx}, {Arakawa}, \&
  {McBride}}]{gendronmarsolais2021}
{Gendron-Marsolais}, M.~L., {Hull}, C.~L.~H., {Perley}, R., {et~al.} 2021,
  \apj, 911, 56

\bibitem[{{Giacintucci} {et~al.}(2019){Giacintucci}, {Markevitch}, {Cassano},
  {Venturi}, {Clarke}, {Kale}, \& {Cuciti}}]{giacintucci2019}
{Giacintucci}, S., {Markevitch}, M., {Cassano}, R., {et~al.} 2019, \apj, 880,
  70

\bibitem[{{Giacintucci} {et~al.}(2011){Giacintucci}, {O'Sullivan}, {Vrtilek},
  {David}, {Raychaudhury}, {Venturi}, {Athreya}, {Clarke}, {Murgia},
  {Mazzotta}, {Gitti}, {Ponman}, {Ishwara-Chandra}, {Jones}, \&
  {Forman}}]{giacintucci2011}
{Giacintucci}, S., {O'Sullivan}, E., {Vrtilek}, J., {et~al.} 2011, \apj, 732,
  95

\bibitem[{{Giacintucci} {et~al.}(2005){Giacintucci}, {Venturi}, {Brunetti},
  {Bardelli}, {Dallacasa}, {Ettori}, {Finoguenov}, {Rao}, \&
  {Zucca}}]{giacintucci2005}
{Giacintucci}, S., {Venturi}, T., {Brunetti}, G., {et~al.} 2005, \aap, 440, 867

\bibitem[{{Girardi} {et~al.}(1996){Girardi}, {Fadda}, {Giuricin},
  {Mardirossian}, {Mezzetti}, \& {Biviano}}]{girardi1996}
{Girardi}, M., {Fadda}, D., {Giuricin}, G., {et~al.} 1996, \apj, 457, 61

\bibitem[{{Govoni} {et~al.}(2001){Govoni}, {En{\ss}lin}, {Feretti}, \&
  {Giovannini}}]{govoni2001}
{Govoni}, F., {En{\ss}lin}, T.~A., {Feretti}, L., \& {Giovannini}, G. 2001,
  \aap, 369, 441

\bibitem[{Gupta {et~al.}(2017)Gupta, Ajithkumar, Kale, Nayak, Sabhapathy,
  Sureshkumar, Swami, Chengalur, Ghosh, Ishwara-Chandra, Joshi, Kanekar, Lal,
  \& Roy}]{gupta2017}
Gupta, Y., Ajithkumar, B., Kale, H., {et~al.} 2017, Current Science, 113, 707

\bibitem[{{Hardcastle}(2013)}]{hardcastle2013b}
{Hardcastle}, M.~J. 2013, \mnras, 433, 3364

\bibitem[{{Hardcastle}(2018)}]{hardcastle2018}
{Hardcastle}, M.~J. 2018, \mnras, 475, 2768

\bibitem[{{Hardcastle} {et~al.}(2019){Hardcastle}, {Croston}, {Shimwell},
  {Tasse}, {G{\"u}rkan}, {Morganti}, {Murgia}, {R{\"o}ttgering}, {van Weeren},
  \& {Williams}}]{hardcastle2019}
{Hardcastle}, M.~J., {Croston}, J.~H., {Shimwell}, T.~W., {et~al.} 2019,
  \mnras, 488, 3416

\bibitem[{{Heesen}(2015)}]{heesen2015}
{Heesen}, V. 2015, in TORUS2015, ed. P.~{Gandhi} \& S.~F. {Hoenig}

\bibitem[{{Hou} {et~al.}(2009){Hou}, {Parker}, {Harris}, \& {Wilman}}]{hou2009}
{Hou}, A., {Parker}, L.~C., {Harris}, W.~E., \& {Wilman}, D.~J. 2009, \apj,
  702, 1199

\bibitem[{{Huarte-Espinosa} {et~al.}(2011){Huarte-Espinosa}, {Krause}, \&
  {Alexander}}]{huarteespinosa2011}
{Huarte-Espinosa}, M., {Krause}, M., \& {Alexander}, P. 2011, \mnras, 417, 382

\bibitem[{{Huchra} {et~al.}(1999){Huchra}, {Vogeley}, \& {Geller}}]{huchra1999}
{Huchra}, J.~P., {Vogeley}, M.~S., \& {Geller}, M.~J. 1999, \apjs, 121, 287

\bibitem[{{Ichinohe} {et~al.}(2021){Ichinohe}, {Simionescu}, {Werner},
  {Markevitch}, \& {Wang}}]{ichinohe2021}
{Ichinohe}, Y., {Simionescu}, A., {Werner}, N., {Markevitch}, M., \& {Wang},
  Q.~H.~S. 2021, \mnras, 504, 2800

\bibitem[{{Ignesti} {et~al.}(2020{\natexlab{a}}){Ignesti}, {Brunetti}, {Gitti},
  \& {Giacintucci}}]{ignesti2020}
{Ignesti}, A., {Brunetti}, G., {Gitti}, M., \& {Giacintucci}, S.
  2020{\natexlab{a}}, \aap, 640, A37

\bibitem[{{Ignesti} {et~al.}(2020{\natexlab{b}}){Ignesti}, {Shimwell},
  {Brunetti}, {Gitti}, {Intema}, {van Weeren}, {Hardcastle}, {Clarke},
  {Botteon}, {Di Gennaro}, {Br{\"u}ggen}, {Browne}, {Mandal}, {R{\"o}ttgering},
  {Cuciti}, {de Gasperin}, {Cassano}, \& {Scaife}}]{ignesti2020b}
{Ignesti}, A., {Shimwell}, T., {Brunetti}, G., {et~al.} 2020{\natexlab{b}},
  \aap, 643, A172

\bibitem[{{Intema} {et~al.}(2009){Intema}, {van der Tol}, {Cotton}, {Cohen},
  {van Bemmel}, \& {R{\"o}ttgering}}]{intema2009}
{Intema}, H.~T., {van der Tol}, S., {Cotton}, W.~D., {et~al.} 2009, \aap, 501,
  1185

\bibitem[{{Islam} {et~al.}(2021){Islam}, {Kim}, {Lin}, {O'Sullivan},
  {Anderson}, {Fabbiano}, {Lauer}, {Morgan}, {Mossman}, {Paggi}, {Trinchieri},
  \& {Vrtilek}}]{islam2021}
{Islam}, N., {Kim}, D.-W., {Lin}, K., {et~al.} 2021, \apjs, 256, 22

\bibitem[{{Jaffe} \& {Perola}(1973)}]{jaffe1973}
{Jaffe}, W.~J. \& {Perola}, G.~C. 1973, \aap, 26, 423

\bibitem[{{Jeltema} {et~al.}(2008){Jeltema}, {Binder}, \&
  {Mulchaey}}]{jeltema2008}
{Jeltema}, T.~E., {Binder}, B., \& {Mulchaey}, J.~S. 2008, \apj, 679, 1162

\bibitem[{{Jurlin} {et~al.}(2021){Jurlin}, {Brienza}, {Morganti}, {Wadadekar},
  {Ishwara-Chandra}, {Maddox}, \& {Mahatma}}]{jurlin2021}
{Jurlin}, N., {Brienza}, M., {Morganti}, R., {et~al.} 2021, \aap, 653, A110

\bibitem[{{Kale} {et~al.}(2018){Kale}, {Parekh}, \& {Dwarakanath}}]{kale2018}
{Kale}, R., {Parekh}, V., \& {Dwarakanath}, K.~S. 2018, \mnras, 480, 5352

\bibitem[{{Kang} \& {Ryu}(2016)}]{kang2016}
{Kang}, H. \& {Ryu}, D. 2016, \apj, 823, 13

\bibitem[{{Kelly}(2007)}]{kelly2007}
{Kelly}, B.~C. 2007, \apj, 665, 1489

\bibitem[{{Kim} {et~al.}(2019){Kim}, {Anderson}, {Burke}, {D'Abrusco},
  {Fabbiano}, {Fruscione}, {Lauer}, {McCollough}, {Morgan}, {Mossman},
  {O'Sullivan}, {Paggi}, {Vrtilek}, \& {Trinchieri}}]{kim2019}
{Kim}, D.-W., {Anderson}, C., {Burke}, D., {et~al.} 2019, \apjs, 241, 36

\bibitem[{{Kim} \& {Fabbiano}(2004)}]{kim2004}
{Kim}, D.-W. \& {Fabbiano}, G. 2004, \apj, 613, 933

\bibitem[{{Kolokythas} {et~al.}(2020){Kolokythas}, {O'Sullivan}, {Giacintucci},
  {Worrall}, {Birkinshaw}, {Raychaudhury}, {Horellou}, {Intema}, \&
  {Loubser}}]{kolokythas2020}
{Kolokythas}, K., {O'Sullivan}, E., {Giacintucci}, S., {et~al.} 2020, \mnras,
  496, 1471

\bibitem[{{Komissarov} \& {Gubanov}(1994)}]{komissarov1994}
{Komissarov}, S.~S. \& {Gubanov}, A.~G. 1994, \aap, 285, 27

\bibitem[{{Kraft} {et~al.}(2004){Kraft}, {Forman}, {Churazov}, {Laslo},
  {Jones}, {Markevitch}, {Murray}, \& {Vikhlinin}}]{kraft2004}
{Kraft}, R.~P., {Forman}, W.~R., {Churazov}, E., {et~al.} 2004, \apj, 601, 221

\bibitem[{{Laing} {et~al.}(2011){Laing}, {Guidetti}, {Bridle}, {Parma}, \&
  {Bondi}}]{laing2011}
{Laing}, R.~A., {Guidetti}, D., {Bridle}, A.~H., {Parma}, P., \& {Bondi}, M.
  2011, \mnras, 417, 2789

\bibitem[{{Lavaux} \& {Hudson}(2011)}]{lavaux2011}
{Lavaux}, G. \& {Hudson}, M.~J. 2011, \mnras, 416, 2840

\bibitem[{{Lovisari} {et~al.}(2021){Lovisari}, {Ettori}, {Gaspari}, \&
  {Giles}}]{lovisari2021}
{Lovisari}, L., {Ettori}, S., {Gaspari}, M., \& {Giles}, P.~A. 2021, Universe,
  7, 139

\bibitem[{{Lovisari} \& {Reiprich}(2019)}]{lovisari2019}
{Lovisari}, L. \& {Reiprich}, T.~H. 2019, \mnras, 483, 540

\bibitem[{{Maccagni} {et~al.}(2020){Maccagni}, {Murgia}, {Serra}, {Govoni},
  {Morokuma-Matsui}, {Kleiner}, {Buchner}, {J{\'o}zsa}, {Kamphuis},
  {Makhathini}, {Moln{\'a}r}, {Prokhorov}, {Ramaila}, {Ramatsoku}, {Thorat}, \&
  {Smirnov}}]{maccagni2020}
{Maccagni}, F.~M., {Murgia}, M., {Serra}, P., {et~al.} 2020, \aap, 634, A9

\bibitem[{{Machacek} {et~al.}(2011){Machacek}, {Jerius}, {Kraft}, {Forman},
  {Jones}, {Randall}, {Giacintucci}, \& {Sun}}]{machacek2011}
{Machacek}, M.~E., {Jerius}, D., {Kraft}, R., {et~al.} 2011, \apj, 743, 15

\bibitem[{{Mahatma} {et~al.}(2018){Mahatma}, {Hardcastle}, {Williams},
  {Brienza}, {Br{\"u}ggen}, {Croston}, {Gurkan}, {Harwood},
  {Kunert-Bajraszewska}, {Morganti}, {R{\"o}ttgering}, {Shimwell}, \&
  {Tasse}}]{mahatma2018}
{Mahatma}, V.~H., {Hardcastle}, M.~J., {Williams}, W.~L., {et~al.} 2018,
  \mnras, 475, 4557

\bibitem[{{Mandal} {et~al.}(2020){Mandal}, {Intema}, {van Weeren}, {Shimwell},
  {Botteon}, {Brunetti}, {de Gasperin}, {Br{\"u}ggen}, {Di Gennaro}, {Kraft},
  {R{\"o}ttgering}, {Hardcastle}, \& {Tasse}}]{manadal2020}
{Mandal}, S., {Intema}, H.~T., {van Weeren}, R.~J., {et~al.} 2020, \aap, 634,
  A4

\bibitem[{{Markevitch} {et~al.}(2000){Markevitch}, {Ponman}, {Nulsen}, {Bautz},
  {Burke}, {David}, {Davis}, {Donnelly}, {Forman}, {Jones}, {Kaastra},
  {Kellogg}, {Kim}, {Kolodziejczak}, {Mazzotta}, {Pagliaro}, {Patel}, {Van
  Speybroeck}, {Vikhlinin}, {Vrtilek}, {Wise}, \& {Zhao}}]{markevitch2000}
{Markevitch}, M., {Ponman}, T.~J., {Nulsen}, P.~E.~J., {et~al.} 2000, \apj,
  541, 542

\bibitem[{{Markevitch} \& {Vikhlinin}(2007)}]{markevitch2007}
{Markevitch}, M. \& {Vikhlinin}, A. 2007, \physrep, 443, 1

\bibitem[{{McKean} {et~al.}(2016){McKean}, {Godfrey}, {Vegetti}, {Wise},
  {Morganti}, {Hardcastle}, {Rafferty}, {Anderson}, {Avruch}, {Beck}, {Bell},
  {van Bemmel}, {Bentum}, {Bernardi}, {Best}, {Blaauw}, {Bonafede},
  {Breitling}, {Broderick}, {Br{\"u}ggen}, {Cerrigone}, {Ciardi}, {de
  Gasperin}, {Deller}, {Duscha}, {Engels}, {Falcke}, {Fallows}, {Frieswijk},
  {Garrett}, {Grie{\ss}meier}, {van Haarlem}, {Heald}, {Hoeft}, {Horst},
  {Iacobelli}, {Intema}, {Juette}, {Karastergiou}, {Kondratiev}, {Koopmans},
  {Kuniyoshi}, {Kuper}, {van Leeuwen}, {Maat}, {Mann}, {Markoff}, {McFadden},
  {McKay-Bukowski}, {Mulcahy}, {Munk}, {Nelles}, {Orru}, {Paas},
  {Pandey-Pommier}, {Pietka}, {Pizzo}, {Polatidis}, {Reich}, {R{\"o}ttgering},
  {Rowlinson}, {Scaife}, {Serylak}, {Shulevski}, {Sluman}, {Smirnov},
  {Steinmetz}, {Stewart}, {Swinbank}, {Tagger}, {Thoudam}, {Toribio},
  {Vermeulen}, {Vocks}, {van Weeren}, {Wucknitz}, {Yatawatta}, \&
  {Zarka}}]{mckean2016}
{McKean}, J.~P., {Godfrey}, L.~E.~H., {Vegetti}, S., {et~al.} 2016, \mnras,
  463, 3143

\bibitem[{{McMullin} {et~al.}(2007){McMullin}, {Waters}, {Schiebel}, {Young},
  \& {Golap}}]{mcmullin2007}
{McMullin}, J.~P., {Waters}, B., {Schiebel}, D., {Young}, W., \& {Golap}, K.
  2007, in Astronomical Society of the Pacific Conference Series, Vol. 376,
  Astronomical Data Analysis Software and Systems XVI, ed. R.~A. {Shaw},
  F.~{Hill}, \& D.~J. {Bell}, 127

\bibitem[{{McNamara} \& {Nulsen}(2007)}]{mcnamara2007}
{McNamara}, B.~R. \& {Nulsen}, P.~E.~J. 2007, \araa, 45, 117

\bibitem[{{McNamara} \& {Nulsen}(2012)}]{mcnamara2012}
{McNamara}, B.~R. \& {Nulsen}, P.~E.~J. 2012, New Journal of Physics, 14,
  055023

\bibitem[{{Menci} \& {Fusco-Femiano}(1996)}]{menci1996}
{Menci}, N. \& {Fusco-Femiano}, R. 1996, \apj, 472, 46

\bibitem[{{Morganti}(2017)}]{morganti2017}
{Morganti}, R. 2017, Nature Astronomy, 1, 596

\bibitem[{{Mulchaey} {et~al.}(2003){Mulchaey}, {Davis}, {Mushotzky}, \&
  {Burstein}}]{mulchaey2003}
{Mulchaey}, J.~S., {Davis}, D.~S., {Mushotzky}, R.~F., \& {Burstein}, D. 2003,
  \apjs, 145, 39

\bibitem[{{Murgia} {et~al.}(2011){Murgia}, {Parma}, {Mack}, {de Ruiter},
  {Fanti}, {Govoni}, {Tarchi}, {Giacintucci}, \& {Markevitch}}]{murgia2011}
{Murgia}, M., {Parma}, P., {Mack}, K.-H., {et~al.} 2011, \aap, 526, A148

\bibitem[{{Nuza} {et~al.}(2017){Nuza}, {Gelszinnis}, {Hoeft}, \&
  {Yepes}}]{nuza2017}
{Nuza}, S.~E., {Gelszinnis}, J., {Hoeft}, M., \& {Yepes}, G. 2017, \mnras, 470,
  240

\bibitem[{{Offringa} {et~al.}(2014){Offringa}, {McKinley}, {Hurley-Walker},
  {Briggs}, {Wayth}, {Kaplan}, {Bell}, {Feng}, {Neben}, {Hughes}, {Rhee},
  {Murphy}, {Bhat}, {Bernardi}, {Bowman}, {Cappallo}, {Corey}, {Deshpande},
  {Emrich}, {Ewall-Wice}, {Gaensler}, {Goeke}, {Greenhill}, {Hazelton},
  {Hindson}, {Johnston-Hollitt}, {Jacobs}, {Kasper}, {Kratzenberg}, {Lenc},
  {Lonsdale}, {Lynch}, {McWhirter}, {Mitchell}, {Morales}, {Morgan},
  {Kudryavtseva}, {Oberoi}, {Ord}, {Pindor}, {Procopio}, {Prabu}, {Riding},
  {Roshi}, {Shankar}, {Srivani}, {Subrahmanyan}, {Tingay}, {Waterson},
  {Webster}, {Whitney}, {Williams}, \& {Williams}}]{offringa2014}
{Offringa}, A.~R., {McKinley}, B., {Hurley-Walker}, N., {et~al.} 2014, \mnras,
  444, 606

\bibitem[{{Orr{\`u}} {et~al.}(2010){Orr{\`u}}, {Murgia}, {Feretti}, {Govoni},
  {Giovannini}, {Lane}, {Kassim}, \& {Paladino}}]{orru2010}
{Orr{\`u}}, E., {Murgia}, M., {Feretti}, L., {et~al.} 2010, \aap, 515, A50

\bibitem[{{O'Sullivan} {et~al.}(2014){O'Sullivan}, {David}, \&
  {Vrtilek}}]{osullivan2014}
{O'Sullivan}, E., {David}, L.~P., \& {Vrtilek}, J.~M. 2014, \mnras, 437, 730

\bibitem[{{O'Sullivan} {et~al.}(2011){O'Sullivan}, {Giacintucci}, {David},
  {Gitti}, {Vrtilek}, {Raychaudhury}, \& {Ponman}}]{osullivan2011}
{O'Sullivan}, E., {Giacintucci}, S., {David}, L.~P., {et~al.} 2011, \apj, 735,
  11

\bibitem[{{O'Sullivan} {et~al.}(2003){O'Sullivan}, {Ponman}, \&
  {Collins}}]{osullivan2003}
{O'Sullivan}, E., {Ponman}, T.~J., \& {Collins}, R.~S. 2003, \mnras, 340, 1375

\bibitem[{{Owen} {et~al.}(2014){Owen}, {Rudnick}, {Eilek}, {Rau}, {Bhatnagar},
  \& {Kogan}}]{owen2014}
{Owen}, F.~N., {Rudnick}, L., {Eilek}, J., {et~al.} 2014, \apj, 794, 24

\bibitem[{{Owers} {et~al.}(2009){Owers}, {Nulsen}, {Couch}, {Markevitch}, \&
  {Poole}}]{owers2009}
{Owers}, M.~S., {Nulsen}, P. E.~J., {Couch}, W.~J., {Markevitch}, M., \&
  {Poole}, G.~B. 2009, \apj, 692, 702

\bibitem[{{Paolillo} {et~al.}(2003){Paolillo}, {Fabbiano}, {Peres}, \&
  {Kim}}]{paolillo2003}
{Paolillo}, M., {Fabbiano}, G., {Peres}, G., \& {Kim}, D.~W. 2003, \apj, 586,
  850

\bibitem[{{Parma} {et~al.}(1986){Parma}, {de Ruiter}, {Fanti}, \&
  {Fanti}}]{parma1986}
{Parma}, P., {de Ruiter}, H.~R., {Fanti}, C., \& {Fanti}, R. 1986, \aaps, 64,
  135

\bibitem[{{Perley} \& {Butler}(2017)}]{perley2017}
{Perley}, R.~A. \& {Butler}, B.~J. 2017, \apjs, 230, 7

\bibitem[{{Piffaretti} {et~al.}(2011){Piffaretti}, {Arnaud}, {Pratt},
  {Pointecouteau}, \& {Melin}}]{piffaretti2011}
{Piffaretti}, R., {Arnaud}, M., {Pratt}, G.~W., {Pointecouteau}, E., \&
  {Melin}, J.~B. 2011, \aap, 534, A109

\bibitem[{{Pinkney} {et~al.}(1996){Pinkney}, {Roettiger}, {Burns}, \&
  {Bird}}]{pinkney1996}
{Pinkney}, J., {Roettiger}, K., {Burns}, J.~O., \& {Bird}, C.~M. 1996, \apjs,
  104, 1

\bibitem[{{R Development Core Team}(2015)}]{R}
{R Development Core Team}. 2015, R: A Language and Environment for Statistical
  Computing, R Foundation for Statistical Computing, Vienna, Austria, {ISBN}
  3-900051-07-0

\bibitem[{{Rajpurohit} {et~al.}(2018){Rajpurohit}, {Hoeft}, {van Weeren},
  {Rudnick}, {R{\"o}ttgering}, {Forman}, {Br{\"u}ggen}, {Croston},
  {Andrade-Santos}, {Dawson}, {Intema}, {Kraft}, {Jones}, \&
  {Jee}}]{rajpurohit2018}
{Rajpurohit}, K., {Hoeft}, M., {van Weeren}, R.~J., {et~al.} 2018, \apj, 852,
  65

\bibitem[{{Rajpurohit} {et~al.}(2021){Rajpurohit}, {Vazza}, {van Weeren},
  {Hoeft}, {Brienza}, {Bonnassieux}, {Riseley}, {Brunetti}, {Bonafede},
  {Br{\"u}ggen}, {Formann}, {Rajpurohit}, {R{\"o}ttgering}, {Drabent},
  {Dom{\'\i}nguez-Fern{\'a}ndez}, {Wittor}, \&
  {Andrade-Santos}}]{rajpurohit2021}
{Rajpurohit}, K., {Vazza}, F., {van Weeren}, R.~J., {et~al.} 2021, \aap, 654,
  A41

\bibitem[{{Ramatsoku} {et~al.}(2020){Ramatsoku}, {Murgia}, {Vacca}, {Serra},
  {Makhathini}, {Govoni}, {Smirnov}, {Andati}, {de Blok}, {J{\'o}zsa},
  {Kamphuis}, {Kleiner}, {Maccagni}, {Moln{\'a}r}, {Ramaila}, {Thorat}, \&
  {White}}]{ramatsoku2020}
{Ramatsoku}, M., {Murgia}, M., {Vacca}, V., {et~al.} 2020, \aap, 636, L1

\bibitem[{{Randall} {et~al.}(2011){Randall}, {Forman}, {Giacintucci}, {Nulsen},
  {Sun}, {Jones}, {Churazov}, {David}, {Kraft}, {Donahue}, {Blanton},
  {Simionescu}, \& {Werner}}]{randall2011}
{Randall}, S.~W., {Forman}, W.~R., {Giacintucci}, S., {et~al.} 2011, \apj, 726,
  86

\bibitem[{{Randall} {et~al.}(2009){Randall}, {Jones}, {Markevitch}, {Blanton},
  {Nulsen}, \& {Forman}}]{randall2009}
{Randall}, S.~W., {Jones}, C., {Markevitch}, M., {et~al.} 2009, \apj, 700, 1404

\bibitem[{{Richard-Laferri{\`e}re} {et~al.}(2020){Richard-Laferri{\`e}re},
  {Hlavacek-Larrondo}, {Nemmen}, {Rhea}, {Taylor}, {Prasow-{\'E}mond},
  {Gendron-Marsolais}, {Latulippe}, {Edge}, {Fabian}, {Sanders}, {Hogan}, \&
  {Demontigny}}]{richardlaferrire2020}
{Richard-Laferri{\`e}re}, A., {Hlavacek-Larrondo}, J., {Nemmen}, R.~S.,
  {et~al.} 2020, \mnras, 499, 2934

\bibitem[{{Roediger} {et~al.}(2011){Roediger}, {Br{\"u}ggen}, {Simionescu},
  {B{\"o}hringer}, {Churazov}, \& {Forman}}]{roediger2011}
{Roediger}, E., {Br{\"u}ggen}, M., {Simionescu}, A., {et~al.} 2011, \mnras,
  413, 2057

\bibitem[{{Roediger} {et~al.}(2012){Roediger}, {Kraft}, {Machacek}, {Forman},
  {Nulsen}, {Jones}, \& {Murray}}]{roediger2012}
{Roediger}, E., {Kraft}, R.~P., {Machacek}, M.~E., {et~al.} 2012, \apj, 754,
  147

\bibitem[{{Saikia} \& {Jamrozy}(2009)}]{saikia2009}
{Saikia}, D.~J. \& {Jamrozy}, M. 2009, Bulletin of the Astronomical Society of
  India, 37, 63

\bibitem[{{Sanders} {et~al.}(2016){Sanders}, {Fabian}, {Taylor}, {Russell},
  {Blundell}, {Canning}, {Hlavacek-Larrondo}, {Walker}, \&
  {Grimes}}]{sanders2016}
{Sanders}, J.~S., {Fabian}, A.~C., {Taylor}, G.~B., {et~al.} 2016, \mnras, 457,
  82

\bibitem[{{Sato} {et~al.}(2009){Sato}, {Matsushita}, {Ishisaki}, {Yamasaki},
  {Ishida}, \& {Ohashi}}]{sato2009}
{Sato}, K., {Matsushita}, K., {Ishisaki}, Y., {et~al.} 2009, \pasj, 61, S353

\bibitem[{{Scaife} \& {Heald}(2012)}]{scaife2012}
{Scaife}, A.~M.~M. \& {Heald}, G.~H. 2012, \mnras, 423, L30

\bibitem[{{Schoenmakers} {et~al.}(2000){Schoenmakers}, {de Bruyn},
  {R{\"o}ttgering}, {van der Laan}, \& {Kaiser}}]{schoenmakers2000}
{Schoenmakers}, A.~P., {de Bruyn}, A.~G., {R{\"o}ttgering}, H.~J.~A., {van der
  Laan}, H., \& {Kaiser}, C.~R. 2000, \mnras, 315, 371

\bibitem[{{Schwarz}(1978)}]{schwartz78}
{Schwarz}, G. 1978, Ann. Statist., 6, 461

\bibitem[{{Sheardown} {et~al.}(2019){Sheardown}, {Fish}, {Roediger}, {Hunt},
  {ZuHone}, {Su}, {Kraft}, {Nulsen}, {Churazov}, {Forman}, {Jones}, {Lyskova},
  {Eckert}, \& {De Grandi}}]{sheardown2019}
{Sheardown}, A., {Fish}, T.~M., {Roediger}, E., {et~al.} 2019, \apj, 874, 112

\bibitem[{{Shimwell} {et~al.}(2017){Shimwell}, {R{\"o}ttgering}, {Best},
  {Williams}, {Dijkema}, {de Gasperin}, {Hardcastle}, {Heald}, {Hoang},
  {Horneffer}, {Intema}, {Mahony}, {Mandal}, {Mechev}, {Morabito}, {Oonk},
  {Rafferty}, {Retana-Montenegro}, {Sabater}, {Tasse}, {van Weeren},
  {Br{\"u}ggen}, {Brunetti}, {Chy{\.z}y}, {Conway}, {Haverkorn}, {Jackson},
  {Jarvis}, {McKean}, {Miley}, {Morganti}, {White}, {Wise}, {van Bemmel},
  {Beck}, {Brienza}, {Bonafede}, {Calistro Rivera}, {Cassano}, {Clarke},
  {Cseh}, {Deller}, {Drabent}, {van Driel}, {Engels}, {Falcke}, {Ferrari},
  {Fr{\"o}hlich}, {Garrett}, {Harwood}, {Heesen}, {Hoeft}, {Horellou},
  {Israel}, {Kapi{\'n}ska}, {Kunert-Bajraszewska}, {McKay}, {Mohan},
  {Orr{\'u}}, {Pizzo}, {Prandoni}, {Schwarz}, {Shulevski}, {Sipior}, {Smith},
  {Sridhar}, {Steinmetz}, {Stroe}, {Varenius}, {van der Werf}, {Zensus}, \&
  {Zwart}}]{shimwell2017}
{Shimwell}, T.~W., {R{\"o}ttgering}, H.~J.~A., {Best}, P.~N., {et~al.} 2017,
  \aap, 598, A104

\bibitem[{{Shimwell} {et~al.}(2019){Shimwell}, {Tasse}, {Hardcastle}, {Mechev},
  {Williams}, {Best}, {R{\"o}ttgering}, {Callingham}, {Dijkema}, {de Gasperin},
  {Hoang}, {Hugo}, {Mirmont}, {Oonk}, {Prandoni}, {Rafferty}, {Sabater},
  {Smirnov}, {van Weeren}, {White}, {Atemkeng}, {Bester}, {Bonnassieux},
  {Br{\"u}ggen}, {Brunetti}, {Chy{\.z}y}, {Cochrane}, {Conway}, {Croston},
  {Danezi}, {Duncan}, {Haverkorn}, {Heald}, {Iacobelli}, {Intema}, {Jackson},
  {Jamrozy}, {Jarvis}, {Lakhoo}, {Mevius}, {Miley}, {Morabito}, {Morganti},
  {Nisbet}, {Orr{\'u}}, {Perkins}, {Pizzo}, {Schrijvers}, {Smith}, {Vermeulen},
  {Wise}, {Alegre}, {Bacon}, {van Bemmel}, {Beswick}, {Bonafede}, {Botteon},
  {Bourke}, {Brienza}, {Calistro Rivera}, {Cassano}, {Clarke}, {Conselice},
  {Dettmar}, {Drabent}, {Dumba}, {Emig}, {En{\ss}lin}, {Ferrari}, {Garrett},
  {G{\'e}nova-Santos}, {Goyal}, {G{\"u}rkan}, {Hale}, {Harwood}, {Heesen},
  {Hoeft}, {Horellou}, {Jackson}, {Kokotanekov}, {Kondapally},
  {Kunert-Bajraszewska}, {Mahatma}, {Mahony}, {Mandal}, {McKean}, {Merloni},
  {Mingo}, {Miskolczi}, {Mooney}, {Nikiel-Wroczy{\'n}ski}, {O'Sullivan},
  {Quinn}, {Reich}, {Roskowi{\'n}ski}, {Rowlinson}, {Savini}, {Saxena},
  {Schwarz}, {Shulevski}, {Sridhar}, {Stacey}, {Urquhart}, {van der Wiel},
  {Varenius}, {Webster}, \& {Wilber}}]{shimwell2019}
{Shimwell}, T.~W., {Tasse}, C., {Hardcastle}, M.~J., {et~al.} 2019, \aap, 622,
  A1

\bibitem[{{Shulevski} {et~al.}(2017){Shulevski}, {Morganti}, {Harwood},
  {Barthel}, {Jamrozy}, {Brienza}, {Brunetti}, {R{\"o}ttgering}, {Murgia},
  {White}, {Croston}, \& {Br{\"u}ggen}}]{shulevski2017}
{Shulevski}, A., {Morganti}, R., {Harwood}, J.~J., {et~al.} 2017, \aap, 600,
  A65

\bibitem[{{Simionescu} {et~al.}(2010){Simionescu}, {Werner}, {Forman},
  {Miller}, {Takei}, {B{\"o}hringer}, {Churazov}, \& {Nulsen}}]{simionescu2010}
{Simionescu}, A., {Werner}, N., {Forman}, W.~R., {et~al.} 2010, \mnras, 405, 91

\bibitem[{{Skrutskie} {et~al.}(2006){Skrutskie}, {Cutri}, {Stiening},
  {Weinberg}, {Schneider}, {Carpenter}, {Beichman}, {Capps}, {Chester},
  {Elias}, {Huchra}, {Liebert}, {Lonsdale}, {Monet}, {Price}, {Seitzer},
  {Jarrett}, {Kirkpatrick}, {Gizis}, {Howard}, {Evans}, {Fowler}, {Fullmer},
  {Hurt}, {Light}, {Kopan}, {Marsh}, {McCallon}, {Tam}, {Van Dyk}, \&
  {Wheelock}}]{skrutskie2006}
{Skrutskie}, M.~F., {Cutri}, R.~M., {Stiening}, R., {et~al.} 2006, \aj, 131,
  1163

\bibitem[{{Slee} {et~al.}(2001){Slee}, {Roy}, {Murgia}, {Andernach}, \&
  {Ehle}}]{slee2001}
{Slee}, O.~B., {Roy}, A.~L., {Murgia}, M., {Andernach}, H., \& {Ehle}, M. 2001,
  \aj, 122, 1172

\bibitem[{{Smirnov} \& {Tasse}(2015)}]{smirnov2015}
{Smirnov}, O.~M. \& {Tasse}, C. 2015, \mnras, 449, 2668

\bibitem[{{Smith} {et~al.}(2001){Smith}, {Brickhouse}, {Liedahl}, \&
  {Raymond}}]{Smith2001}
{Smith}, R.~K., {Brickhouse}, N.~S., {Liedahl}, D.~A., \& {Raymond}, J.~C.
  2001, \apjl, 556, L91

\bibitem[{{Sun} {et~al.}(2007){Sun}, {Jones}, {Forman}, {Vikhlinin}, {Donahue},
  \& {Voit}}]{sun2007}
{Sun}, M., {Jones}, C., {Forman}, W., {et~al.} 2007, \apj, 657, 197

\bibitem[{{Tasse}(2014)}]{tasse2014}
{Tasse}, C. 2014, \aap, 566, A127

\bibitem[{{Tasse} {et~al.}(2018){Tasse}, {Hugo}, {Mirmont}, {Smirnov},
  {Atemkeng}, {Bester}, {Hardcastle}, {Lakhoo}, {Perkins}, \&
  {Shimwell}}]{tasse2018}
{Tasse}, C., {Hugo}, B., {Mirmont}, M., {et~al.} 2018, \aap, 611, A87

\bibitem[{{Tasse} {et~al.}(2021){Tasse}, {Shimwell}, {Hardcastle},
  {O'Sullivan}, {van Weeren}, {Best}, {Bester}, {Hugo}, {Smirnov}, {Sabater},
  {Calistro-Rivera}, {de Gasperin}, {Morabito}, {R{\"o}ttgering}, {Williams},
  {Bonato}, {Bondi}, {Botteon}, {Br{\"u}ggen}, {Brunetti}, {Chy{\.z}y},
  {Garrett}, {G{\"u}rkan}, {Jarvis}, {Kondapally}, {Mandal}, {Prandoni},
  {Repetti}, {Retana-Montenegro}, {Schwarz}, {Shulevski}, \&
  {Wiaux}}]{tasse2021}
{Tasse}, C., {Shimwell}, T., {Hardcastle}, M.~J., {et~al.} 2021, \aap, 648, A1

\bibitem[{{Tully}(2015)}]{tully2015}
{Tully}, R.~B. 2015, \aj, 149, 171

\bibitem[{{van Haarlem} {et~al.}(2013){van Haarlem}, {Wise}, {Gunst}, {Heald},
  {McKean}, {Hessels}, {de Bruyn}, {Nijboer}, {Swinbank}, {Fallows},
  {Brentjens}, {Nelles}, {Beck}, {Falcke}, {Fender}, {H{\"o}randel},
  {Koopmans}, {Mann}, {Miley}, {R{\"o}ttgering}, {Stappers}, {Wijers},
  {Zaroubi}, {van den Akker}, {Alexov}, {Anderson}, {Anderson}, {van Ardenne},
  {Arts}, {Asgekar}, {Avruch}, {Batejat}, {B{\"a}hren}, {Bell}, {Bell}, {van
  Bemmel}, {Bennema}, {Bentum}, {Bernardi}, {Best}, {B{\^i}rzan}, {Bonafede},
  {Boonstra}, {Braun}, {Bregman}, {Breitling}, {van de Brink}, {Broderick},
  {Broekema}, {Brouw}, {Br{\"u}ggen}, {Butcher}, {van Cappellen}, {Ciardi},
  {Coenen}, {Conway}, {Coolen}, {Corstanje}, {Damstra}, {Davies}, {Deller},
  {Dettmar}, {van Diepen}, {Dijkstra}, {Donker}, {Doorduin}, {Dromer}, {Drost},
  {van Duin}, {Eisl{\"o}ffel}, {van Enst}, {Ferrari}, {Frieswijk}, {Gankema},
  {Garrett}, {de Gasperin}, {Gerbers}, {de Geus}, {Grie{\ss}meier}, {Grit},
  {Gruppen}, {Hamaker}, {Hassall}, {Hoeft}, {Holties}, {Horneffer}, {van der
  Horst}, {van Houwelingen}, {Huijgen}, {Iacobelli}, {Intema}, {Jackson},
  {Jelic}, {de Jong}, {Juette}, {Kant}, {Karastergiou}, {Koers}, {Kollen},
  {Kondratiev}, {Kooistra}, {Koopman}, {Koster}, {Kuniyoshi}, {Kramer},
  {Kuper}, {Lambropoulos}, {Law}, {van Leeuwen}, {Lemaitre}, {Loose}, {Maat},
  {Macario}, {Markoff}, {Masters}, {McFadden}, {McKay-Bukowski}, {Meijering},
  {Meulman}, {Mevius}, {Middelberg}, {Millenaar}, {Miller-Jones}, {Mohan},
  {Mol}, {Morawietz}, {Morganti}, {Mulcahy}, {Mulder}, {Munk}, {Nieuwenhuis},
  {van Nieuwpoort}, {Noordam}, {Norden}, {Noutsos}, {Offringa}, {Olofsson},
  {Omar}, {Orr{\'u}}, {Overeem}, {Paas}, {Pandey-Pommier}, {Pandey}, {Pizzo},
  {Polatidis}, {Rafferty}, {Rawlings}, {Reich}, {de Reijer}, {Reitsma},
  {Renting}, {Riemers}, {Rol}, {Romein}, {Roosjen}, {Ruiter}, {Scaife}, {van
  der Schaaf}, {Scheers}, {Schellart}, {Schoenmakers}, {Schoonderbeek},
  {Serylak}, {Shulevski}, {Sluman}, {Smirnov}, {Sobey}, {Spreeuw}, {Steinmetz},
  {Sterks}, {Stiepel}, {Stuurwold}, {Tagger}, {Tang}, {Tasse}, {Thomas},
  {Thoudam}, {Toribio}, {van der Tol}, {Usov}, {van Veelen}, {van der Veen},
  {ter Veen}, {Verbiest}, {Vermeulen}, {Vermaas}, {Vocks}, {Vogt}, {de Vos},
  {van der Wal}, {van Weeren}, {Weggemans}, {Weltevrede}, {White}, {Wijnholds},
  {Wilhelmsson}, {Wucknitz}, {Yatawatta}, {Zarka}, {Zensus}, \& {van
  Zwieten}}]{vanhaarlem2013}
{van Haarlem}, M.~P., {Wise}, M.~W., {Gunst}, A.~W., {et~al.} 2013, \aap, 556,
  A2

\bibitem[{{van Weeren} {et~al.}(2019){van Weeren}, {de Gasperin}, {Akamatsu},
  {Br{\"u}ggen}, {Feretti}, {Kang}, {Stroe}, \& {Zandanel}}]{vanweeren2019}
{van Weeren}, R.~J., {de Gasperin}, F., {Akamatsu}, H., {et~al.} 2019, \ssr,
  215, 16

\bibitem[{{van Weeren} {et~al.}(2021){van Weeren}, {Shimwell}, {Botteon},
  {Brunetti}, {Br{\"u}ggen}, {Boxelaar}, {Cassano}, {Di Gennaro},
  {Andrade-Santos}, {Bonnassieux}, {Bonafede}, {Cuciti}, {Dallacasa}, {de
  Gasperin}, {Gastaldello}, {Hardcastle}, {Hoeft}, {Kraft}, {Mandal},
  {Rossetti}, {R{\"o}ttgering}, {Tasse}, \& {Wilber}}]{vanweeren2020}
{van Weeren}, R.~J., {Shimwell}, T.~W., {Botteon}, A., {et~al.} 2021, \aap,
  651, A115

\bibitem[{{van Weeren} {et~al.}(2016){van Weeren}, {Williams}, {Hardcastle},
  {Shimwell}, {Rafferty}, {Sabater}, {Heald}, {Sridhar}, {Dijkema}, {Brunetti},
  {Br{\"u}ggen}, {Andrade-Santos}, {Ogrean}, {R{\"o}ttgering}, {Dawson},
  {Forman}, {de Gasperin}, {Jones}, {Miley}, {Rudnick}, {Sarazin}, {Bonafede},
  {Best}, {B{\^i}rzan}, {Cassano}, {Chy{\.z}y}, {Croston}, {Ensslin},
  {Ferrari}, {Hoeft}, {Horellou}, {Jarvis}, {Kraft}, {Mevius}, {Intema},
  {Murray}, {Orr{\'u}}, {Pizzo}, {Simionescu}, {Stroe}, {van der Tol}, \&
  {White}}]{vanweeren2016}
{van Weeren}, R.~J., {Williams}, W.~L., {Hardcastle}, M.~J., {et~al.} 2016,
  \apjs, 223, 2

\bibitem[{{Vantyghem} {et~al.}(2014){Vantyghem}, {McNamara}, {Russell}, {Main},
  {Nulsen}, {Wise}, {Hoekstra}, \& {Gitti}}]{vantyghem2014}
{Vantyghem}, A.~N., {McNamara}, B.~R., {Russell}, H.~R., {et~al.} 2014, \mnras,
  442, 3192

\bibitem[{{Vazza} {et~al.}(2021){Vazza}, {Wittor}, {Brunetti}, \&
  {Br{\"u}ggen}}]{vazza2021}
{Vazza}, F., {Wittor}, D., {Brunetti}, G., \& {Br{\"u}ggen}, M. 2021, \aap,
  653, A23

\bibitem[{{Walker} {et~al.}(2017){Walker}, {Hlavacek-Larrondo},
  {Gendron-Marsolais}, {Fabian}, {Intema}, {Sanders}, {Bamford}, \& {van
  Weeren}}]{walker2017}
{Walker}, S.~A., {Hlavacek-Larrondo}, J., {Gendron-Marsolais}, M., {et~al.}
  2017, \mnras, 468, 2506

\bibitem[{{Williams} {et~al.}(2016){Williams}, {van Weeren}, {R{\"o}ttgering},
  {Best}, {Dijkema}, {de Gasperin}, {Hardcastle}, {Heald}, {Prandoni},
  {Sabater}, {Shimwell}, {Tasse}, {van Bemmel}, {Br{\"u}ggen}, {Brunetti},
  {Conway}, {En{\ss}lin}, {Engels}, {Falcke}, {Ferrari}, {Haverkorn},
  {Jackson}, {Jarvis}, {Kapi{\'n}ska}, {Mahony}, {Miley}, {Morabito},
  {Morganti}, {Orr{\'u}}, {Retana-Montenegro}, {Sridhar}, {Toribio}, {White},
  {Wise}, \& {Zwart}}]{williams2016}
{Williams}, W.~L., {van Weeren}, R.~J., {R{\"o}ttgering}, H.~J.~A., {et~al.}
  2016, \mnras, 460, 2385

\bibitem[{{Willingale} {et~al.}(2013){Willingale}, {Starling}, {Beardmore},
  {Tanvir}, \& {O'Brien}}]{Willingale2013}
{Willingale}, R., {Starling}, R.~L.~C., {Beardmore}, A.~P., {Tanvir}, N.~R., \&
  {O'Brien}, P.~T. 2013, \mnras, 431, 394

\bibitem[{{ZuHone} {et~al.}(2013){ZuHone}, {Markevitch}, {Ruszkowski}, \&
  {Lee}}]{zuhone2013}
{ZuHone}, J.~A., {Markevitch}, M., {Ruszkowski}, M., \& {Lee}, D. 2013, \apj,
  762, 69

\bibitem[{{ZuHone} {et~al.}(2021){ZuHone}, {Markevitch}, {Weinberger},
  {Nulsen}, \& {Ehlert}}]{zuhone2021}
{ZuHone}, J.~A., {Markevitch}, M., {Weinberger}, R., {Nulsen}, P., \& {Ehlert},
  K. 2021, \apj, 914, 73

\bibitem[{{Zuhone} \& {Roediger}(2016)}]{zuhone2016}
{Zuhone}, J.~A. \& {Roediger}, E. 2016, Journal of Plasma Physics, 82,
  535820301

\end{thebibliography}

\begin{appendix}

\section{Two kT X-ray spectral analysis}
\label{2kT}

Fitting the spectra with a single temperature model could be too simplistic in the core of a galaxy group either because the central regions may enclose a part of the coronal gas or simply because we are dealing  with projected spectra (i.e. we include all the contributions along the line of sight). Here, we show how the results change when fitting the spectra with two absorbed APEC models. 

During the fit, we link the abundances of the two gas phases, and as in the case of a single APEC, the MOS and pn spectra were fitted simultaneously, linking both temperatures and metallicities  and allowing all normalisations to vary freely. We first checked whether or not a second component substantially improved the fit quality of a spectrum extracted within 3 arcmin (i.e. corresponding roughly to the twenty-fifth magnitude isophote D$_{25}$). Although {\it cstat} does not provide a goodness of the fit, the XSPEC version is defined to tend to $\chi^2$ in the high counts regime (which is our case). This means that the C statistic will tend to a value equal to the number of degrees of freedom for a correct model and can be used as a qualitative check for the improvement of the fit. The value returned by XSPEC decreases significantly from $\sim$2.7 to $\sim$1.4 when using single or double APEC, respectively. This result indeed suggests the presence of multiple temperature components.

\begin{figure}[!ht]
\centering
\vbox{
\includegraphics[width=0.4\textwidth]{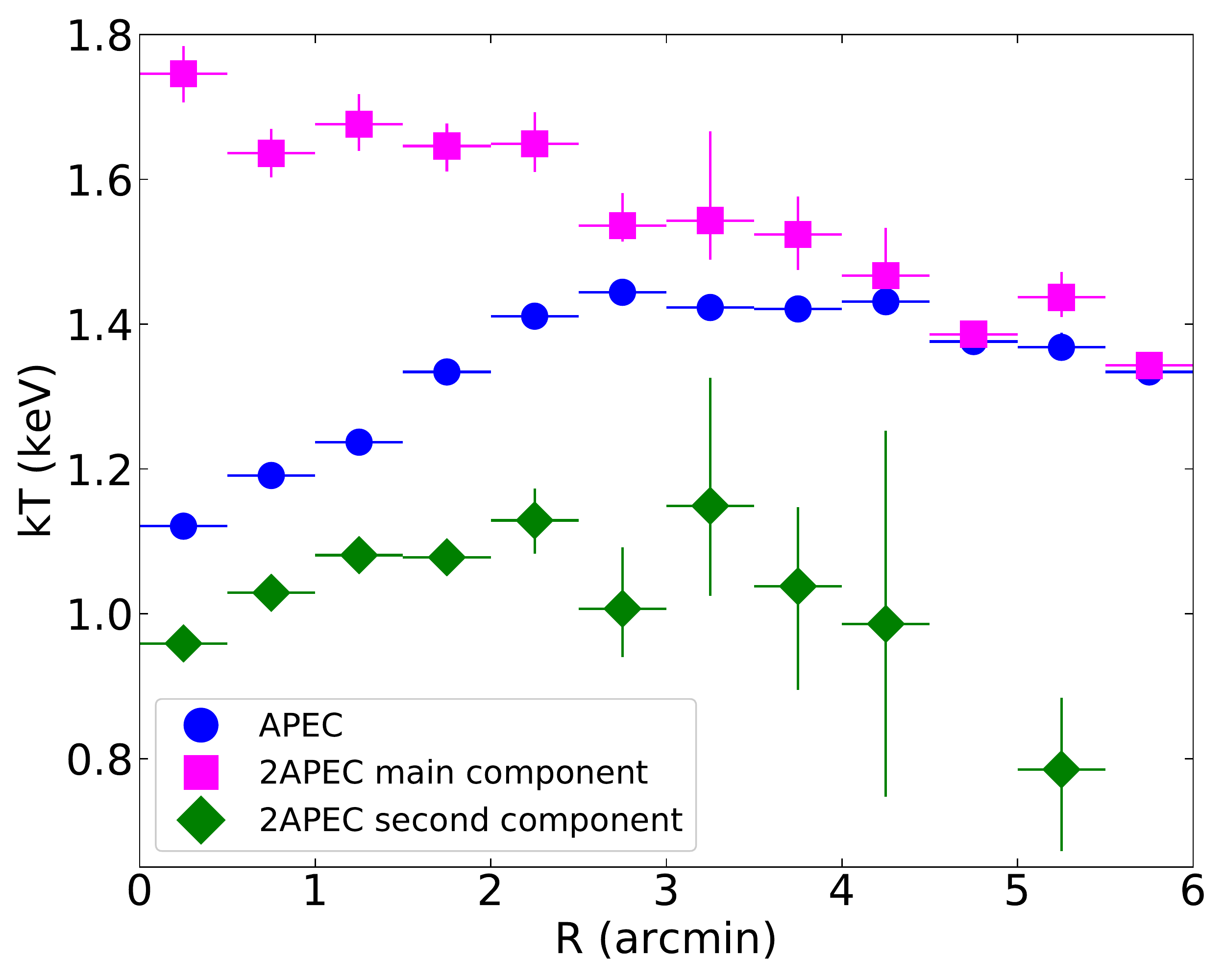}
\includegraphics[width=0.4\textwidth]{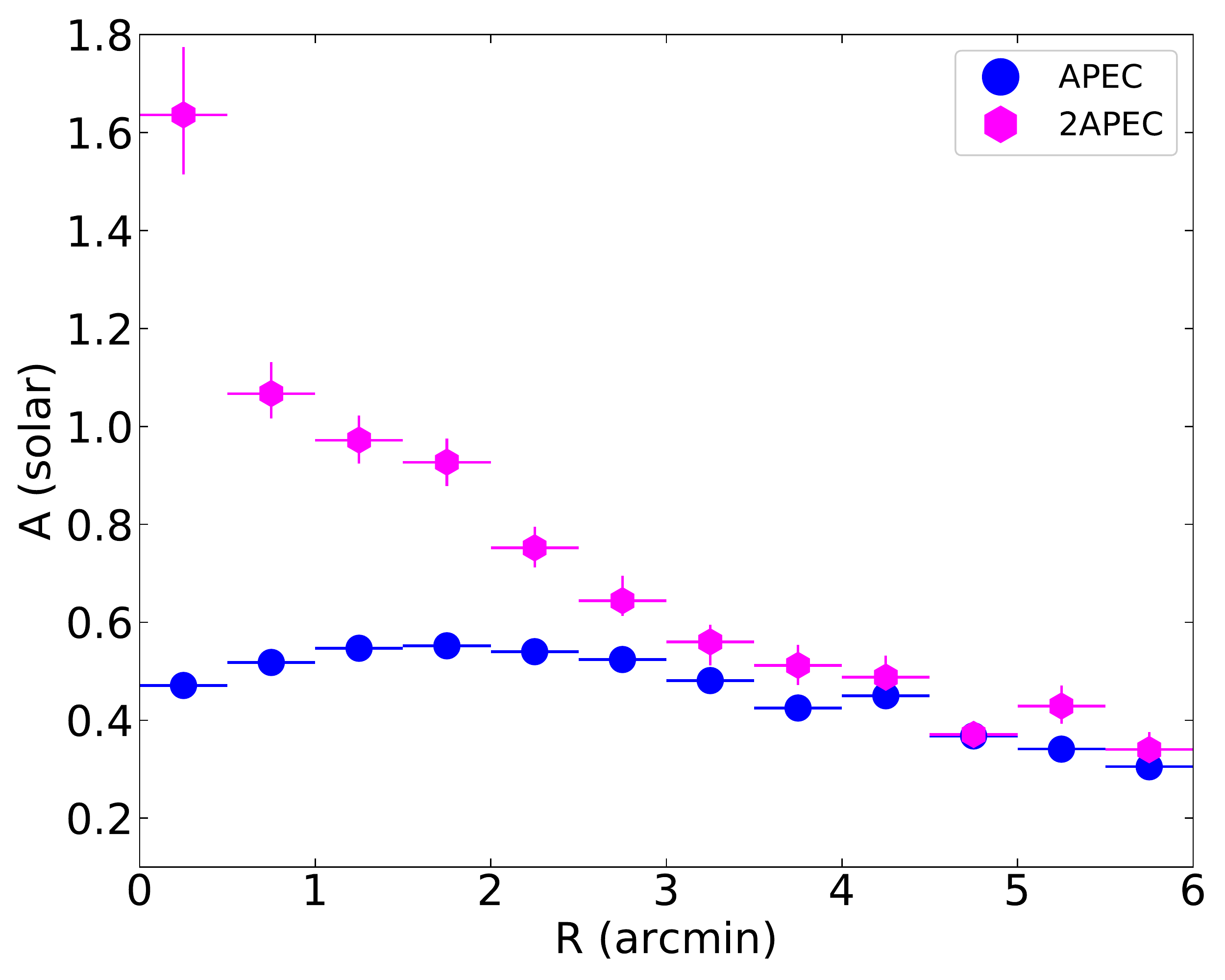}
\includegraphics[width=0.45\textwidth]{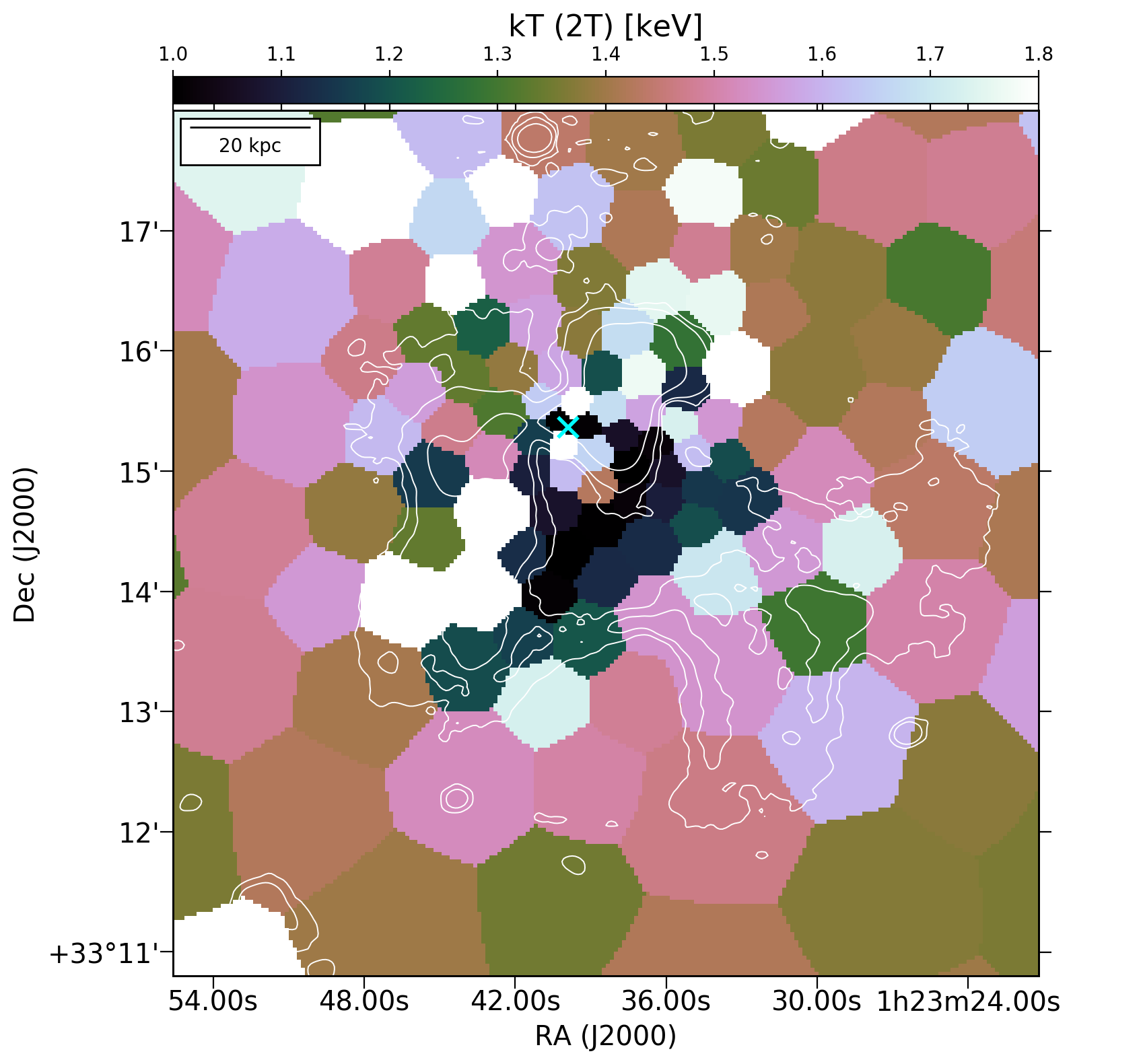}
}
\caption{Comparison between the temperature (top) and abundance (middle) profiles obtained using a single or double APEC model. In the bottom panel we show the temperature map obtained using a 2T model fit as described in the text. A cross marks the centre of the galaxy NGC~507.}
\label{fig:2apec}
\end{figure}

Within the central regions there are at least two spectral components, and so we derived the temperature and abundance profiles by fitting the two-temperature models to the data and compared the result with those obtained with a single component. In Fig. \ref{fig:2apec} we display our findings. The cool component remains nearly constant with just a mild indication of a slight decrease in the core. Instead, the hotter component rises almost monotonically toward the centre. However, it is important to note that the normalisation of the two components is comparable only within $\sim$1.5-2 arcmin and the contribution of the cool component is negligible beyond $\sim$3 arcmin. 

\begin{figure}[!ht]
\centering
\vbox{
\includegraphics[width=0.45\textwidth]{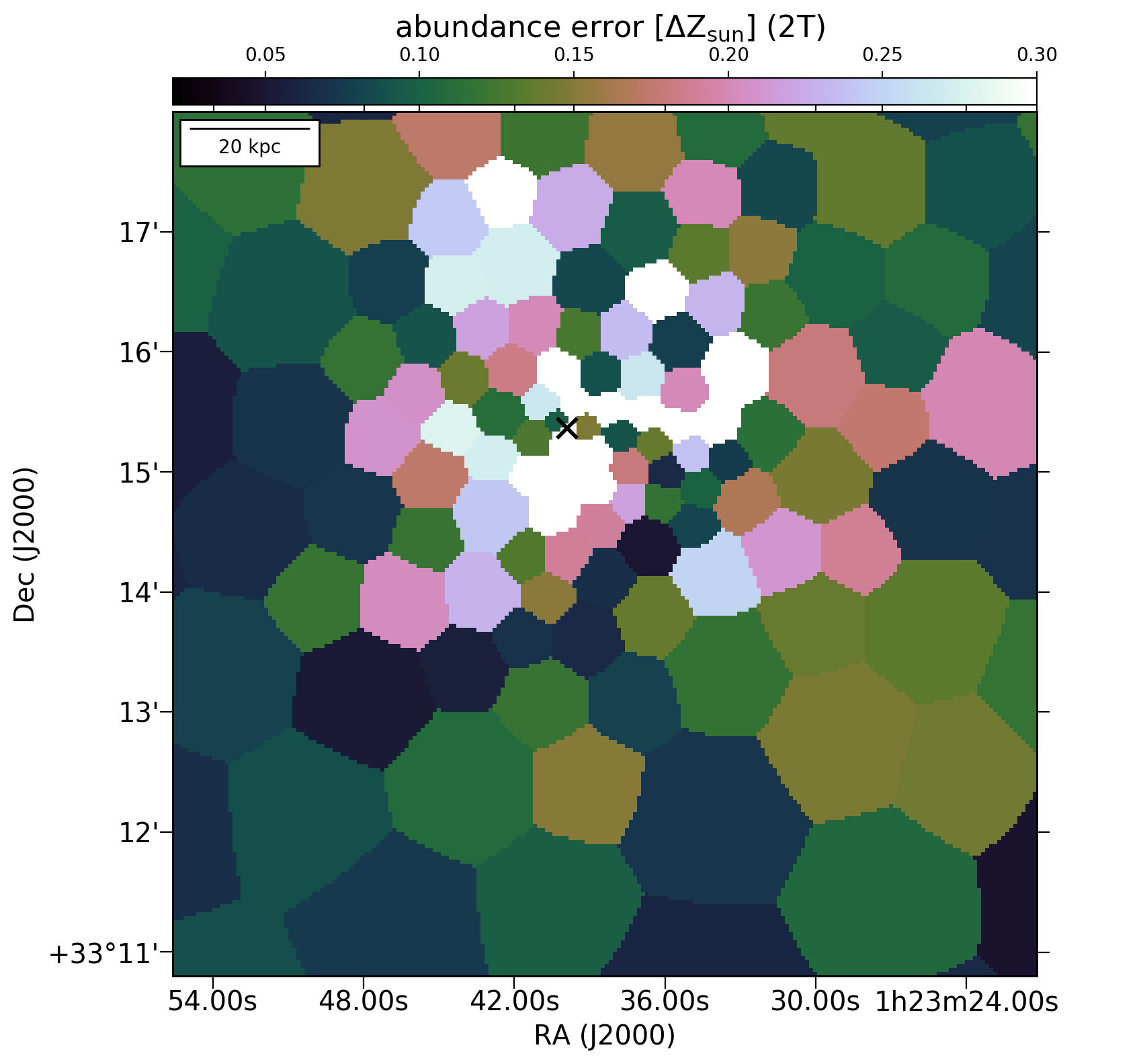}
\includegraphics[width=0.45\textwidth]{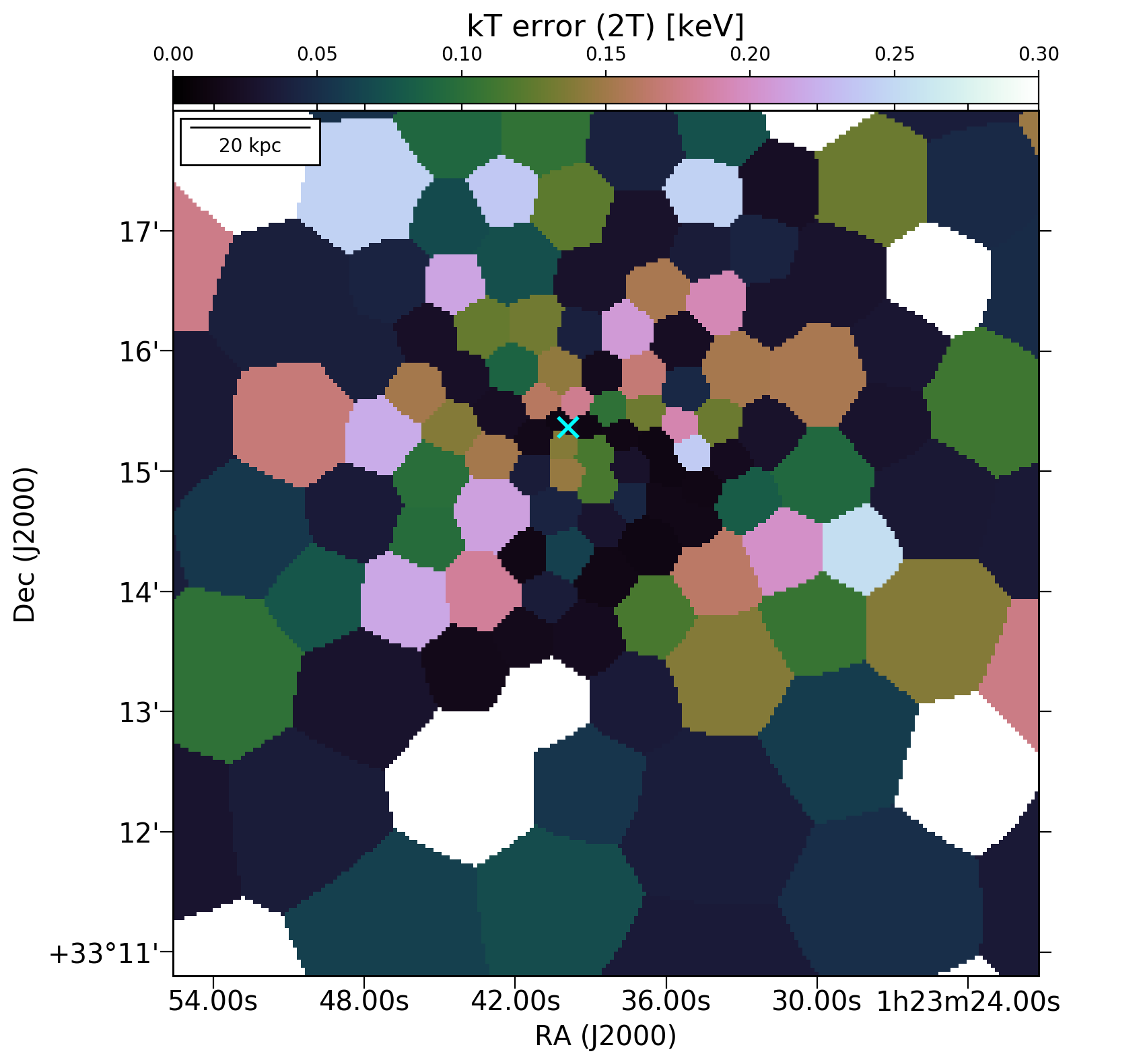}
}
\caption{Top: Abundance (top) and temperature (bottom) 1$\sigma$ error maps obtained using a 2T model as described in the text. A cross marks the centre of the galaxy NGC~507. }
\label{fig:2apecerror}
\end{figure}

In the bottom panel of Fig. \ref{fig:2apec} we show the temperature map obtained using a double APEC model. The two-temperature model that we are using returns two values for temperature and normalisation. The temperature map was built by taking the temperature associated to the gas phase with the highest normalisation (i.e. highest density) for each region. Apart from small changes within $\sim$2 arcmin, the pattern distribution is very similar to that derived with one single component. 
Based on these results, we are confident that our overall interpretation of the system and conclusions are independent of the choice of a 1T or 2T spectral model. For completeness in Fig. \ref{fig:2apecerror} we also show the the temperature and abundance 1-$\sigma$ error maps for the 2T spectral model.

\section{Optical analysis}
\label{opt_app}

In this section we present the details of the optical analysis of the galaxy group NGC~507. We selected all the galaxies with known recession velocities from the NASA/IPAC Extragalactic Database (NED) located within 60 arcmin (1.2 Mpc at the redshift of the source) from the optical position of the galaxy NGC~507. A relatively large radius for the search is considered to select galaxies members of the NGC~507 halo and at the same time investigate the presence of any possible perturbers following the scenario of \citet{ascasibar2006}. We note that NGC~499 is located at 13.7 arcmin from NGC~507 and so falls well within the adopted search radius.

The membership of the group is achieved through a two-stage approach. A first step is performed by using the `velocity gap' method outlined by \citet{depropris2002}, where the galaxies are sorted in redshift space and their velocity gap, defined for the $n$th galaxy as $\Delta v_n\;=\;cz_{n+1}-cz_{n}$, is calculated. 
Clusters and groups appear as well-populated peaks in redshift space, which are well separated by velocity gaps from the nearest foreground and background galaxies. Using this procedure and fixing the velocity gap to 1000 \kms, we clearly detect the NGC~507 group as a peak at the redshift of NGC~507 populated by 74 galaxies.

As a second step we use a slightly modified
version of the `shifting gapper' method, first employed by \citet{fadda1996}, as described by 
\citet{owers2009}. The method uses both radial and peculiar velocity information to separate
interlopers from members as a function of group-centric radius. The data are binned radially such that
each bin contains at least ten objects. Within each bin, galaxies are sorted by peculiar velocity, with velocity
gaps determined in peculiar velocity. Peculiar velocities were determined by estimating the mean group velocity
using the biweight location estimator \citep{beers1990}, which was assumed to represent the cosmological redshift
of the group, $z_{\rm{cos}}$. The peculiar redshift of the galaxy is 
$z_{\rm{pec}} = (z_{\rm{gal}} - z_{\rm{cos}})/(1 + z_{\rm{cos}})$ 
and the peculiar velocity is derived using the special re\-la\-ti\-vi\-stic formula 
$v_{\rm{pec}} = c((1 + z_{\rm{pec}})^2 - 1)/((1 + z_{\rm{pec}})^2 + 1)$, where $c$ is the speed of light.
The `f pseudosigma' \citep{beers1990} derived from the first and third quartiles of the peculiar velocity distribution
was used as the fixed gap to separate the group from the interlopers.
However, this procedure did not reject any further galaxies  with respect to the sample obtained above.

In conclusion, the final group sample consists of 74 members (see left panel of Fig.\ref{fig:507optical}). The value of the biweight location estimator of the
mean group velocity is $5030\pm74$ \kms, which corresponds to $z_{\rm{cos}}=0.01678\pm0.00025$.
We used the biweight scale estimator to estimate a velocity dispersion of $611\pm50$ \kms.
The errors for the redshift and velocity dispersion are at 1$\sigma$ and are estimated using the jackknife resampling technique.

We calculated the peculiar velocity of the central galaxy NGC 507 which is $-96\pm74$ \kms, simply adding in quadrature the errors on the galaxy \citep[using the error of 7 \kms\ as reported in the CfA redshift survey,][]{huchra1999} and group velocities. An offset velocity must lie outside a range defined by the cluster velocity dispersion to be significant \citep{gebhardt1991}.
We therefore calculated the Z-score for the central galaxy \citep[eq. (7) of][]{gebhardt1991} which is -0.175 and its
68\% confidence limit is 0.127. The confidence interval has been calculated using 10000 bootstrap resamplings and allowing 
for the measurement error of the central galaxy velocity by sampling from a Gaussian distribution with standard deviation set to the reported 
error as part of the bootstrap. The velocity offset is therefore not significant.

We then applied a series of substructure tests from 1D in the velocity space to 3D involving both spatial and velocity spaces, because the existence of correlations between position and velocities of clusters galaxies is a signature of true substructure.
This is essential for the investigation: the sensitivity of individual diagnostics depends on the line of sight relative to 
the merger axis and therefore no single substructure test can be considered the most sensitive in all situations \citep{pinkney1996}.

We started with 1D tests, analysing the velocity distribution to look for departures from Gaussianity, which can be attributed to
dynamical activity. We applied the Anderson-Darling (AD)
test as the most reliable and powerful test for detecting real departures from an underlying Gaussian distribution following \citet{hou2009}. We used the AD test as implemented in the task \emph{ad.test} of the package \emph{nortest} in version 4.1 of the R statistical software environment\footnote{http://www.r-project.org}\citep{R}. The AD test returns a  $A^{2}$ statistic of 0.42432, which corresponds to a $p$-value of $0.3101,$ 
therefore consistent with having a Gaussian distribution.
Furthermore, we estimated four shape indicators: the skewness, the kurtosis, and 
their robust counterparts, that is, the asymmetry index and the scaled tail index, the indicator comparing the spread of the dataset at the 90\% level with the spread at the 75\% level \citep{bird1993}. The values for the sample are 0.177 and -0.274 for skewness and kurtosis, respectively, and they show no departure from a Gaussian distribution \citep[see Table 2 of][for 75 data-points]{bird1993}. 
The value of the scaled tail index is 1.023 and is consistent with a Gaussian distribution.
We investigated the presence of significant gaps in the velocity distribution following the weighted gap analysis of 
\citet{beers1990} looking for normalised gaps larger than 2.25, because in random draws of a Gaussian distribution they 
arise at most in about 3\% of the cases, independently of the sample size.
We did find a significant gap with a normalised value of 2.65 between the velocities of 4761 \kms\ and 4868 \kms, separating the range of low velocities of the sample (including NGC~499) and the rest of the member galaxies.

To detect possible multiple components in velocity space we also applied a Gaussian mixture model implemented in the package \textsc{mclust} \citep{mclust1,mclust2} in R. This is an algorithm for fitting normal mixture models, i.e. maximum likelihood fits are performed assuming that between 1 and 9 normal components are present in
the data. The model selection is performed by comparing the
Bayesian information criterion \citep[BIC,][]{schwartz78} defined as ${\rm{BIC}} = 2lnL - k log(n),$ where $L$
is the likelihood, $k$  the number of parameters of the model, and $n$ the number of data points; $klog(n)$ is the penalty term which compensates the difference in likelihood due to an increase in the number of fitting
parameters. The best model is the one that maximises the BIC.
The result of the mixture model favours a single Gaussian with a BIC of -1164 and there were no statistically significant partitions of the sample: for example a two-Gaussian-component model has a BIC of -1176 for unequal variances.

We then applied the 3D tests. First we adopted the Dressler-Shectman $\Delta$ statistic \citep{dressler1988}, which tests
for differences in the local mean and dispersion compared to the global mean and dispersion and is recommended by
\citet{pinkney1996} as the most sensitive 3D test.
Calculation of the $\Delta$ statistic involves the summation of local velocity anisotropy, $\delta$,
for each galaxy in the group, defined as

\begin{equation}
\label{eq.delta}
\delta^2 =
\left(\frac{N_{\mathrm{nn}}+1}{\sigma^2}\right)[(\bar{v}_{\mathrm{local}}-\bar{v}_{\mathrm{gro
up}})^2-(\sigma_{\mathrm{local}}-\sigma)^2],
\end{equation}

\noindent where $N_{\mathrm{nn}}$ is the number of nearest neighbours over which the local recession velocity 
($\bar{v}_{\mathrm{local}}$) and velocity dispersion ($\sigma_{\mathrm{local}}$) are calculated.
Here we adopt $N_{\mathrm{nn}}=\sqrt{N}$, following \citet{pinkney1996}. The significance of $\Delta$
is estimated using 10000 Montecarlo simulations randomly shuffling the galaxy velocities.
We did not find suggestions of substructures with the DS test (significance level of 63.5\%). 
We then applied the $\kappa$ test of \citet{colless1996}. 
The $\kappa$-test searches for local departures from the global velocity distribution around each cluster member galaxy
by using the Kolmogorov-Smirnov (KS) test which determines the likelihood that the velocity distribution of the
$\sqrt{N}$ nearest neighbours around the galaxy of interest and the global cluster velocity distribution are drawn from the same parent distribution.
Calculation of the $\kappa$ statistic involves the summation of the negative log likelihood for each galaxy defined as 

\begin{equation}
\label{eq.kappa}
\kappa_{i} = -log P_{KS} ( D > D_{obs,i}).
\end{equation}
 
\noindent Also the $\kappa$ test returned no significant substructure (at the 91.3\% level).

Finally, we applied a 3D mixture model with \textsc{mclust} using the spatial coordinates plus the velocity of each galaxy \citep[e.g.][]{einasto2012}  as input
. Similarly, in this case we did not find any significant partition in subclusters of the member galaxies.

\end{appendix}

\end{document}